\documentclass[prd,aps,preprint,tightenlines,superscriptaddress,nofootinbib,showpacs,showkeys,floatfix]{revtex4-1}
\usepackage{amsfonts}
\usepackage{array}
\usepackage{epsfig}
\usepackage{rotating}
\usepackage{multirow}
\usepackage{soul}
\usepackage{amsmath}    
\usepackage{verbatim}   
\usepackage{color}      
\usepackage{colordvi}
\usepackage{hyperref}   

\usepackage{graphicx}

\usepackage[normalem]{ulem}   

\newcommand{\nn}{\nonumber}

\graphicspath{{../pics/}}
\DeclareGraphicsExtensions{.eps,.ps}

\usepackage{pslatex}
\usepackage{txfonts}
\usepackage[latin1]{inputenc}
\usepackage[T1]{fontenc}

\begin{document}
\begin{titlepage}
\thispagestyle{empty}
\noindent
DESY 24-073
\hfill
May 2024 \\
CERN-TH-2024-068\\
ZU-TH 26/24\\
MPP-2024-106
\vspace{0.6cm}

\begin{center}
  {\bf \Large
    Status of QCD precision predictions for Drell-Yan \\[0.5ex] rapidity distributions
    }
  \vspace{1.0cm}

 {\large
   S.~Alekhin$^{\, a}$,
   S.~Amoroso$^{\, b}$,
   L.~Buonocore$^{\, c}$,
   A.~Huss$^{\, c}$,
   S.~Kallweit$^{\, d}$,
   A.~Kardos$^{\, e,f}$, 
   J.~Michel$^{\, g,h}$,
   S.~Moch$^{\, a}$,
   F.~Petriello$^{\, i,j}$,
   L.~Rottoli$^{\, d,k}$,
   Z.~Tr\'ocs\'anyi$^{\, f,e}$,
    and
   M.~Wiesemann$^{\, l}$
   \\
 }
 \vspace{0.8cm}
 {\footnotesize{
     \begin{flushleft}       
 {\it
   $^a$ II. Institut f\"ur Theoretische Physik, Universit\"at Hamburg,
   Luruper Chaussee 149, 22761 Hamburg, Germany \\

   $^b$ Deutsches Elektronen-Synchrotron DESY, Notkestr. 85, 22607 Hamburg, Germany\\

   $^c$ Theoretical Physics Department, CERN, CH-1211 Geneva 23, Switzerland\\

   $^d$ Physik Institut, Universit\"at Z\"urich, CH-8057 Z\"urich, Switzerland\\

   $^e$ Department of Experimental Physics, Institute of Physics, Faculty of Science and Technology, University of Debrecen, 
   4010 Debrecen, PO Box 105, Hungary \\

   $^f$ Institute for Theoretical Physics, ELTE E\"otv\"os Lor\'and University,
   P\'azm\'any P\'eter 1/A, H--1117 Budapest, Hungary \\

   $^g$ Institute for Theoretical Physics Amsterdam and Delta Institute for Theoretical Physics, University of Amsterdam, Science Park 904, 1098 XH Amsterdam, The Netherlands \\

   $^h$ Nikhef, Theory Group, Science Park 105, 1098 XG, Amsterdam, The Netherlands \\
   
   $^i$ Department of Physics \& Astronomy, Northwestern University, Evanston, Illinois 60208, USA \\

   $^j$ HEP Division, Argonne National Laboratory, Argonne, Illinois 60439, USA\\

   $^k$ Dipartimento di Fisica G. Occhialini, Universit\`a degli Studi di Milano-Bicocca and INFN, Sezione di Milano-Bicocca,
   Piazza della Scienza, 3, 20126 Milano, Italy \\
   
   $^l$ Max-Planck-Institut f\"{u}r Physik, Boltzmannstra\ss{}e 8, 85748 Garching, Germany\\

 }
     \end{flushleft}
}}
  \vspace{0.8cm}
  \large {\bf Abstract}
  \vspace{-0.2cm}
\end{center}
We compute differential distributions for Drell-Yan processes at the LHC and the Tevatron colliders at next-to-next-to-leading order in perturbative QCD, 
including fiducial cuts on the decay leptons in the final state.
The comparison of predictions obtained with four different codes shows excellent agreement, once linear power corrections from the fiducial cuts are included in those codes that rely on phase-space slicing subtraction schemes.
For $Z$-boson production we perform a detailed study of the symmetric cuts on the transverse momenta of the decay leptons.
Predictions at fixed order in perturbative QCD for those symmetric cuts, typically imposed in experiments, suffer from an instability. 
We show how this can be remedied by an all-order resummation of the fiducial transverse momentum spectrum, 
and we comment on the choice of cuts for future experimental analyses.
\end{titlepage}

\newpage
\setcounter{footnote}{0}
\setcounter{page}{1}

\section{Introduction}
\label{sec:intro}

Drell-Yan lepton pair production processes are among the most important hard scattering events at the LHC. The measured final state contains only leptons. As a result, the corresponding cross sections are known experimentally very precisely. For example, the transverse momentum distribution of Drell-Yan lepton pairs reaches a precision of 0.2\,\% for the normalized spectra at low values of $p_T^{(\ell\ell)}$ \cite{ATLAS:2019zci,CMS:2018mdl}. 
The importance of the process is shown by its frequent use in precision extraction of the parameters in the Standard Model, such as parton distribution functions (PDFs) and the strong coupling constant \cite{Alekhin:2017kpj,Bailey:2020ooq,Ball:2018iqk}. It is also used in the determination of the mass of the $W^{\pm}$-boson \cite{ATLAS:2017rzl,LHCb:2021bjt}. 
All these measurements require a reduction of the theoretical uncertainties to match the experimental ones. The current state of the art in the theory description has advanced significantly in recent years. 
It has reached next-to-next-to-next-to-leading order (N$^3$LO) accuracy at fixed order in quantum chromodynamics (QCD) perturbation theory \cite{Duhr:2020sdp,Duhr:2021vwj} at the inclusive level. 
Transverse momentum resummation is known at the next-to-next-to-next-to-leading logarithmic (N$^3$LL) level \cite{Camarda:2021ict,Re:2021con,Ju:2021lah,Chen:2022cgv} and even at approximate N$^4$LL \cite{Neumann:2022lft,Camarda:2023dqn} accuracy for differential distributions in perturbation theory. 
At present, the fully differential calculations at N$^3$LO accuracy employ transverse momentum subtractions~\cite{Catani:2007vq} that neglect power corrections and necessarily rely on predictions for vector boson + jet production at the N$^2$LO accuracy. Hence, the reliable computation of cross sections at N$^2$LO in QCD is a prerequisite for further developments in the field.

In Ref.~\cite{Alekhin:2021xcu} a subset of the present authors published a detailed comparison of the publicly available codes \cite{Catani:2007vq,Catani:2009sm,Gavin:2012sy,Boughezal:2015dva} for $W^{\pm}$- and $Z$-boson production, including their decay. They found differences among the predictions at the NNLO level, whose size depended on the observable. 
The differences were estimated as similar to and sometimes even larger than the sizes of the NNLO QCD corrections themselves. This observation suggested that the neglected power corrections in transverse momentum subtraction, and other methods that rely upon phase-space slicing to regulate real emissions, could be the source of the differences. Depending on the fiducial cuts, those power corrections become linear, and hence not negligible.

The publication triggered discussions among the authors of the relevant codes, which resulted in a better understanding of the neglected terms and improvements in the computations. In this paper we provide an update of the comparisons carried out in Ref.~\cite{Alekhin:2021xcu}. The authors of the codes have provided new predictions for the benchmark calculations that we present in Sec.~\ref{sec:benchmark}, now showing excellent agreement. Based on this validation of fixed-order perturbative QCD calculations through NNLO we then study the impact of fiducial cuts on the decay leptons.
To that end, we focus on the case of symmetric cuts on the transverse momenta of the leptons, as they are routinely imposed in experimental analyses but display certain unphysical features~\cite{Frixione:1997ks}.
In Sec.~\ref{sec:lepton-cuts}, we study cuts on the transverse momenta staggered in a range from a few tens of MeV to a few GeV in perturbative QCD, both at fixed order and applying all-orders resummation of large lepton-pair transverse momentum logarithms. We comment on proposed modifications of the fiducial cuts put forward recently~\cite{Salam:2021tbm}. Since the pathology observed in fixed-order perturbation theory arises from the region of small lepton-pair transverse momentum we can cure it with resummation of the small $p_T$ region. We do so in Section~\ref{sec:resummation} and find only small differences between the resummed results and the available fixed-order codes. We provide a short discussion of the experimental resolution to gauge the impact of our findings on current analyses in Section~\ref{sec:exp-resolution}. We summarize in Sec.~\ref{sec:summary} and conclude with our comments on the choice of cuts for experimental Drell-Yan analyses.
Cross sections and information on the set-up and input parameters for some of the codes used for the benchmark comparisons are listed in the Appendices~\ref{sec:appXS}--\ref{sec:appE}.

\section{Benchmark computations}
\label{sec:benchmark}

\subsection{Set-up and code validation}

The set-up and validation for benchmarking theory predictions for $W^{\pm}$- and
$Z$-boson hadro-production cross sections up to NNLO in QCD were described in Ref.~\cite{Alekhin:2021xcu}. In order to study the effect of cuts on the fiducial phase space in the experimental measurement 
two sets of data on $W^{\pm}$- and $Z$-boson production have been chosen, 
collected at the LHC by the ATLAS experiment and at the Tevatron by the D{\O} experiment, respectively. 
\begin{itemize}
\item 
  (Pseudo-)rapidity distributions for the decay leptons for the $W^\pm$- and $Z/\gamma^*$-production cross
  sections \cite{Aaboud:2016btc} measured by the ATLAS experiment at a center-of-mass energy of
  $\sqrt{s}=7$\ TeV, where
  the leptonic transverse momenta $p_T^\ell$ and pseudo-rapidities are subject to fiducial cuts.

\item Distributions in the electron pseudo-rapidity for the electron charge 
  asymmetry measured by the D{\O} experiment in $W^\pm$-boson production at $\sqrt{s}=1.96$\ TeV at the 
  Tevatron~\cite{D0:2014kma}.
  The D{\O} data is taken with fiducial cuts on the transverse momenta
  $p_T^{e,\nu_e}$ of the electron and the missing energy, both symmetric as well as staggered, and on their pseudo-rapidities.
\end{itemize}
The above choices are representative of the numerous data sets collected at the LHC and the Tevatron in which symmetric cuts on the final-state lepton phase space are imposed.
In our theoretical predictions we use the $G_\mu$ scheme with input values $G_F$, $M_Z$, $M_W$.
The QED coupling $\alpha(M_Z)$ and $\sin^2(\theta_W)$ are then output values, 
which minimizes the impact of NLO electroweak corrections, see e.g.\ Ref.~\cite{Dittmaier:2009cr}.
The SM input parameters are~\cite{pdg2020}
\begin{equation}
    \begin{array}{ll}
\label{eq:input-masses}
~~G_{\mu} = 1.16637\times 10^{-5} \, {\rm GeV}^{-2} \, , 
\\ 
~M_Z = 91.1876 \, {\rm GeV}\, , \quad\quad   &~\Gamma_{Z} =  2.4952  \, {\rm GeV} \, , 
\\ 
M_W = 80.379\, {\rm GeV}\, , \quad\quad  
&\Gamma_W = 2.085\, {\rm GeV} \, ,
\end{array}
\end{equation}
and the relevant CKM parameters are
\begin{eqnarray}
\label{eq:input-ckm}
|V_{ud}| \,=\, 0.97401\, , & \quad & |V_{us}| \,=\, 0.2265\phantom{0}  \, ,
\nonumber \\
|V_{cd}| \,=\, 0.2265\phantom{0} \, , & \quad & |V_{cs}| \,=\, 0.97320 \, ,
\nonumber \\
|V_{ub}| \,=\, 0.00361\, , & \quad & |V_{cb}| \,=\, 0.04053 \, .
\end{eqnarray}
The computations are performed in the $\overline{\rm MS}$ factorization scheme
with $n_f=5$ light flavors with the $n_f=5$ flavor PDFs of
ABMP16~\cite{Alekhin:2017kpj,Alekhin:2018pai} as an input 
and the value of the strong coupling, $\alpha_s^{(n_f=5)}(M_Z) = 0.1147$.
The renormalization and factorization scales $\mu_R$ and $\mu_F$ are taken to
be $\mu_R = \mu_F = M_V$, with $M_V$ being the mass of the gauge boson $V = W^\pm, Z$. We note that the main results of our study are insensitive to these parameter choices.

The following codes for the computation of the fully differential NNLO QCD predictions for the lepton
rapidity distributions are considered:
\begin{itemize}
\item {\tt DYTURBO} (version 1.2)~\cite{Camarda:2019zyx}%
  \footnote{Code available from \url{https://dyturbo.hepforge.org/}.}
  
\item {\tt FEWZ} (version 3.1)~\cite{Li:2012wna,Gavin:2012sy}%
  \footnote{Code available from \url{https://www.hep.anl.gov/fpetriello/FEWZ.html}.}
  
\item {\tt MATRIX} (version 2.1)~\cite{Buonocore:2021tke};%
  \footnote{Code available from \url{https://matrix.hepforge.org/}.}
  {\tt MATRIX} uses the scattering amplitudes from {\tt OpenLoops} \cite{Cascioli:2011va}. This version supersedes the previous one (version 1.0.4)~\cite{Grazzini:2017mhc}.
  
\item {\tt NNLOJET}~\cite{Gehrmann-DeRidder:2023urf}%
  \footnote{Private code. Results provided from the authors upon request.} 
\end{itemize}

These codes differ by the subtraction schemes used, which are either fully
local or based on final-state phase space slicing in a resolution variable. {\tt FEWZ} and {\tt NNLOJET} employ fully local subtraction schemes, with 
{\tt FEWZ} using sector decomposition~\cite{Anastasiou:2003gr} and {\tt NNLOJET} using antenna subtraction~\cite{Gehrmann-DeRidder:2005btv}.
{\tt NNLOJET} is a Monte Carlo parton-level event generator, providing fully differential QCD predictions at NNLO for a number of LHC observables~\cite{Gauld:2019ntd}. 
Codes based on phase space slicing methods are {\tt DYTurbo} and {\tt MATRIX}.
{\tt DYTurbo} features an improved reimplementation of the {\tt DYNNLO}
code~\cite{Catani:2007vq,Catani:2009sm}\footnote{The {\tt DYNNLO} program, considered in the previous comparison~\cite{Alekhin:2021xcu}, 
is a legacy code now superseded by {\tt DYTurbo}.}
for fast predictions for Drell-Yan processes~\cite{Camarda:2019zyx}. It also includes the resummation of large logarithmic corrections.
{\tt DYTurbo} and {\tt MATRIX} both use $q_T$-subtraction~\cite{Catani:2007vq} at NNLO.
The slicing parameters are $q_{T, \rm cut}$ for {\tt DYTurbo} and 
$r_{\rm cut} = q_{T,\rm cut}/M$ for {\tt MATRIX}
where $M$ is the mass of the two-body final state.

Subtraction schemes based on phase space slicing are susceptible to power corrections in $q_T$. 
While these power corrections for vector-boson mediated process are known to be quadratic in the absence of fiducial cuts~\cite{Grazzini:2016ctr,Ebert:2018gsn,Buonocore:2019puv,Cieri:2019tfv,Oleari:2020wvt}, 
they become linear in the presence of  cuts applied on the transverse momenta of the two final state particles~\cite{Grazzini:2017mhc,Ebert:2019zkb,Alekhin:2021xcu,Salam:2021tbm}. 
The appearance of linear power corrections in the fiducial phase space is a purely kinematic effect, which allows for their efficient computation via a suitable recoil prescription~\cite{Catani:2015vma,Grazzini:2015wpa,Ebert:2020dfc}.

{\tt DYTurbo} (version 1.2)~\cite{Camarda:2019zyx} and the 
new release {\tt MATRIX} (version 2.1)~\cite{Buonocore:2021tke} include the computation of linear power corrections in $q_T$ for $2 \rightarrow 2$ processes mediated by a vector boson. 
This reduces significantly the dependence of the predictions on the parameter for the $q_T$ cut-off. 
Additionally, {\tt MATRIX} (version 2.1) also includes a bin-wise $r_{\rm cut}$ extrapolation, which allows the user to obtain a yet more robust prediction than the one obtained with a finite value of $r_{\rm cut}$. All the predictions shown below and obtained with {\tt MATRIX} (version 2.1) include both the inclusion of linear power corrections as well as the new bin-wise extrapolation feature.

The initial benchmark \cite{Alekhin:2021xcu} also considered {\tt MCFM} (version 9.0)~\cite{Campbell:2019dru} and found large effects from linear power corrections in comparison to {\tt FEWZ} results (local subtraction).
The {\tt MCFM} code\footnote{Code available from \url{https://mcfm.fnal.gov/}.} implements the NNLO computation of Ref.~\cite{Boughezal:2016wmq} and applies $N$-jettiness subtraction~\cite{Boughezal:2015dva,Gaunt:2015pea} with $\tau_{\rm cut}$ as the jettiness slicing parameter.
Since the recoil prescription~\cite{Catani:2015vma,Grazzini:2015wpa,Ebert:2020dfc} to remove the linear power corrections cannot be easily adapted to $N$-jettiness slicing we refer to 
Ref.~\cite{Alekhin:2021xcu} for predictions 
with this version of the code.
As of {\tt MCFM} (version 10.0)~\cite{Neumann:2022lft} the code also allows for all NNLO calculations implemented (as well as
for the N$^3$LO calculations for charged- and neutral-current Drell-Yan) to be performed using $q_T$-subtraction,
also accounting for fiducial power corrections, 
    see, e.g. Ref.~\cite{Campbell:2023lcy,Campbell:2024hjq}.

\subsection{NNLO benchmark predictions}

We computed the QCD predictions at NNLO accuracy for $W^\pm$- and $Z/\gamma^*$-production cross sections at $\sqrt{s}=7$\ TeV 
with the cuts imposed by ATLAS~\cite{Aaboud:2016btc} using the ABMP16 PDFs~\cite{Alekhin:2017kpj}.
The conclusions do not depend on this choice.
The ATLAS data were not included in the ABMP16 PDFs fit and are shown to illustrate the accuracy of the experimental measurements.
Following the set-up of the previous study~\cite{Alekhin:2021xcu} the predictions from {\tt FEWZ} are chosen as the baseline, to which other predictions are compared.

\begin{figure}[t!]
\begin{center}
\includegraphics[width=16.5cm]{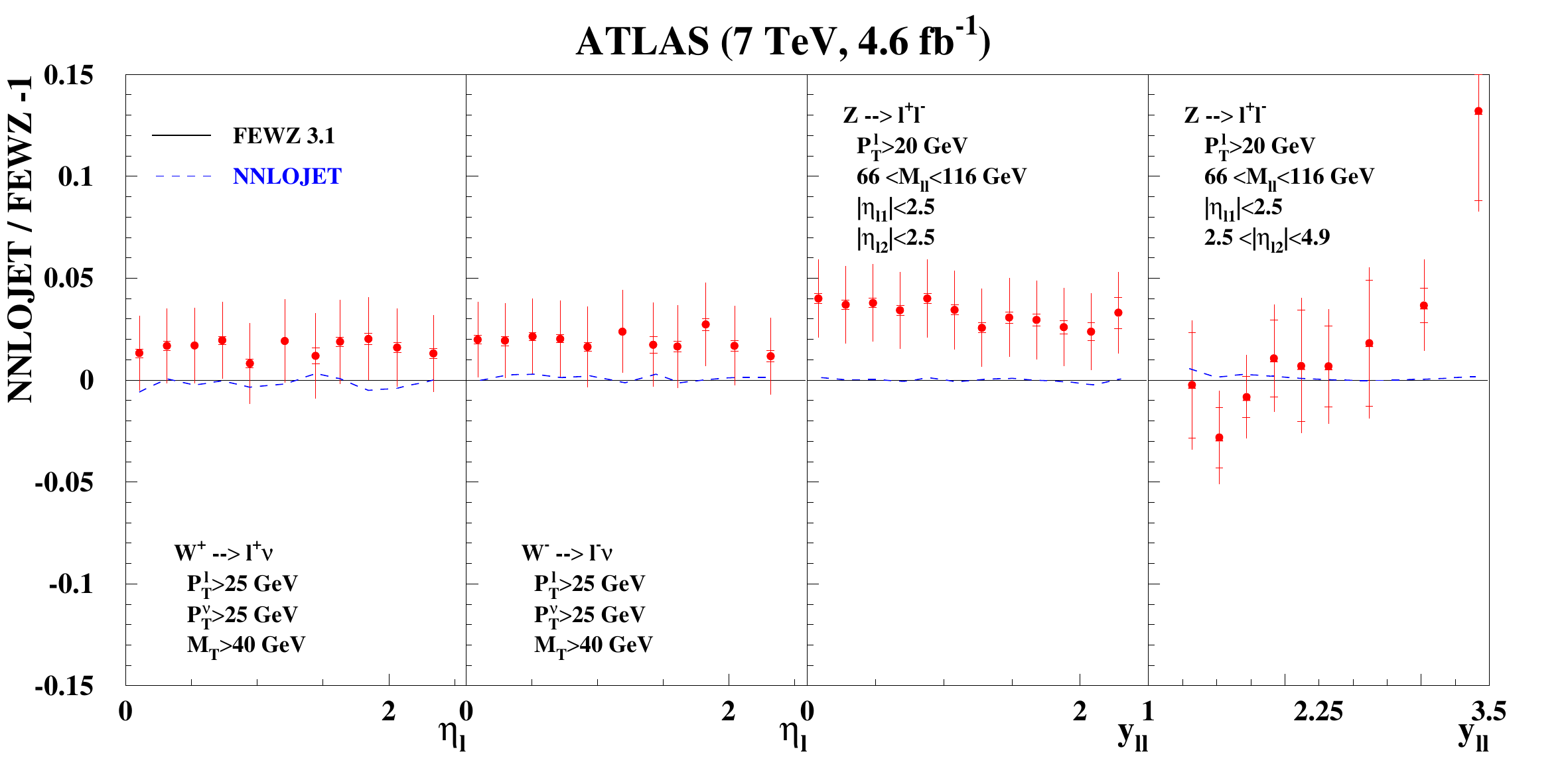}
\caption{\small
  \label{fig:nnlojet-atlas7data}
  Relative deviation of ATLAS data
  measured in inclusive $pp \to W^\pm+X \to l^\pm \nu + X$ 
  and $pp \to Z/\gamma^*+X \to l^+l^-  + X$ production 
  at $\sqrt{s}=7$\ TeV~\cite{Aaboud:2016btc} with the 
  statistical (inner bar) and the total uncertainties, including the 
  systematic ones. The fiducial cuts on the decay leptons in the final state
  are indicated in the figure.
  The ABMP16 central predictions at NNLO are obtained
  with {\tt FEWZ} and the deviations of the predictions from {\tt NNLOJET} 
  are shown (dashed) for comparison. 
}
\end{center}
\end{figure}

We begin by comparing the two codes based on a local subtraction scheme,  {\tt FEWZ} and {\tt NNLOJET}, with the 7\ TeV ATLAS data.
The comparison of both codes with the data as a function of lepton $p_T$ is shown in Fig.~\ref{fig:nnlojet-atlas7data}. We note that both codes are in excellent agreement with each other, with relative deviations between them at the per-mille level. Both codes are in reasonable agreement with the $W^{\pm}$- and $Z$-boson data, with the largest differences of 3--4\% occurring for $Z$-boson production with central leptons.

\begin{figure}[t!]
\begin{center}
\includegraphics[width=16.5cm]{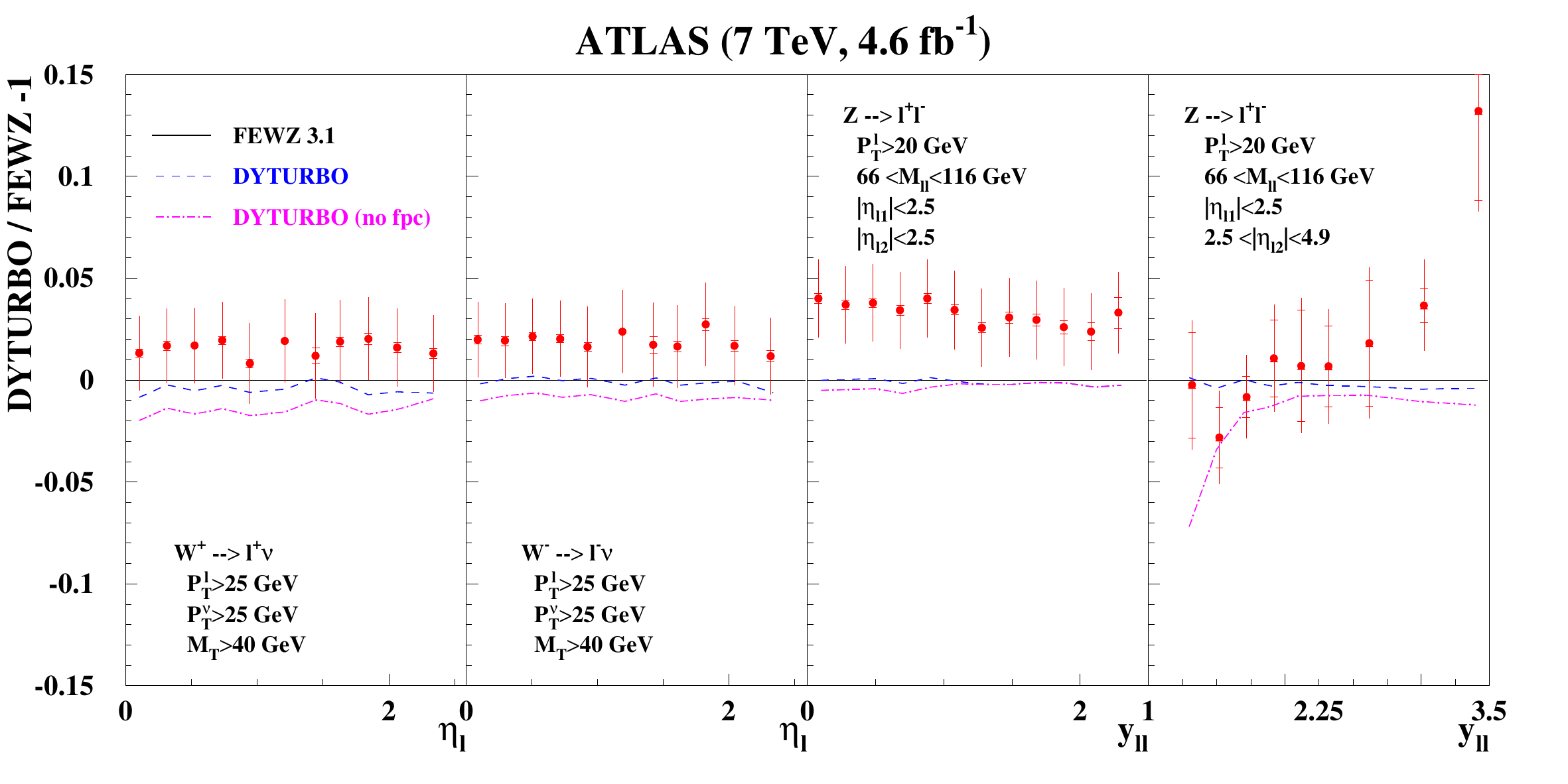}
\caption{\small
  \label{fig:dyturbo-atlas7data}
  Same as Fig.~\ref{fig:nnlojet-atlas7data} using predictions by the 
  {\tt DYTURBO} code with (dashed) and without (dashed-dotted) the linear power corrections labeled ``(no fpc)'' for comparison.
}
\end{center}
\end{figure}
\begin{figure}[t!]
\begin{center}
\includegraphics[width=16.5cm]{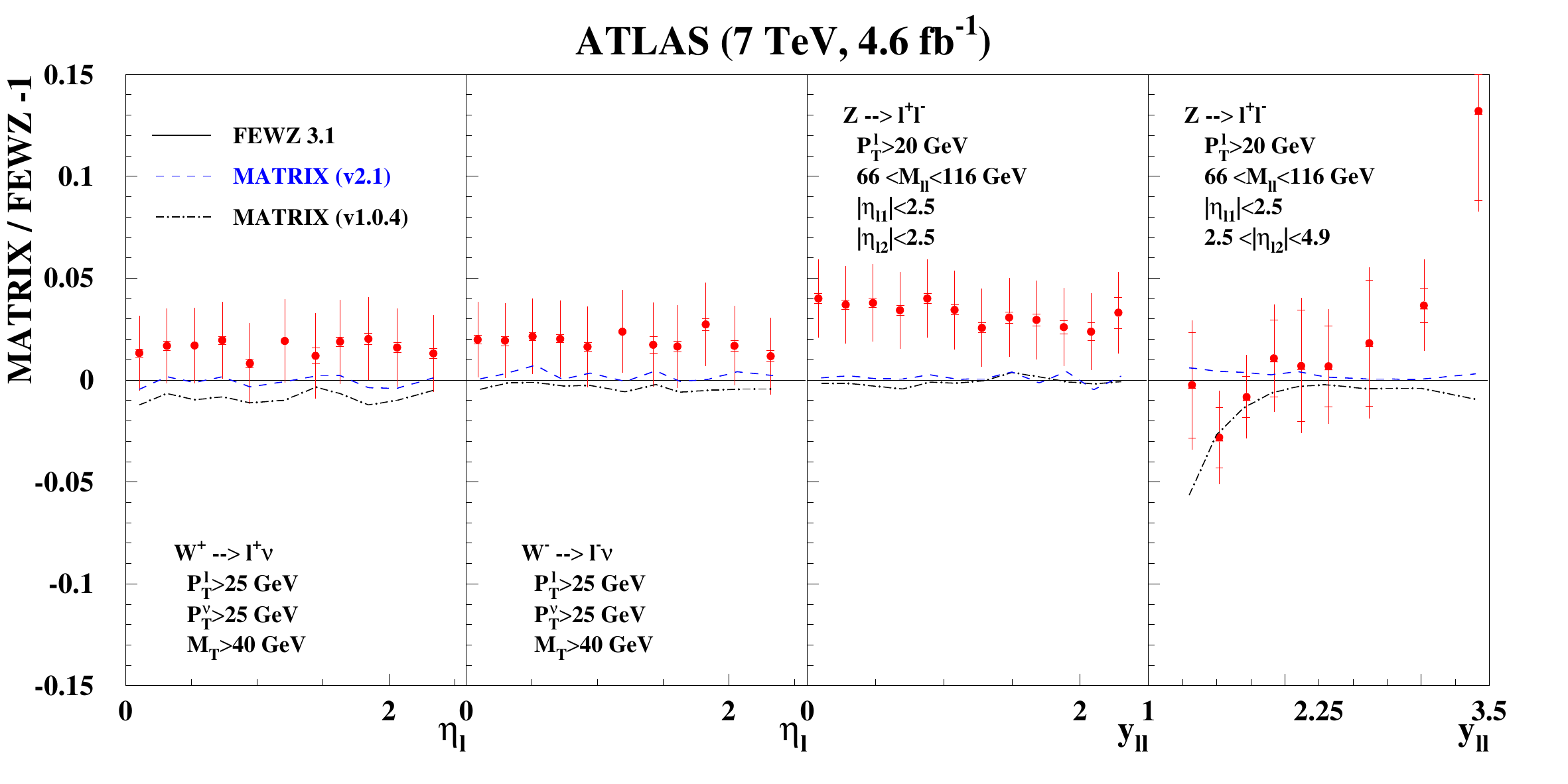}
\caption{\small
  \label{fig:matrix-atlas7data}
  Same as Fig.~\ref{fig:nnlojet-atlas7data} using predictions by the 
  {\tt MATRIX} code with version 1.0.4 (dashed-dotted) 
  and a value for the $q_T$-slicing cut $r^{\rm min}_{\rm cut}=0.15\%$
  as well as predictions with the improvements for the linear power corrections 
  in version 2.1 (dashed) using $r^{\rm min}_{\rm cut}=0.5\%$.
}
\end{center}
\end{figure}

Having established agreement between both codes based on local subtraction schemes, as well as their agreement with the 7\ TeV data, we now compare them to codes dependent on the $q_T$ slicing parameter.
Fig.~\ref{fig:dyturbo-atlas7data} illustrates the agreement with {\tt DYTURBO} (version 1.2), which accounts for linear power corrections as discussed earlier in the text.
All {\tt DYTURBO} predictions agree with those of {\tt FEWZ} at the level of a few per-mille.
The {\tt DYTURBO} predictions labeled ``no fpc'' in Fig.~\ref{fig:dyturbo-atlas7data} reproduce the {\tt DYNNLO} predictions obtained previously~\cite{Alekhin:2021xcu},
and differ from {\tt FEWZ} by up to 2\% for $W^{\pm}$-boson production, and by more than 5\% for $Z$-boson production with forward leptons.
This comparison clearly demonstrates the quality of the improvements achieved with the new version 1.2 of {\tt DYTURBO}.

In Fig.~\ref{fig:matrix-atlas7data} we show predictions for two versions of the {\tt MATRIX} code normalized to the {\tt FEWZ} predictions.
The new {\tt MATRIX} version 2.1 is distinguished by its inclusion of linear power corrections.
As evident from Fig.~\ref{fig:matrix-atlas7data} the accounting of these power corrections has a significant impact, improving the agreement with {\tt FEWZ}, see also Ref.~\cite{Buonocore:2021tke}.
{\tt MATRIX (v2.1)} agrees at the level of a few per-mille with {\tt FEWZ},
with the largest deviation for $Z$ production with forward leptons still significantly below 1\%. 
The older version 1.0.4 of {\tt MATRIX} without linear power corrections, considered previously~\cite{Alekhin:2021xcu},
differs from {\tt FEWZ} by 1\% for $W^{\pm}$-boson production, and by up to 5\% for $Z$-boson production with forward leptons.

\begin{figure}[t!]
\begin{center}
\includegraphics[width=16.5cm]{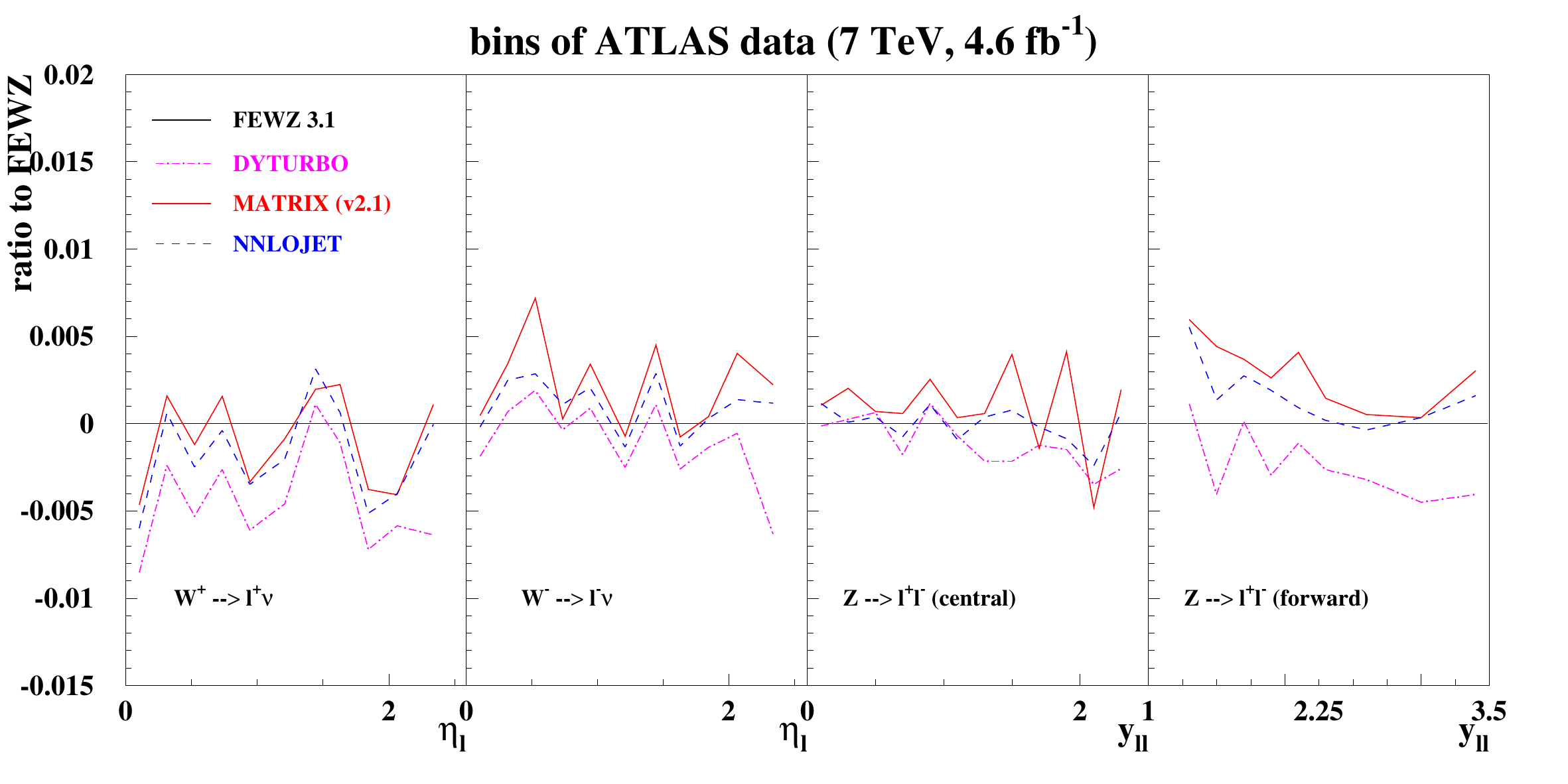}
\caption{\small
  \label{fig:all-ratios-atlas}
Zoom on the per-mille level agreement of the predictions from 
the codes 
{\tt DYTURBO} (dashed-dotted) 
{\tt MATRIX (v2.1)} (solid)
and {\tt NNLOJET} (dashed) 
relative to the central predictions at NNLO with {\tt FEWZ} 
  for the fiducial cuts and the bins of the ATLAS at $\sqrt{s}=7$\ TeV~\cite{Aaboud:2016btc}.
}
\end{center}
\end{figure}
The findings of this benchmark study are summarized
Fig.~\ref{fig:all-ratios-atlas}, which compares in each case the best predictions of {\tt DYTURBO}, {\tt MATRIX (v2.1)} and {\tt NNLOJET}.
The predictions of {\tt MATRIX (v2.1)} and {\tt NNLOJET} are in excellent
agreement, with deviations being often less than one per-mille, i.e.\ the target uncertainty from the numerical integration.
Also the {\tt DYTURBO} predictions agree very well, typically within two per-mille with the {\tt NNLOJET} ones, except for the $Z$-boson production with forward leptons, where agreement is at the level of a few per-mille.
Fig.~\ref{fig:all-ratios-atlas} also shows, that the predictions from {\tt DYTURBO}, {\tt MATRIX (v2.1)} and {\tt NNLOJET} are all aligned, 
especially for the case of $W^{\pm}$-boson production.
The normalization to the {\tt FEWZ} predictions introduces some fluctuations,
which are due to the numerical integration uncertainties in the {\tt FEWZ} results, being at the level of a few per-mille. 
The predictions obtained with {\tt NNLOJET} are listed in App.~\ref{sec:appXS}.

\begin{figure}[t!]
\begin{center}
\includegraphics[width=8.1cm]{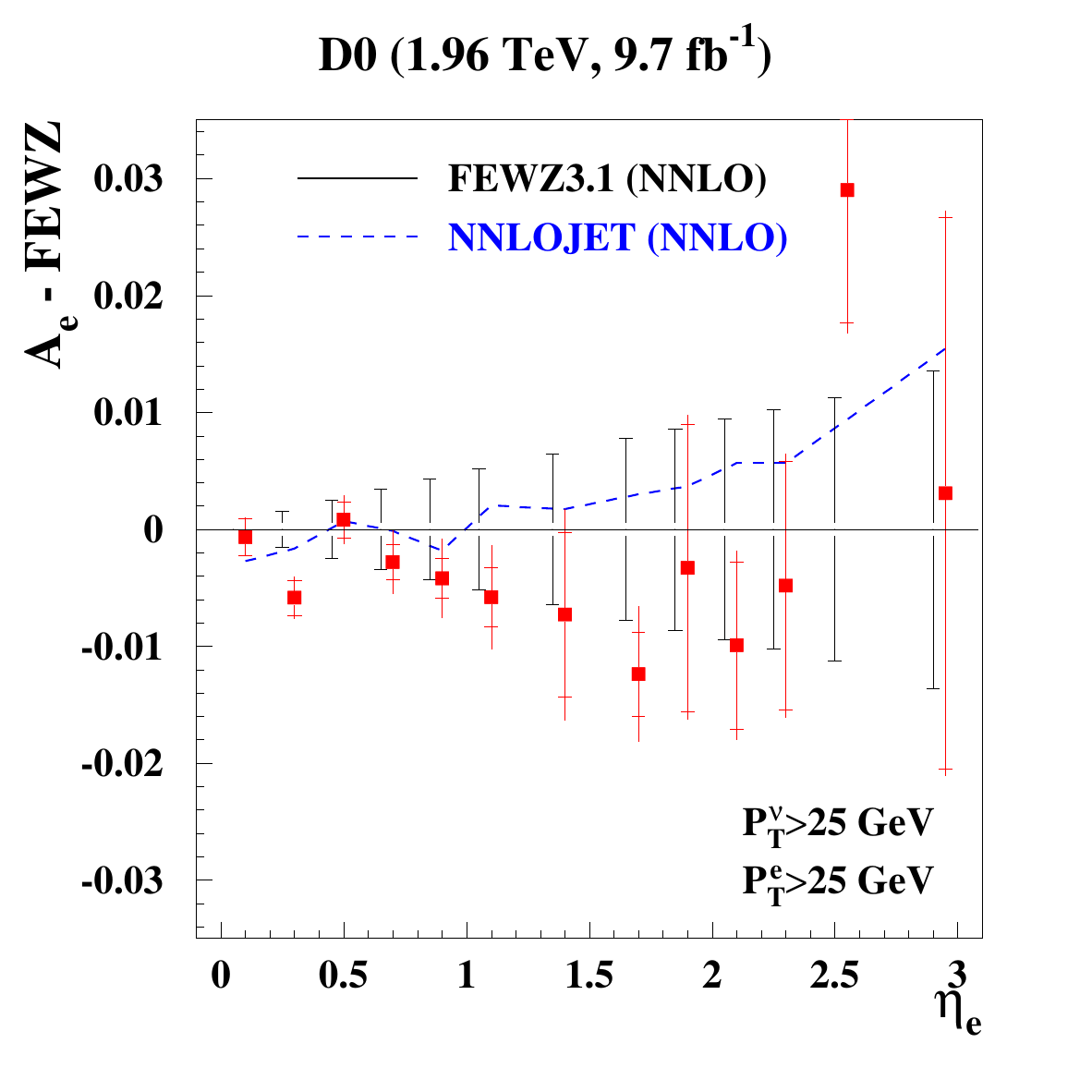}
\includegraphics[width=8.1cm]{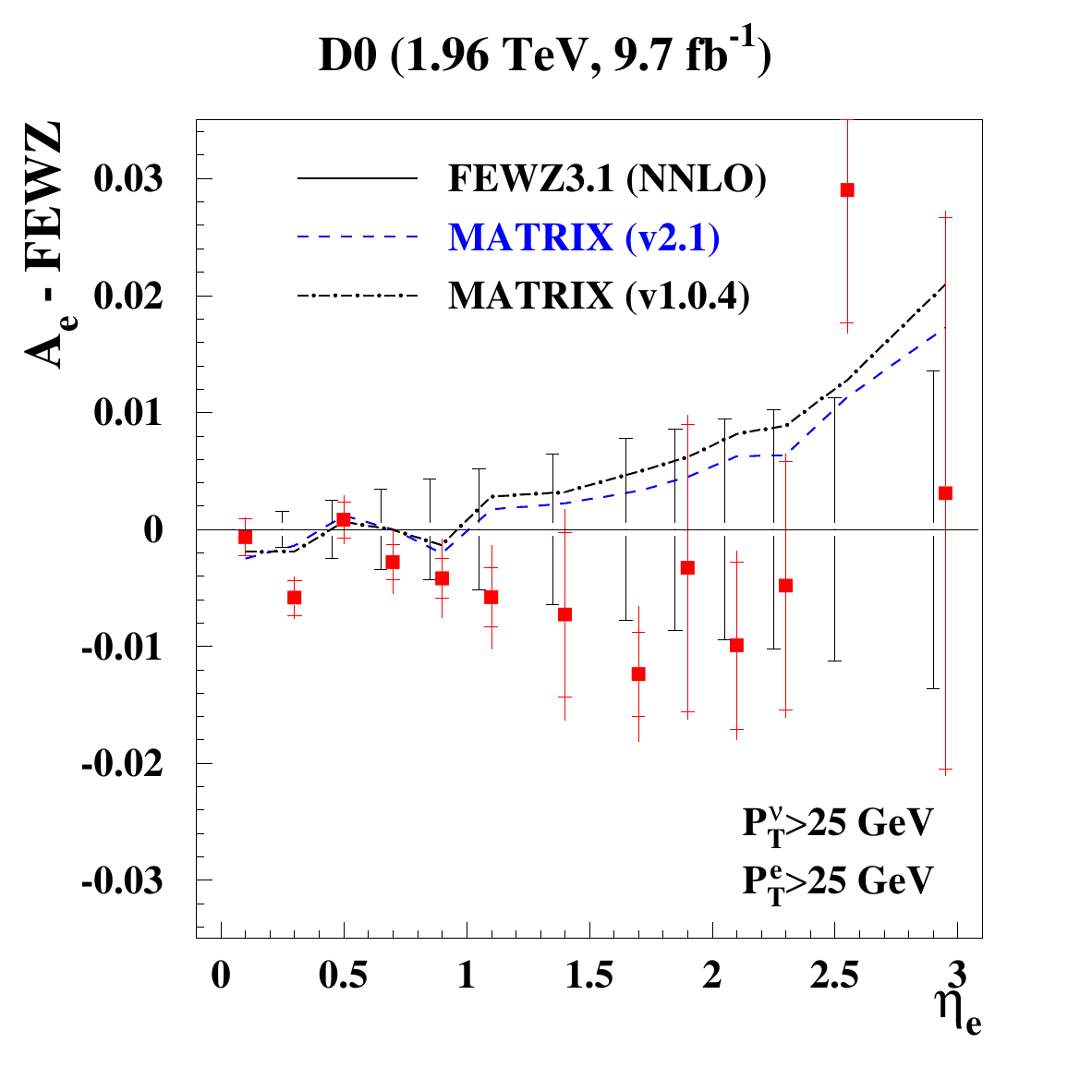}
\caption{\small
  \label{fig:dynnlo-d0}
  The D{\O} data on the electron charge
  asymmetry distribution $A_e$ in $W^{\pm}$-boson production at $\sqrt{s}=1.96$\ TeV with the 
  statistical (inner bar) and the total uncertainties, including the 
  systematic ones. 
  The difference of $A_e$ to the ABMP16 central predictions at NNLO obtained with {\tt FEWZ} is shown together with the numerical integration uncertainties (black vertical solid lines) of the {\tt FEWZ} predictions.
  The symmetric $p_T^{e,\nu}$-cuts of the decay leptons are indicated in the figure. 
  The NNLO predictions by 
  the {\tt NNLOJET} code (dashed lines, left plot) and by the versions of the {\tt MATRIX} code (dashed and dashed-dotted, right plot)
  are displayed for comparison.
}
\end{center}
\end{figure}

The D{\O} data on the electron charge asymmetry distribution $A_e$ has been obtained as a function of the electron pseudo-rapidity from $W^\pm$-boson production at $\sqrt{s}=1.96$\ TeV at the Tevatron~\cite{D0:2014kma}.
Theoretical predictions for the benchmark studies are numerically challenging due to large cancellations in the asymmetry.
With predictions from {\tt NNLOJET} and the update of {\tt MATRIX} (version 2.1) we are in a position to check the {\tt FEWZ} results, already presented in 
Ref.~\cite{Alekhin:2021xcu}.
This is illustrated in Fig.~\ref{fig:dynnlo-d0}, where we plot the NNLO predictions obtained with those codes.
The numbers obtained with {\tt NNLOJET} and {\tt MATRIX} (version 2.1) are in excellent agreement and are also compatible with the {\tt FEWZ} predictions within the substantially larger uncertainties from the numerical integration of the latter. 
The numerical integration uncertainties of the {\tt NNLOJET} and {\tt MATRIX} numbers are negligible on the scale of Fig.~\ref{fig:dynnlo-d0}, while those of {\tt FEWZ} are indicated in the plot.
All numbers computed with {\tt NNLOJET} are also given in App.~\ref{sec:appXS}.

\newpage

\section{Theoretical formalism for lepton fiducial cuts}
\label{sec:lepton-cuts}

To further investigate the role of lepton fiducial cuts, which lead to discrepancies between the local-subtraction codes and the slicing ones if not properly accounted for theoretically, we consider $Z$-boson production with central leptons. We stagger the $p_T$ cuts on the two leptons by a small parameter $\Delta p_T$. We review here the expected form of the cross section both in fixed-order perturbation theory and when including all-orders resummation.

\subsection{Definition of linear asymmetric fiducial cuts}
\label{sec:cuts_staggered}

Experimental analyses commonly use linear cuts on final transverse momenta. 
Let $p_T^{\ell_1}$ and $p_T^{\ell_2}$ be the leading and subleading lepton
transverse momenta. We consider the following linear asymmetric fiducial cuts
parametrized by $\Delta p_T$ (which can take either sign) and staggered on the leading and subleading leptons:
\begin{equation}
\label{eq:cuts_asy}
p_T^{\ell_1} \geq
\begin{cases}
20 \,\text{GeV} \,, \quad &\Delta p_T < 0
\,, \\
20 \,\text{GeV} + |\Delta p_T|\,, \quad &\Delta p_T > 0
\,,\end{cases}
\qquad
p_T^{\ell_2} \geq
\begin{cases}
20 \,\text{GeV} - |\Delta p_T|\,, \quad &\Delta p_T < 0
\,, \\
20 \,\text{GeV} \,, \quad &\Delta p_T > 0
\,.\end{cases}
\end{equation}

%
\begin{figure}[htbp]
\begin{center}
\includegraphics[width=8.15cm]{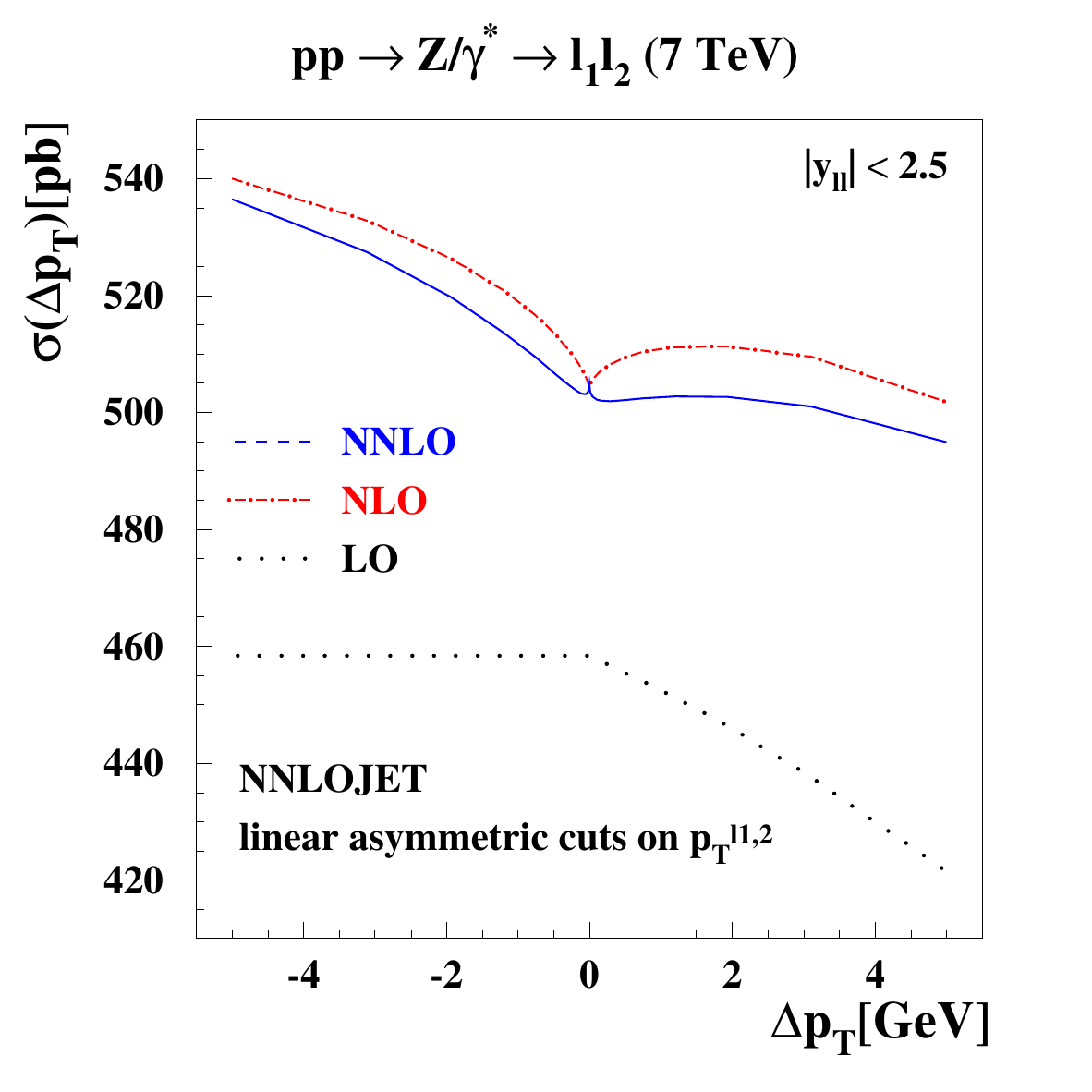}
\includegraphics[width=8.15cm]{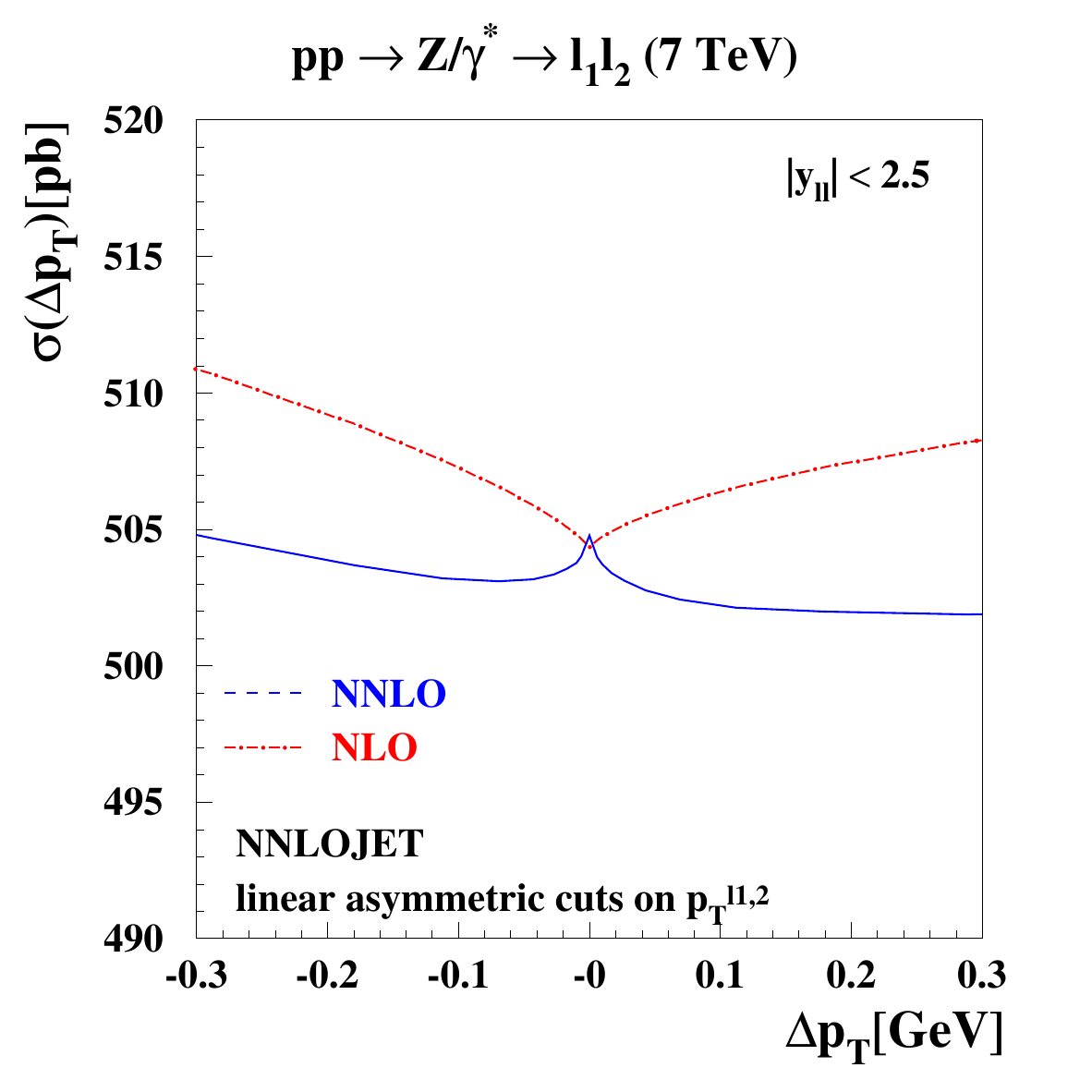}
\caption{\small
 \label{fig:pTcuts}
  The cross sections for $pp \to Z/\gamma^*+X \to l^+l^-  + X$ production 
  at $\sqrt{s}=7$\ TeV at LO (dotted), NLO (dashed) and NNLO (solid) in QCD 
  with ABMP16 PDFs and $y_{ll} \leq 2.5$ computed with {\tt NNLOJET} as a function of $\Delta p_T \in [-5,5]$\ GeV defined in Eq.~(\ref{eq:cuts_asy}) for the linear asymmetric fiducial cuts on the decay leptons in the final state (left plot) and zoom 
  on the range $\Delta p_T \in [-0.3,0.3]$\ GeV (right plot). 
}
\end{center}
\end{figure}

\begin{figure}[htbp]
\begin{center}
\includegraphics[width=8.15cm]{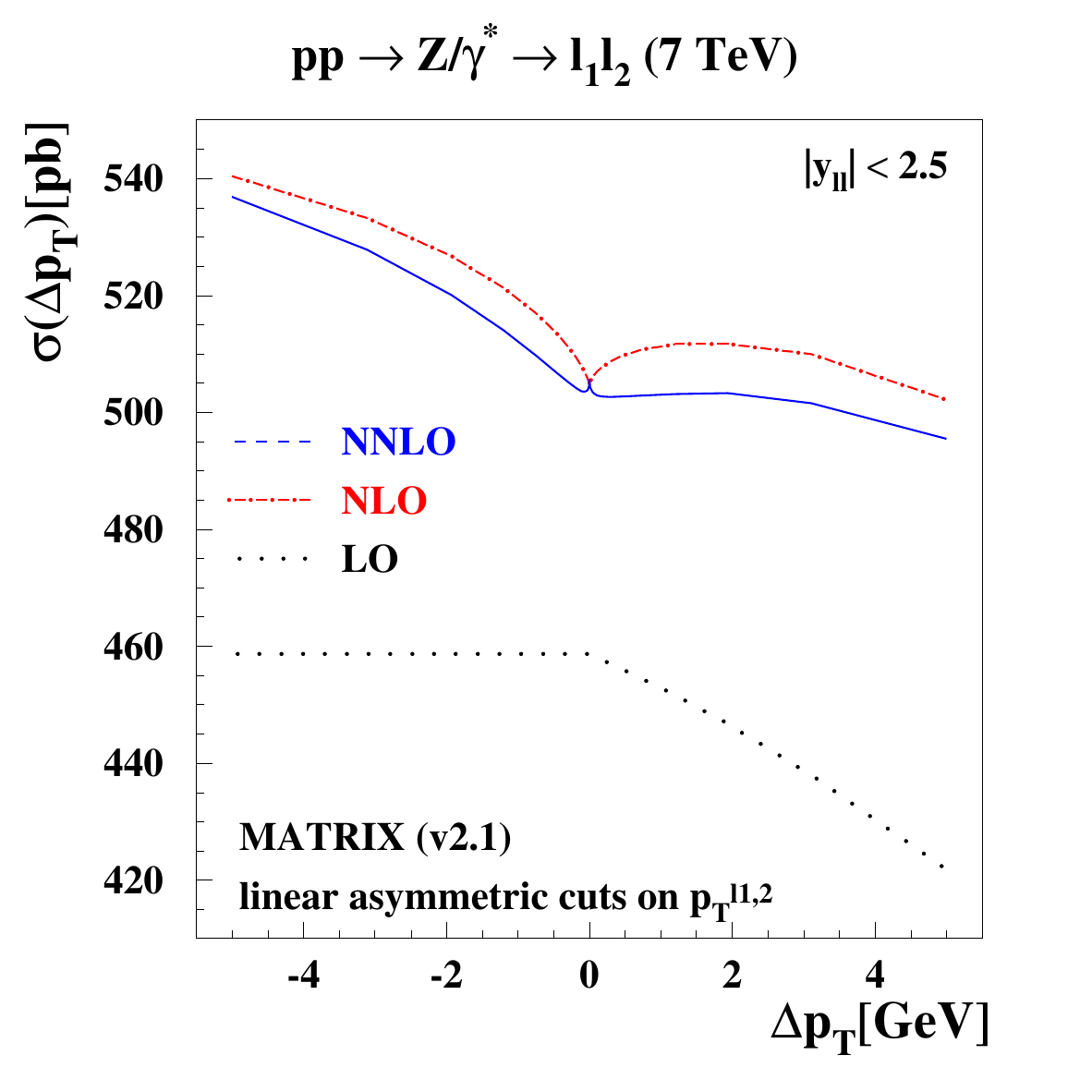}
\includegraphics[width=8.15cm]{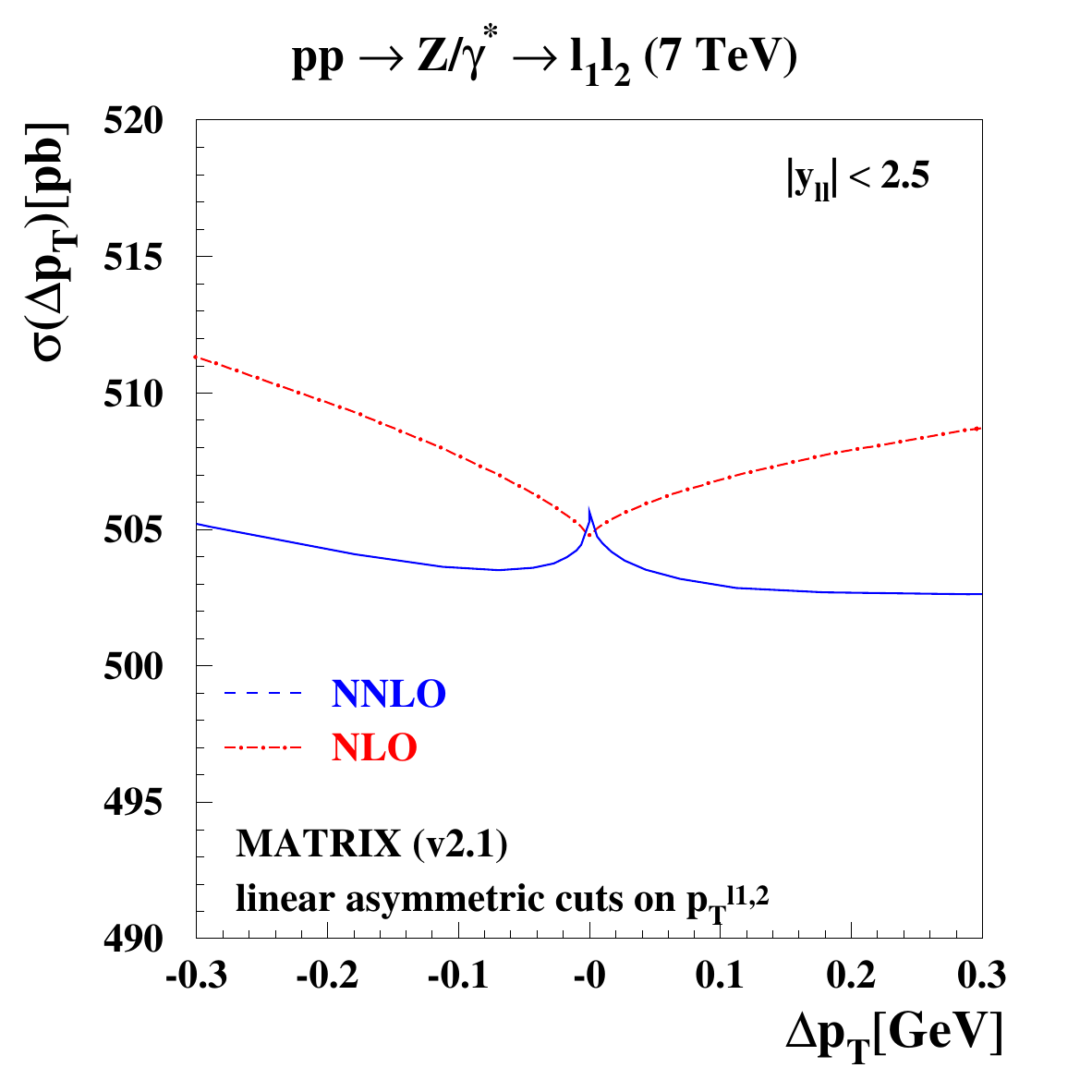}
\caption{\small
 \label{fig:matrix-pTcuts}
 Same as Fig.~\ref{fig:pTcuts}, now showing the predictions obtained 
 with {\tt MATRIX (v2.1)}.
}
\end{center}
\end{figure}
We note that in the literature so-called staggered cuts have also been  defined by
applying cuts on identified leptons (electron and positron in neutral-current
Drell-Yan or lepton and neutrino in charged current Drell-Yan). In this case
it can be shown~\cite{Grazzini:2017mhc,Alekhin:2021xcu,Buonocore:2021tke} that
the performance of slicing methods are much improved compared to symmetric
cuts or to staggered cuts imposed on leading and subleading leptons, as they
lead to a quadratic sensitivity in the region $q_T < \Delta p_T$. However, we
expect a similar behavior of the fixed-order cross section as a function of
$\Delta p_T$ as that discussed for our definition of linear asymmetric cuts. 

The phase space available in the final
state decreases with increasing $\Delta p_T$, hence the
cross section should monotonically decrease with this cut. We see 
in Figs.~\ref{fig:pTcuts} and \ref{fig:matrix-pTcuts} that this 
occurs only at LO. A kink appears for $\Delta p_T=0$ at higher 
orders. The effect leading to this unphysical result was first 
explained in the context of photo-production of jets at 
HERA~\cite{Frixione:1997ks}, and can be explained with the simple 
one-loop example presented there. Real-emission corrections 
collinear to the initial-state partons in the hard scattering 
process contain physical singularities that can be regulated with a 
cutoff $\delta$. Taking $\Delta p_T \geq 0$ at first,
the NLO real-emission cross section $\sigma_R$ and its derivative
have the following functional dependence on $\Delta p_T$ and the 
collinear cutoff $\delta$:
\begin{eqnarray}
    \sigma_R(\Delta p_T,\delta) &=& A(\Delta p_T,\delta)+B \, \ln \frac{\delta}{m_Z} +C(\Delta p_T+\delta)\ln \frac{\Delta p_T+\delta}{m_Z}
    \,, \nonumber \\
    \frac{d\sigma_R(\Delta p_T,\delta)}{d\delta} &=& \frac{dA}{d\delta}+\frac{B}{\delta}+C \Bigl[ 1+\ln \frac{\Delta p_T+\delta}{m_Z} \Bigr]
    \,.
    \label{eq:cutoff}
\end{eqnarray}
Here, $B$ and $C$ are coefficients with no dependence on either $\delta$ or $\Delta p_T$,
while both $A$ and its derivative are regular for all parameter values.
We learn two things from the above formulae.
First, in the final cross section, the $B \log \delta$ term
cancels against the NLO virtual diagram at this order,
but still leaves behind the $C$ term which is a purely real effect. Second,
setting the collinear cutoff $\delta=0$ we can see that the first derivative is singular
(specifically, logarithmically divergent) when $\Delta p_T \to 0$
even after adding the virtual contribution. More generally, allowing for either sign of $\Delta p_T$,
including the virtual and higher-order corrections, and taking $\delta \to 0$,
the fixed-order cross section at fixed $\mu_R$, $\mu_F$ and small 
$x = \Delta p_T/m_Z$ takes the form
\begin{align} \label{eq:delta_pt_fo_xsec_all_orders}
\sigma(x)
= \sum_{n=0}^\infty \alpha_s^n \biggl\{
   c_n
   + x \theta(-x) \sum_{m=0}^{2n} a_{n,m} \ln^m (-x)
   + x \theta(x) \sum_{m=0}^{2n} b_{n,m} \ln^m x
   + \mathcal{O}(x^2)
\,\biggr\}.
\end{align}
As in the one-loop example above, the logarithmic terms
are expected to arise from real emissions
close to the Born configuration,
i.e., a residual mis-cancellation against the corresponding virtual parts. We will make this expectation more precise in Sec.~\ref{sec:resummation}.

To investigate the impact of these considerations on the cross section for $Z$ production at $\sqrt{s}=7$\ TeV 
with a cut on the lepton pseudorapidities $|\eta_{\ell_1,\ell_2}| \leq 2.5$ as a function of $\Delta p_T$, 
we present the fixed-order predictions of {\tt NNLOJET} in Fig.~\ref{fig:pTcuts} and {\tt MATRIX} in Fig.~\ref{fig:matrix-pTcuts}. 
Results at LO, NLO and NNLO in perturbative QCD 
for the linear asymmetric lepton decay cuts defined in Eq.~(\ref{eq:cuts_asy}) are shown. 
Excellent agreement between the {\tt NNLOJET} and the {\tt MATRIX (v2.1)} numbers is observed for all values of $\Delta p_T$. 
The latter have been obtained accounting for the linear power corrections in the $q_T$-subtraction and bin-wise $r_{\rm cut}$ extrapolation, as discussed previously.
The logarithmic divergences described above lead to the kink at 
$\Delta p_T =0$ and the resulting non-monotonic behavior as $\Delta p_T$ is varied in these plots. This result is a pathology of fixed-order perturbation theory when symmetric cuts are applied.

\subsection{Structure of the physical cross section}
\label{sec:structure_physical}

It is useful to discuss the general form that the \emph{physical} cross section
should take as a function of $x = \Delta p_T/m_Z$, ignoring QED and weak corrections (but assuming that Nature exactly solved QCD for us).
In this case we expect that the distribution
$d \sigma/d \Phi_{\ell\ell} \equiv d^6 \sigma/(d^3 \vec{p}_{\ell^+} \, d^3 \vec{p}_{\ell^-})$
in the lab frame is non-negative and smooth in all limits, unlike the fixed-order cross section
with its singular ridge along the surface $\vec{p}_{T,\ell^+} = - \vec{p}_{T,\ell^-}$.
However, even in the case of the physical cross section
we still generically find
a discontinuous derivative as $x  \to 0$ from above vs.\ from below.
Letting $p_T^\mathrm{cut} = 20 \,\text{GeV}$,
this can be seen as follows:
\begin{align}
\label{eq:dsig(x>0)}
\frac{1}{m_Z} \frac{d \sigma(x > 0)}{d x}
&= -\int \! d \Phi_{\ell\ell} \, \frac{d \sigma}{d \Phi_{\ell\ell}}
\,
\delta\Big(p_T^{\ell_1} - p_T^\mathrm{cut} - x m_Z\Big) \,
\Theta\Big(p_T^{\ell_2} - p_T^\mathrm{cut}\Big)
\\ \nn 
&= - \int \! d \Phi_{\ell\ell} \, \frac{d \sigma}{d \Phi_{\ell\ell}}
\,
\delta\Big(p_T^{\ell_1} - p_T^\mathrm{cut}\Big) \,
\Theta\Big(p_T^{\ell_2} - p_T^\mathrm{cut}\Big)
+ \mathcal{O}(x)
\,, \\
\label{eq:dsig(x<0)}
\frac{1}{m_Z} \frac{d \sigma(x < 0)}{d x}
&= -\int \! d \Phi_{\ell\ell} \, \frac{d \sigma}{d \Phi_{\ell\ell}}
\,
\Theta\Big(p_T^{\ell_1} - p_T^\mathrm{cut}\Big) \,
\delta\Big(p_T^{\ell_2} - p_T^\mathrm{cut} - x m_Z\Big)
\\ \nn 
&= -\int \! d \Phi_{\ell\ell} \, \frac{d \sigma}{d \Phi_{\ell\ell}}
\,
\Theta\Big(p_T^{\ell_1} - p_T^\mathrm{cut}\Big) \,
\delta\Big(p_T^{\ell_2} - p_T^\mathrm{cut}\Big)
+ \mathcal{O}(x)
\,,\end{align}
where $p_T^{\ell_1} = \max \Big\{ p_T^{\ell^+}, p_T^{\ell^-}\Big\}$
and $p_T^{\ell_2} = \min \Big\{ p_T^{\ell^+}, p_T^{\ell^-}\Big\}$ are functions of $\Phi_{\ell\ell}$.
We see that the derivatives cannot be positive in either 
case, since in each case the integrand is non-negative. However, the cross sections in Eqs.~\eqref{eq:dsig(x>0)} and \eqref{eq:dsig(x<0)} do not coincide.
For $x > 0$ we have
\begin{align}
\delta(p_T^{\ell_1} - p_T^\mathrm{cut}\Big) \,
\Theta\Big(p_T^{\ell_2} - p_T^\mathrm{cut}\Big)
&= \Theta\Big(p_T^{\ell_-} - p_T^{\ell_+}\Big) \,
\delta\Big(p_T^{\ell_-} - p_T^\mathrm{cut}\Big) \,
\Theta\Big(p_T^{\ell_+} - p_T^\mathrm{cut}\Big)
\nn \\ & \quad
+ \Theta\Big(p_T^{\ell_+} - p_T^{\ell_-}\Big) \,
\delta\Big(p_T^{\ell_+} - p_T^\mathrm{cut}\Big) \,
\Theta\Big(p_T^{\ell_-} - p_T^\mathrm{cut}\Big)
\nn \\
&= \Theta\Big(p_T^\mathrm{cut} - p_T^{\ell_+}\Big) \,
\delta\Big(p_T^{\ell_-} - p_T^\mathrm{cut}\Big) \,
\Theta\Big(p_T^{\ell_+} - p_T^\mathrm{cut}\Big)
\nn \\ & \quad
+ \Theta\Big(p_T^\mathrm{cut} - p_T^{\ell_-}\Big) \,
\delta\Big(p_T^{\ell_+} - p_T^\mathrm{cut}\Big) \,
\Theta\Big(p_T^{\ell_-} - p_T^\mathrm{cut})
= 0
\,,\end{align}
which implies
\begin{align}
\frac{1}{m_Z} \frac{d \sigma(x > 0)}{d x} = 0 + \mathcal{O}(x)
\,,
\end{align}
because the integration region is a lower-dimensional manifold
(assuming some precise limit-taking treatment of $\theta(y) \theta(-y)$),
whereas the physical cross section is free of distributional terms
and bounded on every subdomain
such that we can drop this null set.
By contrast, for $x < 0$ we find a genuinely 5-dimensional integration region that is not a null set,
\begin{align}
\Theta\Big(p_T^{\ell_1} - p_T^\mathrm{cut}\Big) \,
\delta\Big(p_T^{\ell_2} - p_T^\mathrm{cut}\Big)
&= \Theta\Big(p_T^{\ell_-} - p_T^{\ell_+}\Big) \,
\Theta\Big(p_T^{\ell_-} - p_T^\mathrm{cut}\Big) \,
\delta\Big(p_T^{\ell_+} - p_T^\mathrm{cut}\Big)
\nn \\ & \quad
+ \Theta\Big(p_T^{\ell_+} - p_T^{\ell_-}\Big) \,
\Theta\Big(p_T^{\ell_+} - p_T^\mathrm{cut}\Big) \,
\delta\Big(p_T^{\ell_-} - p_T^\mathrm{cut}\Big)
\nn \\
&= \Theta\Big(p_T^{\ell_-} - p_T^\mathrm{cut}\Big) \,
\delta\Big(p_T^{\ell_+} - p_T^\mathrm{cut}\Big)
\nn \\ & \quad
+ \Theta\Big(p_T^{\ell_+} - p_T^\mathrm{cut}\Big) \,
\delta\Big(p_T^{\ell_-} - p_T^\mathrm{cut}\Big)
\,.\end{align}
Thus, we find a non-vanishing derivative at $x < 0$,
\begin{align} \label{eq:physical_xsec_result_derivative_x_neg}
\frac{1}{m_Z} \frac{d \sigma(x < 0)}{d x}
&= - \left[
   \frac{d \sigma_\mathrm{fid}}{d p_T^{\ell^-}}
   + \frac{d \sigma_\mathrm{fid}}{d p_T^{\ell^+}}
\right]_{p_T = p_T^\mathrm{cut}}
+ \mathcal{O}(x)
\,,\end{align}
which we can identify as the sum over the lepton $p_T$ spectra
evaluated at $p_T = p_T^\mathrm{cut}$,
applying a fiducial $p_T \geq p_T^\mathrm{cut}$ on the respective other one.
Importantly, the derivative is $\leq 0$,
as expected because the physical fiducial cross section must decrease with $\Delta p_T$.

It is useful to note how a fixed-order calculation
of $d \sigma/d \Phi_{\ell\ell}$ modifies the above conclusions.
Specifically, unlike the physical cross section,
\begin{enumerate}
   \item
   the fixed-order calculation is not guaranteed to be positive,
   so the unphysical scenario of a cross section \emph{increasing}
   with $\Delta p_T$ (as the phase space is being constrained further)
   is possible; and
   \item
   the fixed-order calculation generically
   contains terms with a finite integral but supported on a null set,
   which is easiest to see from the tree-level $\delta^{(2)}(\vec{p}_{T}^{\,\,\ell^-} - \vec{p}_{T}^{\,\,\ell^+})$,
   and thus the derivative in the limit $x\to 0^+$ need not vanish.
\end{enumerate}
Both of these properties are addressed by resummation
at the level of the hadronic structure functions,
and we thus expect matched predictions to have the physical properties
derived above, which will be addressed in Sec.~\ref{sec:resummation}.

\subsection{Definition of product cuts}
\label{sec:product_cuts}

\begin{figure}[tbp]
\begin{center}
\includegraphics[width=8.15cm]{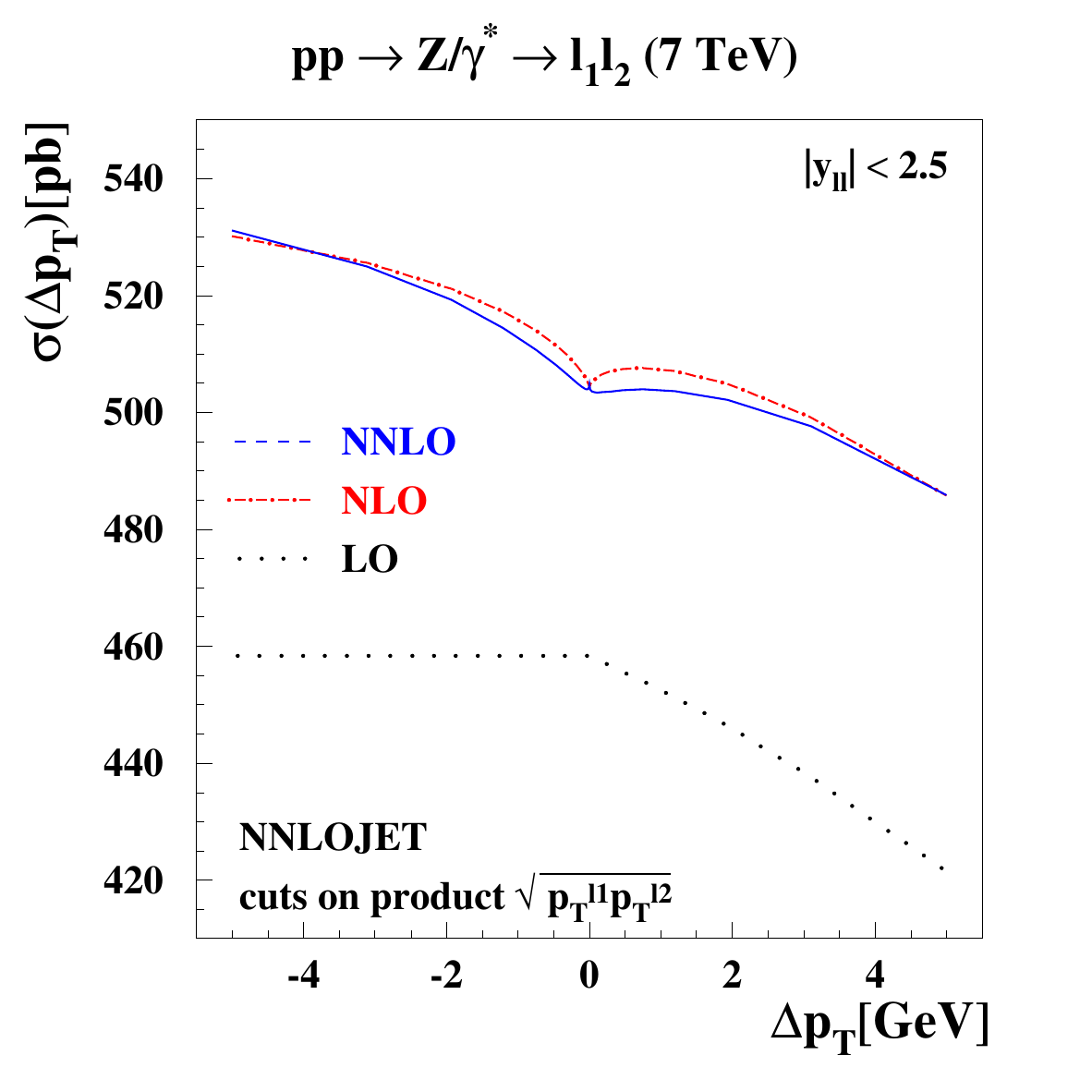}
\includegraphics[width=8.15cm]{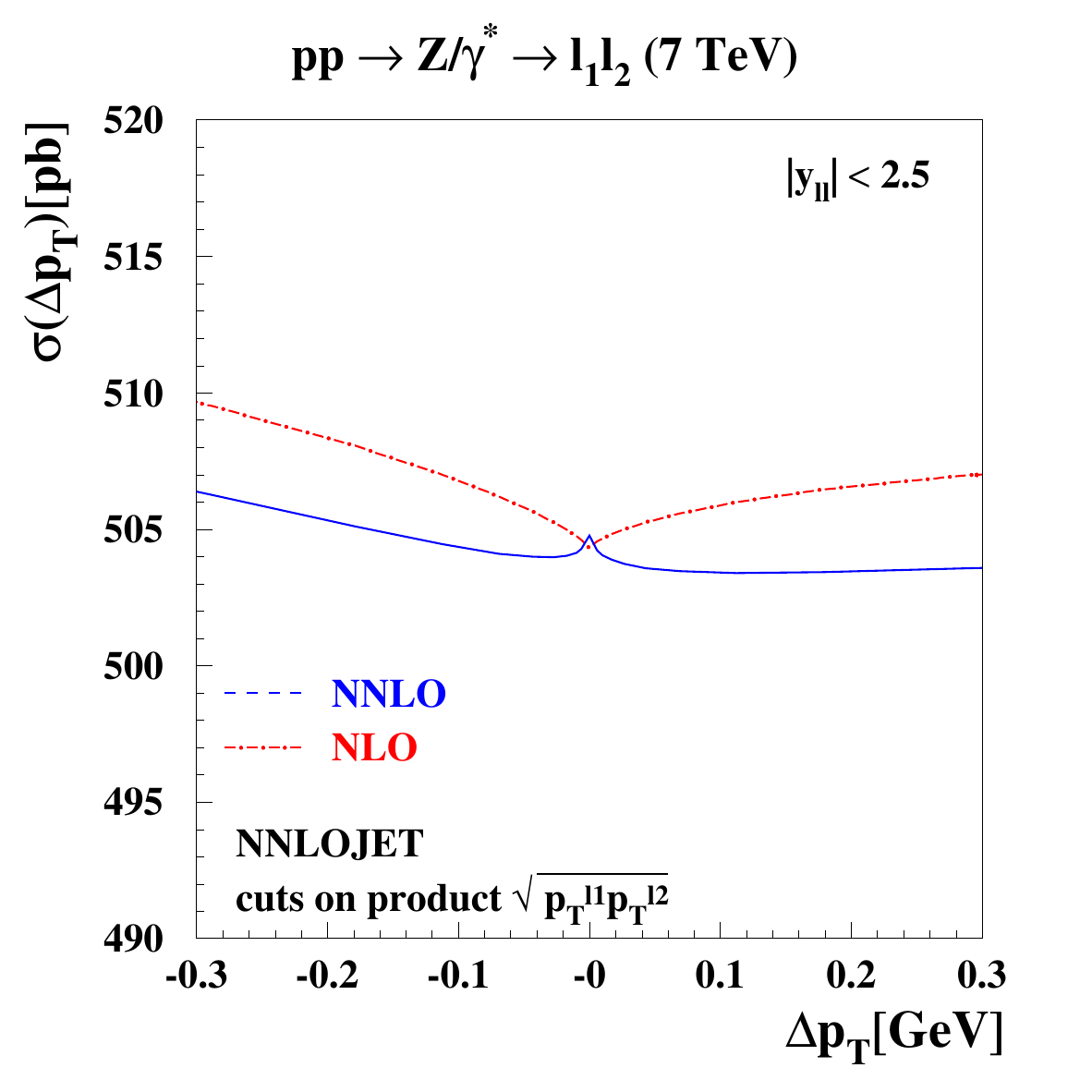}
\caption{\small
  \label{fig:prodcuts}
  Same as in Fig.~\ref{fig:pTcuts}, now with product cuts on the transverse momenta of the decay leptons 
  in the final state. 
  The product cuts are defined in Eq.~(\ref{eq:cuts_prod}) and ranges 
  $\Delta p_T \in [-5,5]$\ GeV (left plot) and $\Delta p_T \in [-0.3,0.3]$\ GeV (right plot) are shown.
}
\end{center}
\end{figure}
\begin{figure}[htbp]
\begin{center}
\includegraphics[width=8.15cm]{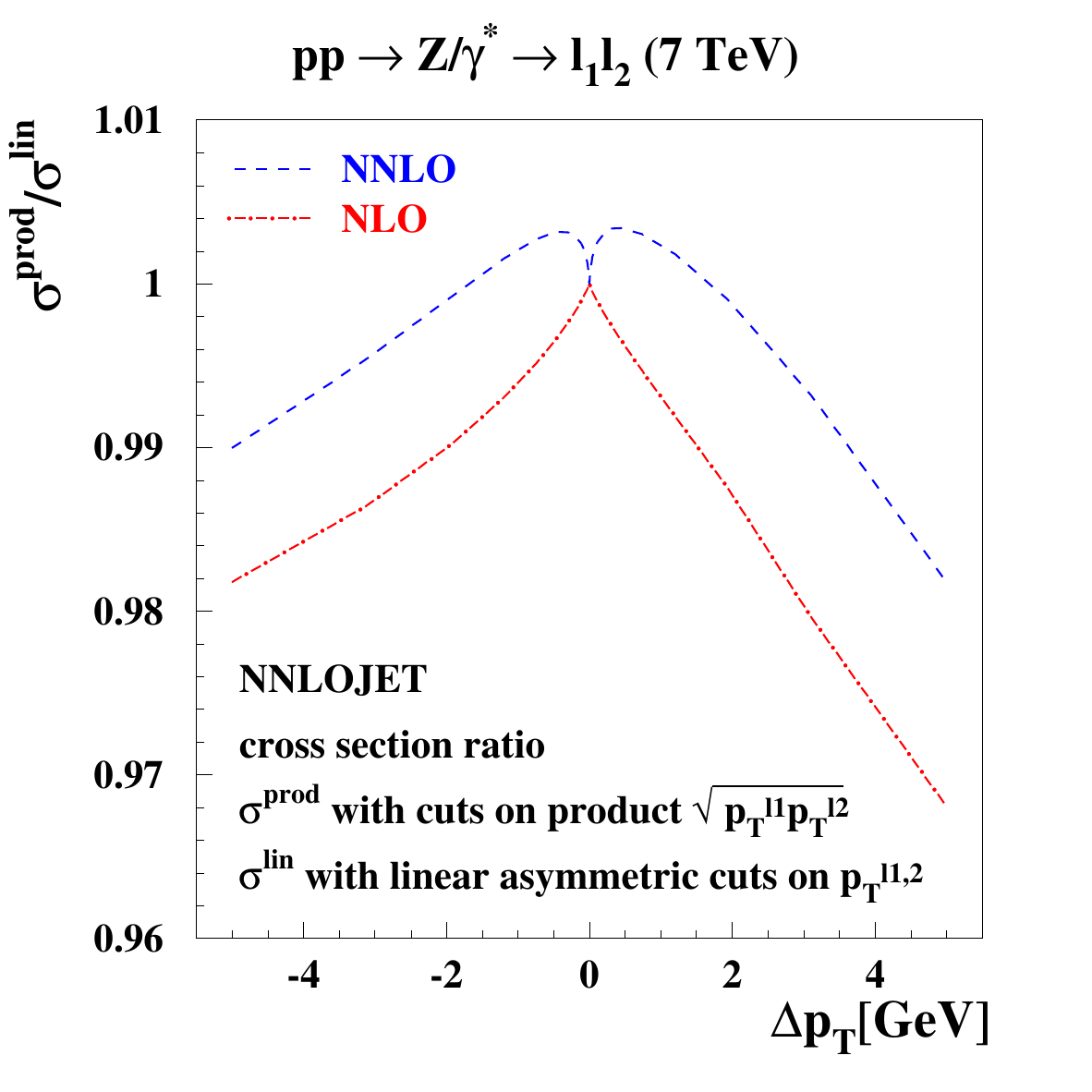}
\includegraphics[width=8.15cm]{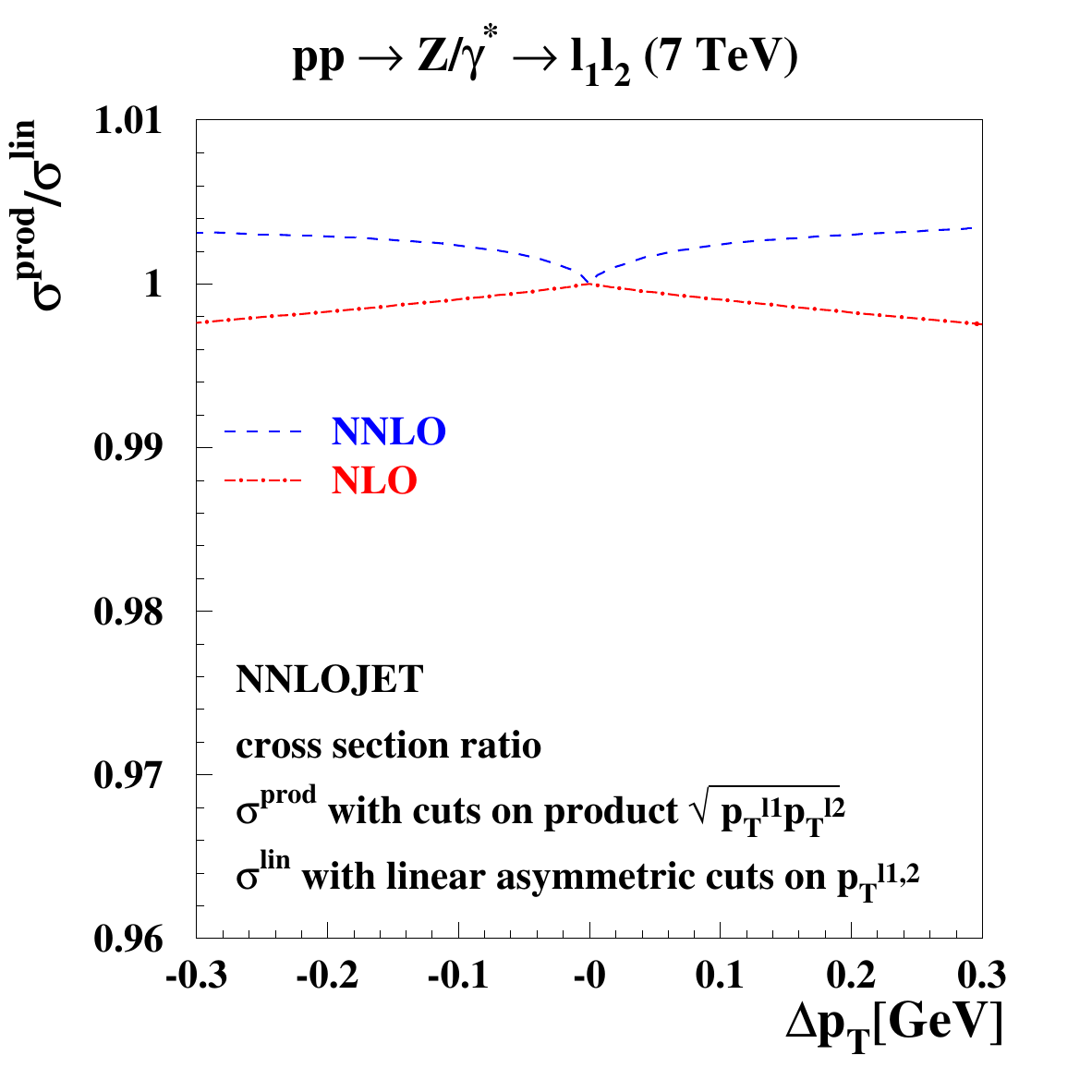}
\caption{\small
  \label{fig:relative}
  The ratio of cross sections for $pp \to Z/\gamma^*+X \to l^+l^-  + X$ production 
  from Fig.~\ref{fig:prodcuts} with the product cuts of Eq.~(\ref{eq:cuts_prod}) 
  over those from Fig.~\ref{fig:pTcuts} with linear asymmetric cuts of Eq.~(\ref{eq:cuts_asy})  
  as a function of $\Delta p_T$ at NLO (dashed) and NNLO (solid) in QCD.
  The ranges $\Delta p_T \in [-5,5]$\ GeV (left plot) and $\Delta p_T \in [-0.3,0.3]$\ GeV (right plot) are shown.
}
\end{center}
\end{figure}

Several resolutions to the problem of symmetric cuts have been suggested in the literature. In Ref.~\cite{Salam:2021tbm}, the origin of this behavior
was traced back to terms in the perturbative expansion of the cross section linear
in the transverse momentum $p_{T}$ of the decaying $Z$-boson that produces the lepton pair. It was suggested there to replace the cuts on the separate leptons with a cut on the product of the transverse momenta of the leptons, which we implement as follows:
\begin{equation}
\label{eq:cuts_prod}
\sqrt{p_T^{\ell_1} p_T^{\ell_2}} \geq
\begin{cases}
20 \,\text{GeV} \,, \quad &\Delta p_T < 0
\,, \\
20 \,\text{GeV} + |\Delta p_T|\,, \quad &\Delta p_T > 0
\,,\end{cases}
\qquad
p_T^{\ell_2} \geq
\begin{cases}
20 \,\text{GeV} - |\Delta p_T|\,, \quad &\Delta p_T < 0
\,, \\
20 \,\text{GeV} \,, \quad &\Delta p_T > 0
\,.\end{cases}
\end{equation}

The perturbative expansion for the cross section in this case depends only quadratically on the $Z$-boson transverse momentum, and it is interesting to study whether the unphysical behavior as $\Delta p_T\to 0$ improves in this case.
We show the predictions for these product cuts using {\tt NNLOJET} in Fig.~\ref{fig:prodcuts}.
A comparison of the relative difference between the regular cuts and the product cuts at NLO and NNLO is shown in Fig.~\ref{fig:relative}.
While the unphysical dependence on $\Delta p_T$ is lessened by switching to product cuts it is still clearly visible on the plots. We note that both the fixed-order and all-orders structure of the cross section
with product cuts defined in Eq.~(\ref{eq:cuts_prod}) are analogous to those discussed in the previous subsections,
including the possibility of a mismatch between linear slopes
due to the way $\Delta p_T$ is defined here.

\subsection{Rapidity distributions}
\label{sec:rapidity}
\begin{figure}[tbp]
\begin{center}
\includegraphics[width=8.15cm]{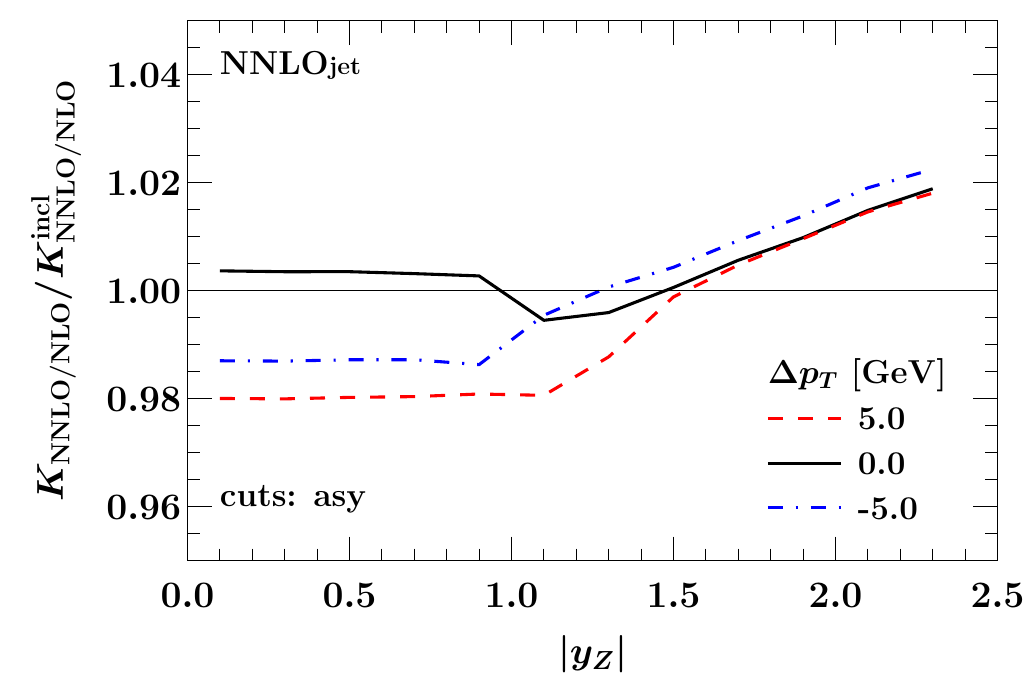}
\includegraphics[width=8.15cm]{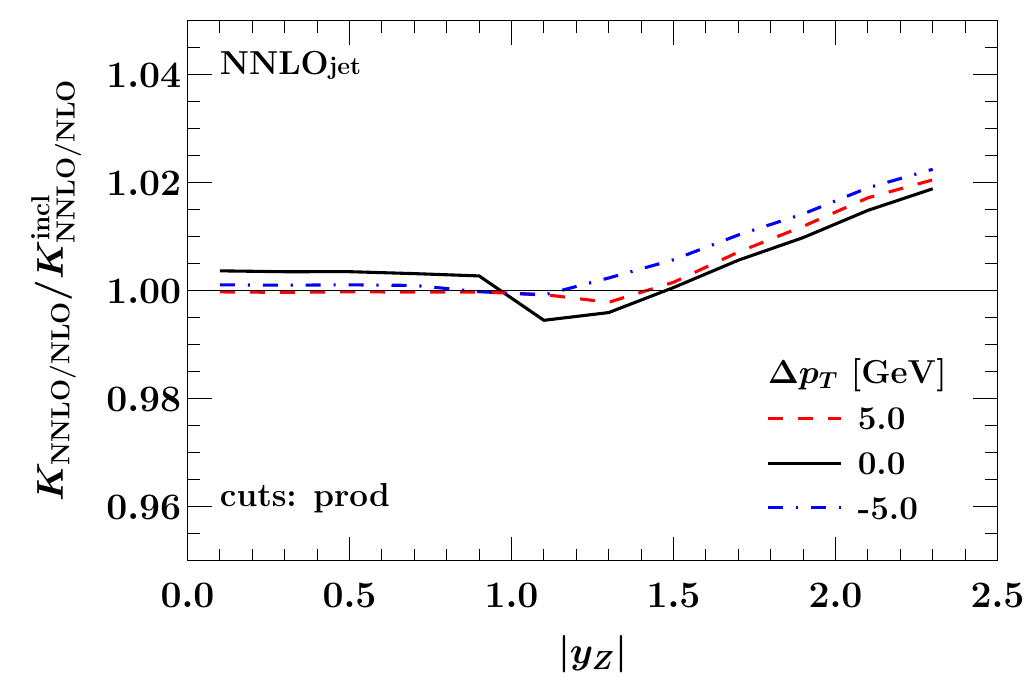}
\caption{\small
  \label{fig:asy-prodcuts-y}
  Deviation of the rapidity distribution 
  for $pp \to Z/\gamma^*+X \to l^+l^-  + X$ production 
  at $\sqrt{s}=7$\ TeV as a function of $\Delta p_T$ relative 
  to the results for $\Delta p_T=0$ at NNLO in QCD computed with {\tt NNLOJET}. 
  Shown are selected $\Delta p_T$ values (indicated by color) using the definition in Eq.~(\ref{eq:cuts_asy}) for linear asymmetric cuts (left plot) and 
  the one of Eq.~(\ref{eq:cuts_prod}) for product cuts (right plot).
  Solid lines denote negative values of $\Delta p_T$ as indicated in the plots, dotted lines of the same color display the corresponding $\Delta p_T$ values with positive sign.
  }
\end{center}
\end{figure}
\begin{figure}[tbp]
\begin{center}
\includegraphics[width=8.15cm]{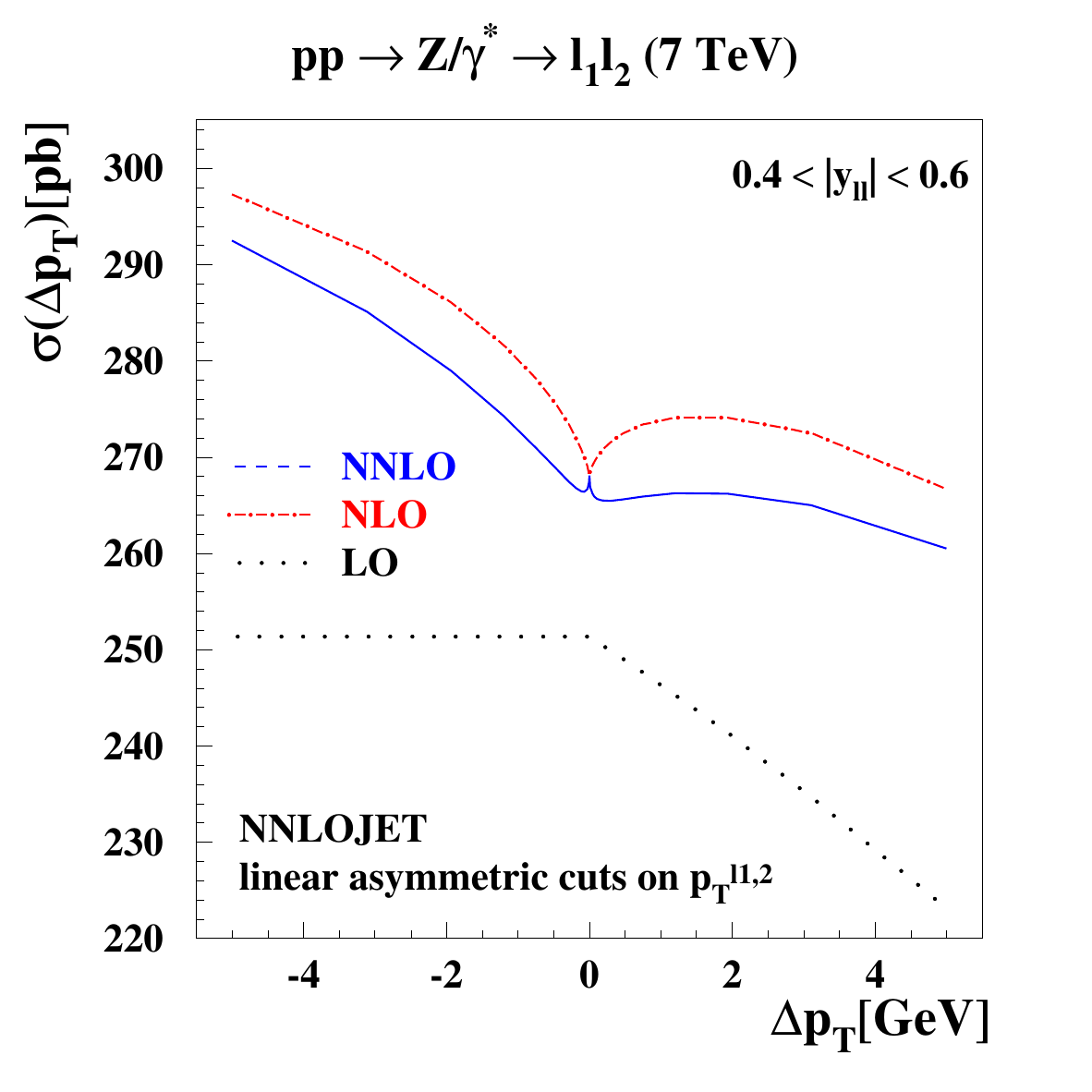}
\includegraphics[width=8.15cm]{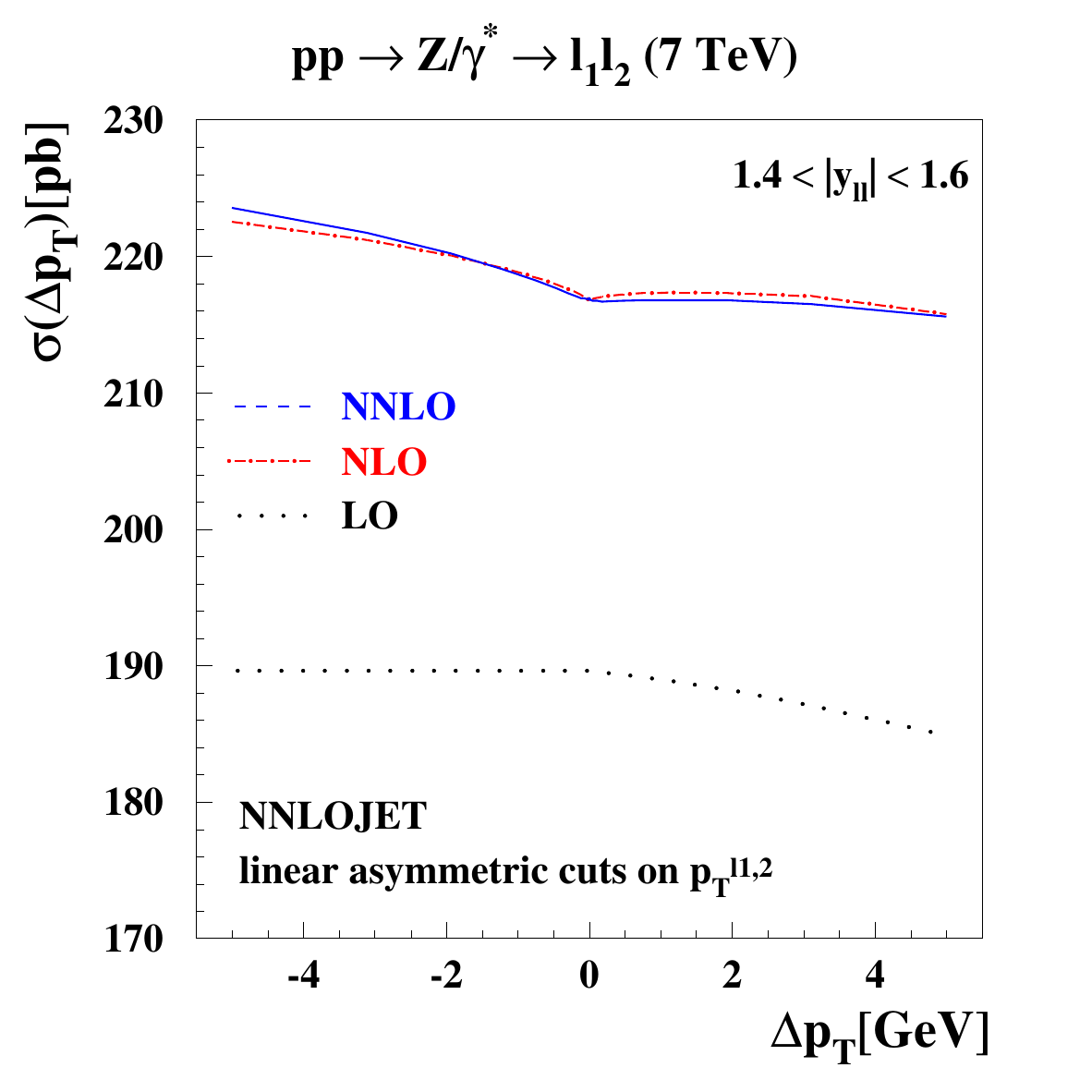}
\caption{\small
  \label{fig:asycuts-y05-y15}
    The cross sections for $pp \to Z/\gamma^*+X \to l^+l^-  + X$ production 
  at $\sqrt{s}=7$\ TeV at LO (dotted), NLO (dashed) and NNLO (solid) in QCD 
  with ABMP16 PDFs computed with {\tt NNLOJET} as a function of $\Delta p_T \in [-5,5]$\ GeV defined in Eq.~(\ref{eq:cuts_asy}) for the linear asymmetric fiducial cuts on the decay leptons in the final state.
    The rapidity bins $|y_{ll}| \in [0.4,0.6]$ (left plot) and $|y_{ll}| \in [1.4,1.6]$ (right plot) are shown.
}
\end{center}
\end{figure}
\begin{figure}[htbp]
\begin{center}
\includegraphics[width=8.15cm]{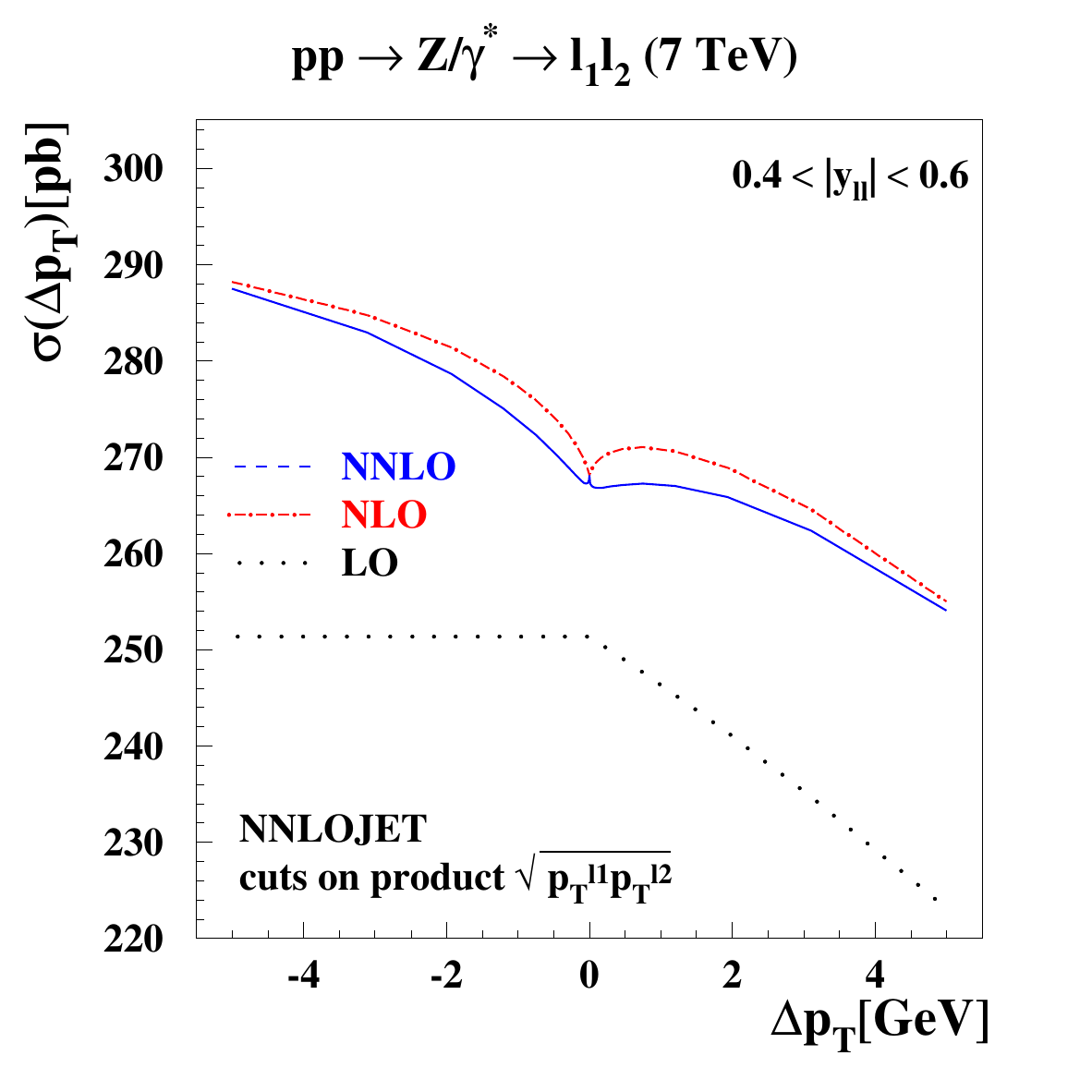}
\includegraphics[width=8.15cm]{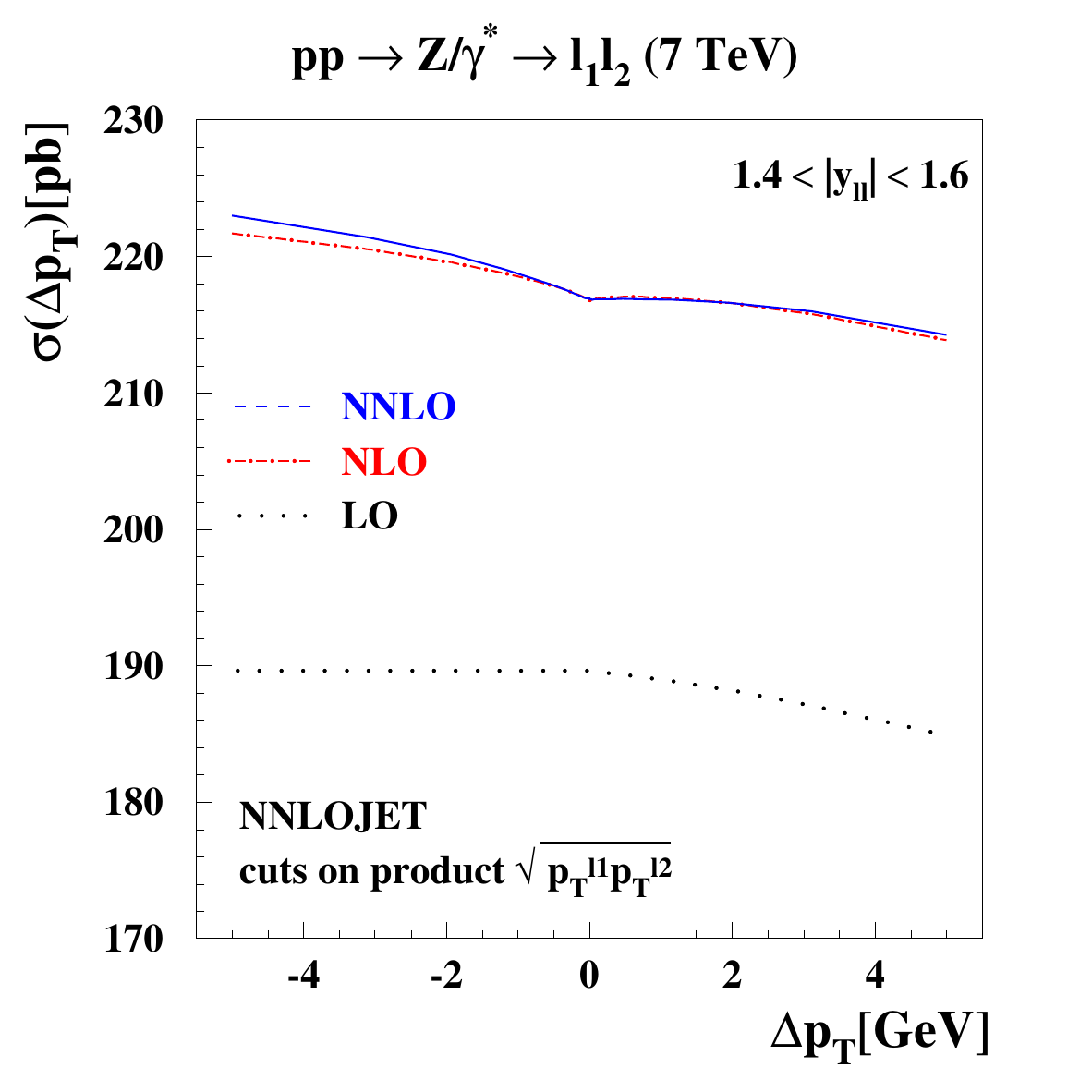}
\caption{\small
  \label{fig:prodcuts-y05-y15}
Fiducial NNLO to NLO $K$-factor as a function of $|y_{\ell \ell}|$
for linear asymmetric cuts (left)
and product cuts (right) for different values of $\Delta p_T$, normalised to the inclusive  NNLO to NLO $K$-factor.
}
\end{center}
\end{figure}

It is instructive to revisit the study of the $\Delta p_T$ dependence of the cross section for $Z$-boson production with central leptons, which directly relates to the numerical studies presented in Sec.~\ref{sec:benchmark}.
For selected fixed values of $\Delta p_T$ we display in Fig.~\ref{fig:asy-prodcuts-y} the $K$-factor $\sigma_{\rm NNLO}/\sigma_{\rm NLO}$ relative to the inclusive case for asymmetric cuts (left) and product cuts (right) as a function of the $Z$-boson rapidity.
In both figures we also display the $K$-factor with symmetric cuts in black.

For the selected values of $\Delta p_T$, we observe a difference at the 2\% level across the whole range in rapidity between the asymmetric case and the inclusive case.
There is a turning point halfway through the rapidity range considered, with the ratio moving from $\simeq -2 \%$ at central rapidity to $\simeq + 2 \%$ at $|y_{\ell \ell}| = 2.2$, with an abrupt change in slope starting around $|y_{\ell \ell}| \simeq 1$.
The symmetric cuts case displays a similar trend at large $|y_{\ell \ell}|$, while it is closer to the inclusive case at central rapidities.
Also in this case, we observe sudden changes in slope at central rapidities, even though the overall deviation with respect to the inclusive case is between [-0.5\%, +0.5\%] for $|y_{\ell \ell}|$ below 1.5.

On the other hand, the product cut curves display a smoother behaviour, and they are almost indistinguishable from the inclusive result for $|y_{\ell \ell}|$ below 1.2.
Above this value, the $K$-factors start to depart from the inclusive case with a slope similar to the one observed in the asymmetric and the symmetric case, with differences up to  $\simeq +2 \%$.
For the kinematics under consideration ($pp \to Z/\gamma^*+X \to l^+l^-  + X$ production at $\sqrt{s}=7$\ TeV with cuts on the decay leptons $p_T^{\ell} \geq 20 \,\text{GeV}$), linear power corrections in $\Delta p_T$ appear for central rapidities, $y_{ll} \lesssim 1.0$, due to the fiducial cuts breaking the azimuthal symmetry in the integral over the lepton decay phase space, as discussed in Refs.~\cite{Ebert:2019zkb}, \cite{Ebert:2020dfc}.
For larger rapidities $y_{ll} \gtrsim 1.5 $ the rapidity constraints dominate the integral over the lepton phase space and azimuthal symmetry is restored, resulting in quadratic power corrections in $\Delta p_T$. This leads to only small deviations of all predictions from the case of symmetric cuts.
The impact of the cuts on the lepton decay phase space has been illustrated in our previous study~\cite{Alekhin:2021xcu}, including the region around around $y_{ll} \simeq 1.2$, where the transition occurs~\footnote{
The impact of the fiducial cuts applied to $Z/\gamma^*$-boson production for different values of the gauge-boson rapidity $y_{ll}$ on the real emission phase space of the decay leptons has been illustrated in Fig.~15 of Ref.~\cite{Alekhin:2021xcu}.}.

The fact that the product cut $K$-factors display a better behaviour at central rapidities shows that, also at the differential level, product cuts are free of linear power corrections, which are instead responsible for the larger differences observed in the asymmetric case and, to some extent, in the symmetric case.
The effect observed are, however, limited to a few percent, and do not display the much larger differences observed in the Higgs case when asymmetric cuts are enforced~\cite{Chen:2021isd}.
Moreover, our analysis is performed at NNLO, while larger shape distortions were observed in the Higgs case at N$^3$LO.
A better assessment of the impact of linear power corrections associated with the choice of cuts requires to consider resummation effects at NNLL and beyond, which we will discuss in the next section.

\newpage
\clearpage

\section{Resummation}
\label{sec:resummation}

Using resummation techniques, e.g.\ 
in the framework of soft-collinear effective theory (SCET), we can quantify the structure of the cross section further. We produce matched predictions of the form
\begin{align} \label{eq:matched_xsec}
\sigma_\mathrm{match}
= \sigma_\mathrm{FO}
+ \int_{0}^{q_T^\mathrm{off}} \! d q_T \,
\left[
   \frac{d \sigma_\mathrm{res}}{d q_T}
   - \frac{d \sigma_\mathrm{sing}}{d q_T}
\right]
\,,\end{align}
where $\sigma_\mathrm{FO}$ is the fixed-order result,
$d \sigma_\mathrm{res}/d q_T$ is the all-orders resummed fiducial $q_T$ spectrum,
and $d \sigma_\mathrm{sing}/d q_T$ is its fixed-order expansion. The latter spectra can be computed by \texttt{SCETlib} using the settings of Ref.~\cite{Ebert:2020dfc}.
The upper limit $q_T^\mathrm{off}$ of the integral is chosen high enough
such that all profile scale functions (and thus the resummation)
are turned off exactly at $q_T > q_T^\mathrm{off}$,
and the resummed and singular cross sections cancel exactly beyond that point.
In practice we pick $q_T^\mathrm{off} = 150 \,\text{GeV}$
such that this criterion is fulfilled
even for the highest $m_{\ell\ell} \leq 150 \,\text{GeV}$ values that we keep.
For our central choice of profile scale transition points, the resummation
is actually fully off already for $q_T \geq 0.9 \, m_{\ell\ell}$.
For most of the $Z$ resonance the cancellation is therefore exact much earlier.
It is useful to further define the non-singular cross section
\begin{align} \label{eq:def_nons}
\sigma_\mathrm{nons}
\equiv \sigma_\mathrm{FO}
- \int_{0}^{q_T^\mathrm{off}} \! d q_T \,
   \frac{d \sigma_\mathrm{sing}}{d q_T}
\,.\end{align}
The relevant physical settings and cuts on $Q \equiv m_{\ell\ell}$ and $Y \equiv Y_{\ell\ell}$
are given by the input file in App.~\ref{sec:appD}.
Note that we always cut on the lepton pseudorapidities
$|\eta_{\ell_1,\ell_2}| \leq 2.5$ in addition.
The cuts on the lepton transverse momenta are given in Eqs.~(\ref{eq:cuts_asy}) and (\ref{eq:cuts_prod}) for linear asymmetric and product cuts, respectively.

\subsection{Structure of the fixed-order cross section and terms captured by resummation}
\label{sec:structure_fo}

Returning to the general all-orders form of the fixed-order cross section
in Eq.~(\ref{eq:delta_pt_fo_xsec_all_orders}),
it is important to ask which terms are actually being captured and resummed
by the singular cross section $d \sigma_\mathrm{sing}$
when dividing up the fixed-order cross section as in Eq.~(\ref{eq:def_nons}).
In general, we expect that logarithms ($m > 0$ for $\ln^m(\pm x)$ in Eq.~(\ref{eq:delta_pt_fo_xsec_all_orders}))
arise from corresponding logarithmic terms
in the expansion of the hadronic structure functions $W_i$
to some (high) power in $q_T/Q$,
\begin{align}
2\pi q_T \, W_i(q^\mu) = \sum_{n=0}^\infty \alpha_s^n \biggl\{ &
   A_{i,n,0}(Q, Y) \, \delta(q_T)
   + \sum_{m > 0}^{2n} A_{i,n,m}(Q,Y) \, \frac{1}{Q} \left[ \frac{\ln^{m-1} (q_T/Q)}{q_T/Q} \right]_+
\nn \\ & \quad
+ \sum_{k = 0}^\infty (q_T/Q)^k \sum_{m = 0}^{2n} B_{i,n,m}^{(k)}(Q, Y) \ln^m(q_T/Q)
\biggr\}.
\end{align}
Here we work in the notation of Ref.~\cite{Ebert:2020dfc},
where the inclusive cross section $d \sigma/d^4 q$ is proportional to $W_{-1} + W_0/2$,
while for $i \geq 0$ the $W_i \propto A_i \, d \sigma/d^4 q$ are in direct correspondence
to the standard angular coefficients $A_i$ up to leptonic prefactors
like electroweak charges and electroweak gauge boson propagators.
The limit $q_T \ll Q$ is the only relevant source of large logarithms
as a function of $q^\mu$, i.e., the momentum transfer between the hadronic and leptonic systems,
because other large (threshold) logarithms present in the partonic structure functions
are cut off by PDF suppression and the proton-proton kinematics at the hadronic level.

It is easy to verify that for the observable at hand,
only the hadronic structure functions $i = -1, 0, 2$ contribute.
The leptonic phase space integral vanishes for $i = 3 \dots 7$,
while for $i = 1$ it is an odd function of $Y$ and thus vanishes when integrated
against the even structure function.
For these we have
\begin{align}
A_{i, n, m} &= 0 \,, \quad i = 0, 2
\,, \nn \\
B^{(0)}_{i, n, m} &= 0 \,, \quad i = -1, 0, 2
\,,\end{align}
where the latter relation holds exactly
for $i = -1, 0$ and within twist-two collinear factorization for $i = 2$.
All the relevant hadronic power corrections
to the leading-power factorization predicting the $A_{-1, n, m}$
are thus suppressed by at least two relative powers of $(q_T/Q)^2$
and scale as $(q_T/Q) \ln^m (q_T/Q)$.
Since we expect the leptonic phase space integral
to lower the degree of divergence by at least one power,
cf.\ the leading-power and tree-level $\delta(q_T) \sim q_T^{-1}$ being mapped
onto the coefficients 
$c_0 \sim x^0 $, $b_{0,0} \, x \sim x^1$, and $a_{0,0} = 0$ 
in Eq.~(\ref{eq:delta_pt_fo_xsec_all_orders}), we conclude that these at most contribute
to the terms of $\mathcal{O}(x^2)$,
while all coefficients $a_{n,m}$ and $b_{n,m}$ for $m > 0$
are predicted in terms of the $A_{-1, n, m}$.
These are fully captured by the leading-power (LP) factorized singular cross section,
\begin{align} \label{eq:lp_singular_xsec}
\frac{d\sigma_\mathrm{sing}}{d Q^2 \, d Y\, d q_T} 
&\propto 2 \pi q_T W_{-1}^\mathrm{LP}(q^\mu)
\nn \\
&=  \sum_{n=0}^\infty \alpha_s^n \biggl\{ A_{-1,n,0}(Q, Y) \, \delta(q_T)
   + \sum_{m > 0}^{2n} A_{-1,n,m}(Q,Y) \, \frac{1}{Q} \biggl[ \frac{\ln^{m-1} (q_T/Q)}{q_T/Q} \biggr]_+
\biggl\} 
\,,\end{align}
where the proportionality again means up to leptonic prefactors.
The fiducial singular $q_T$ spectrum entering Eqs.~(\ref{eq:matched_xsec}) and (\ref{eq:def_nons}),
is then obtained by integrating Eq.~(\ref{eq:lp_singular_xsec})
over $Q^2$, $Y$, and the leptonic decay phase space,
while weighting by $1 + \cos^2 \theta$ and the fiducial acceptance.
In the resummed case, the second line of Eq.~(\ref{eq:lp_singular_xsec})
is evaluated using the all-orders resummation instead,
which we perform as described in Ref.~\cite{Ebert:2020dfc}.

Importantly, the total offset $c_n$ and
the remaining linear slope terms $a_{n,0}$ and $b_{n,0}$
on either side of Eq.~(\ref{eq:delta_pt_fo_xsec_all_orders}) are not predicted by this argument
because they can arise from rational integrals
over contributions from the cross section at any $q_T$. They do not have to arise from integrals over the $1/q_T$ singularity at small $q_T \ll Q$
because they are not transcendental.
This is analogous to the case of Ref.~\cite{Bhattacharya:2023qet},
where resummation predicted the second derivative
of the spectrum of interest,
meaning that a constant and a slope term
had to be obtained as boundary conditions
in a double integral through fixed-order matching.
(The discussion in Sec.~\ref{sec:structure_physical} implies that in the case at hand,
$a_{n,0}$ and $b_{n,0}$ are not actually connected to each other
due to the definition of $\Delta p_T$.)
For fiducial Drell-Yan production,
the more complicated double integral functional acting
on the underlying hadronic structure functions predicted by resummation
is instead given by the $q_T$ integral and the constrained leptonic phase-space integral.
As in Ref.~\cite{Bhattacharya:2023qet}, a natural consequence of this double integral
is the presence of ``non-singular'' terms
that feature the same or even higher power counting
as the singular terms predicted by factorization,
the only distinction being whether they are enhanced by logarithms as $x \to 0$,
i.e., whether their derivative is bounded (non-singular)
or logarithmically divergent (singular).

In conclusion, we predict that
\begin{align} \label{eq:structure_sing_nons_fid_xsecs}
\sigma_\mathrm{sing}(x)
&= \sum_{n=0}^\infty \alpha_s^n \biggl\{
   c_n^\mathrm{sing}
   + x \theta(-x) \, a_{n,0}^\mathrm{sing}
   + x \theta(x) \, b_{n,0}^\mathrm{sing}
\nn \\ & \qquad \qquad
   + x \theta(-x) \sum_{m = 1}^{2n} a_{n,m} \ln^m (-x)
   + x \theta(x) \sum_{m = 1}^{2n} b_{n,m} \ln^m x
   + \mathcal{O}(x^2)
\biggr\}
\,, \nn \\
\sigma_\mathrm{nons}(x)
&= \sum_{n=0}^\infty \alpha_s^n \biggl\{
   c_n^\mathrm{nons}
   + x \theta(-x) \, a_{n,0}^\mathrm{nons}
   + x \theta(x) \, b_{n,0}^\mathrm{nons}
   + \mathcal{O}(x^2)
\biggr\}
\,.\end{align}
Here we have moved the unique prediction of LP factorization
($a_{n,m}$ and $b_{n,m}$ with $m > 0$)
to the second line for the singular cross section.
In contrast, the precise breakdown of $a_{n,0}$, $b_{n,0}$, and $c_{n}$
between singular and non-singular cross section
at fixed order depends on $q_T^\mathrm{off}$,
the dependence on which cancels between the two terms.
Once the singular cross section is resummed,
the associated ambiguity is quantified by profile scale variations 
quoted as $\Delta_\mathrm{match}$ in Ref.~\cite{Ebert:2020dfc} and below,
because they determine down to which values of $q_T$
the resummed singular cross section is equal to its fixed-order counterpart.

In summary, the logarithmic terms $a_{n,m}$ and $b_{n,m}$ for $m \geq 1$
which lead to the unphysical non-monotonic behavior of the fixed-order cross section for $\Delta p_T \neq 0$ appear in the singular cross section.
This behavior is therefore cured by resummation. The discontinuity of the cross section derivative for $\Delta p_T = 0$ as encoded
in the coefficients $a_{n,0}$, $b_{n,0}$ remains present even after the resummation, in agreement with the general argument of Sec.~\ref{sec:structure_physical}.
Here the two contributions to the cross section separately obey monotonicity,
i.e., $a_{n,0}^\mathrm{sing} < 0$, $a_{n,0}^\mathrm{nons} < 0$ and $b_{n,0}^\mathrm{sing} = b_{n,0}^\mathrm{nons} = 0$.

\subsection{Numerical results for the non-singular cross section}
\label{sec:nonsingular-numerics}

\begin{figure*}
\centering
\includegraphics[width=8.15cm]{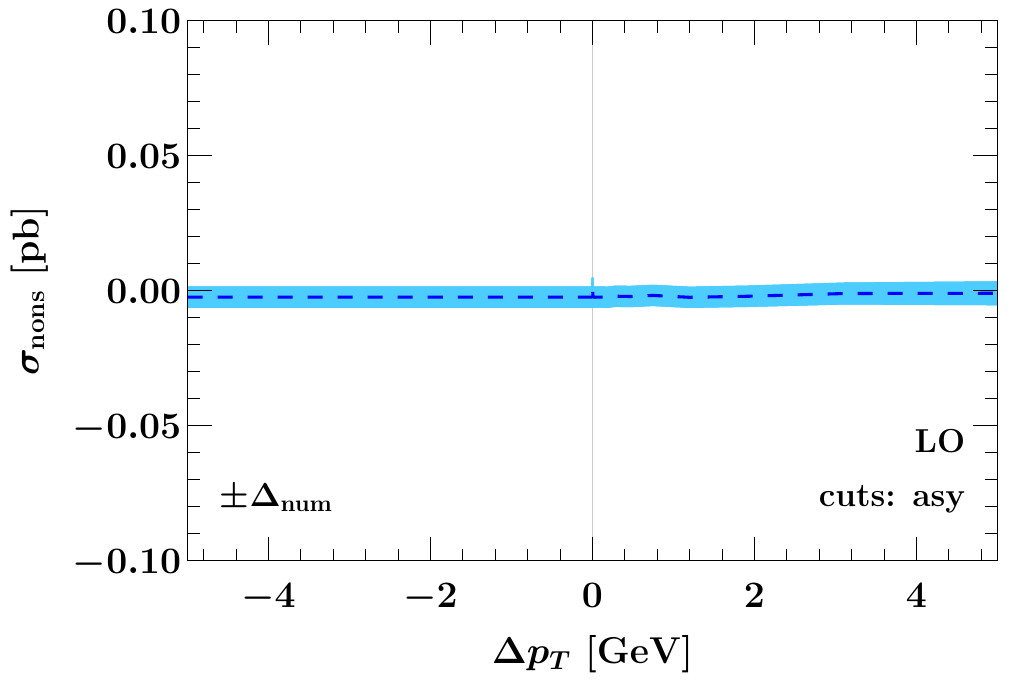}
\hfill
\includegraphics[width=8.15cm]{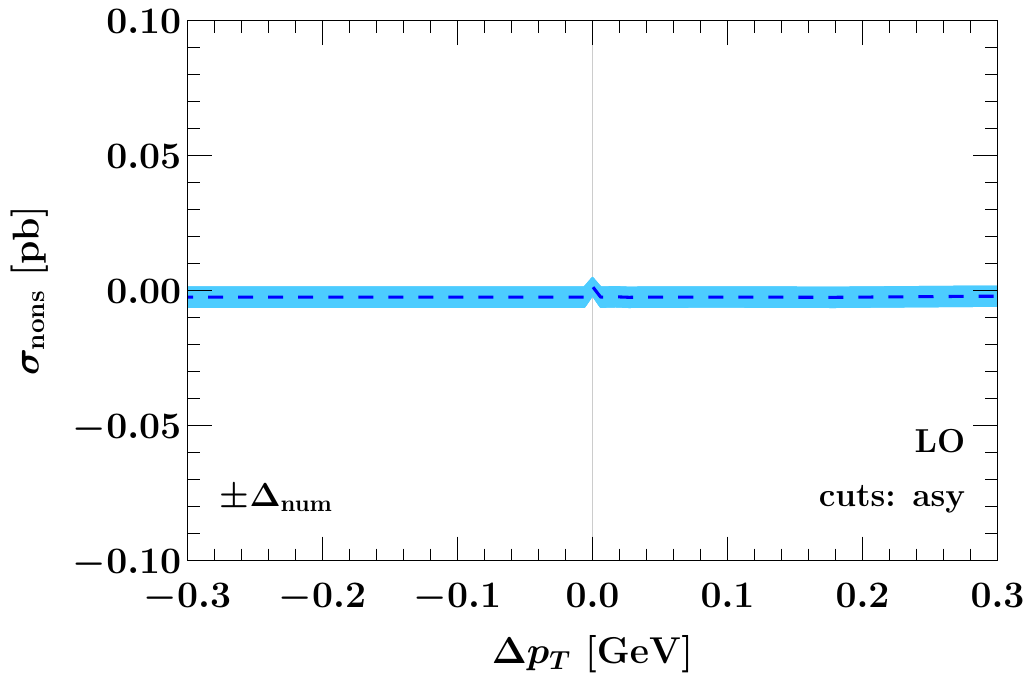}
\\
\includegraphics[width=8.15cm]{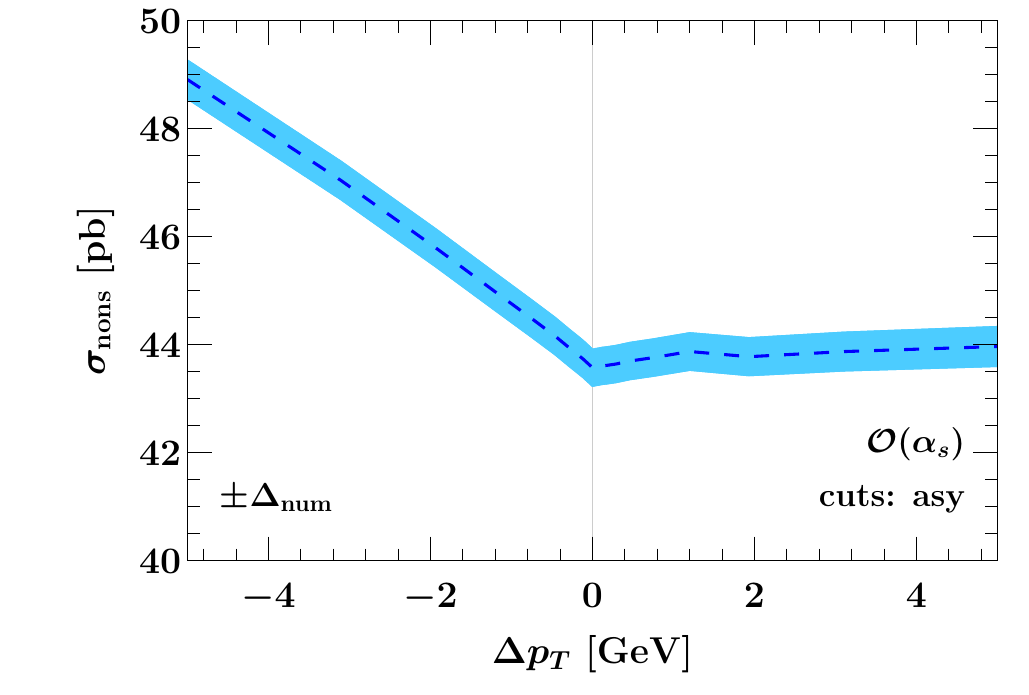}
\hfill
\includegraphics[width=8.15cm]{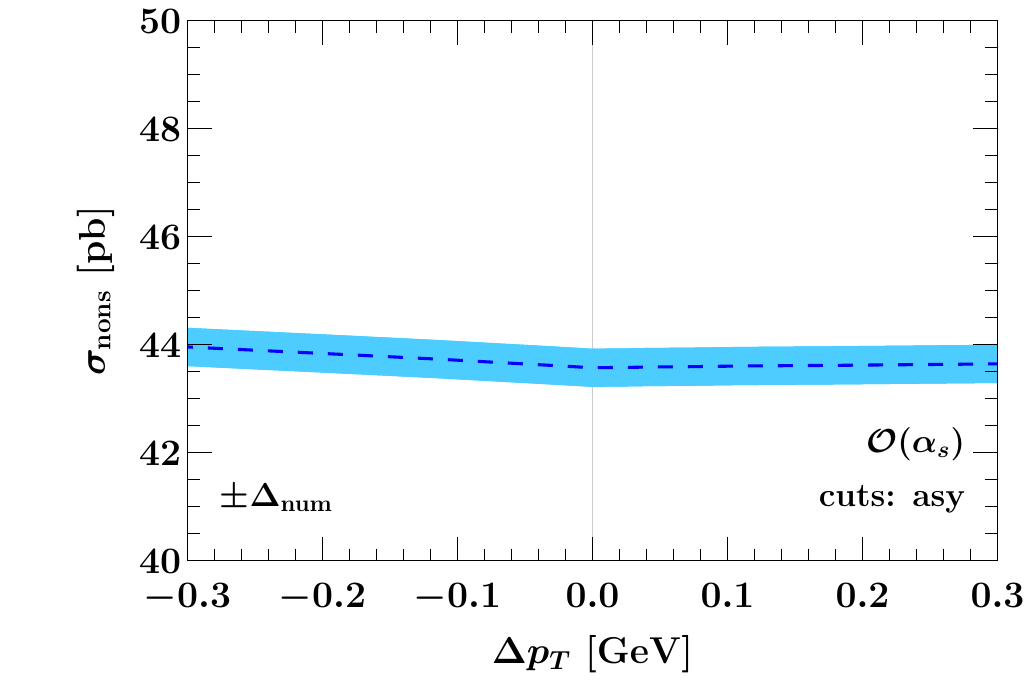}
\\
\includegraphics[width=8.15cm]{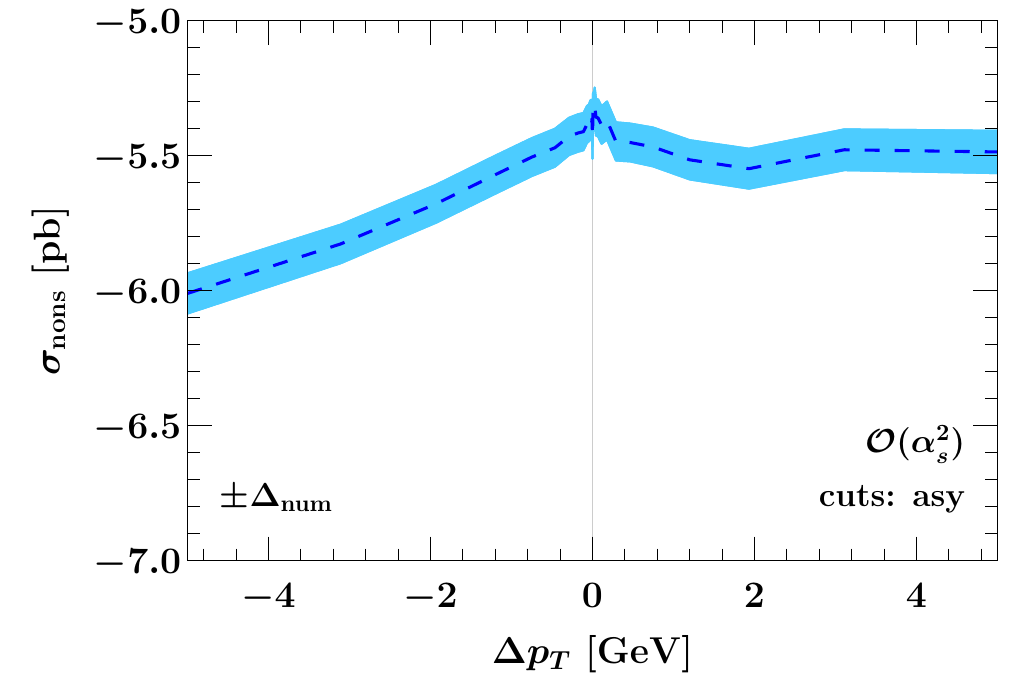}
\hfill
\includegraphics[width=8.15cm]{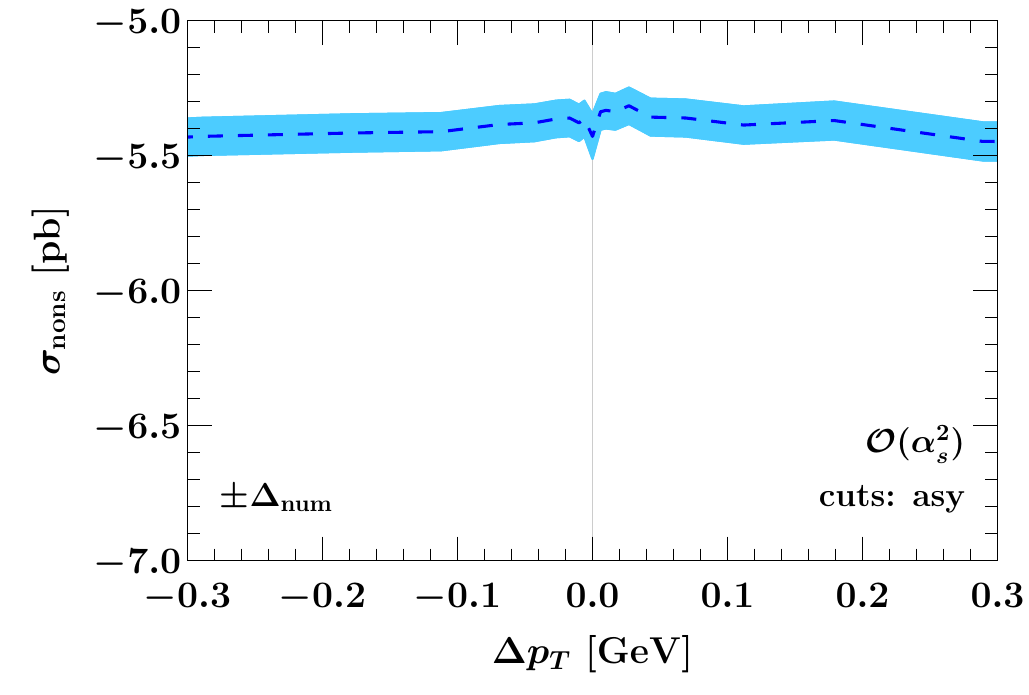}
\caption{\small
\label{fig:nons_asy}
The non-singular cross section with linear asymmetric cuts according to Eq.~(\ref{eq:cuts_asy}) 
as a function of $\Delta p_T$ on a wide (left) and zoomed-in scale (right).
The blue band indicates the numerical uncertainty, which is predominantly due to the \texttt{SCETlib} integration.
}
\end{figure*}
\begin{figure*}
\centering
\includegraphics[width=8.15cm]{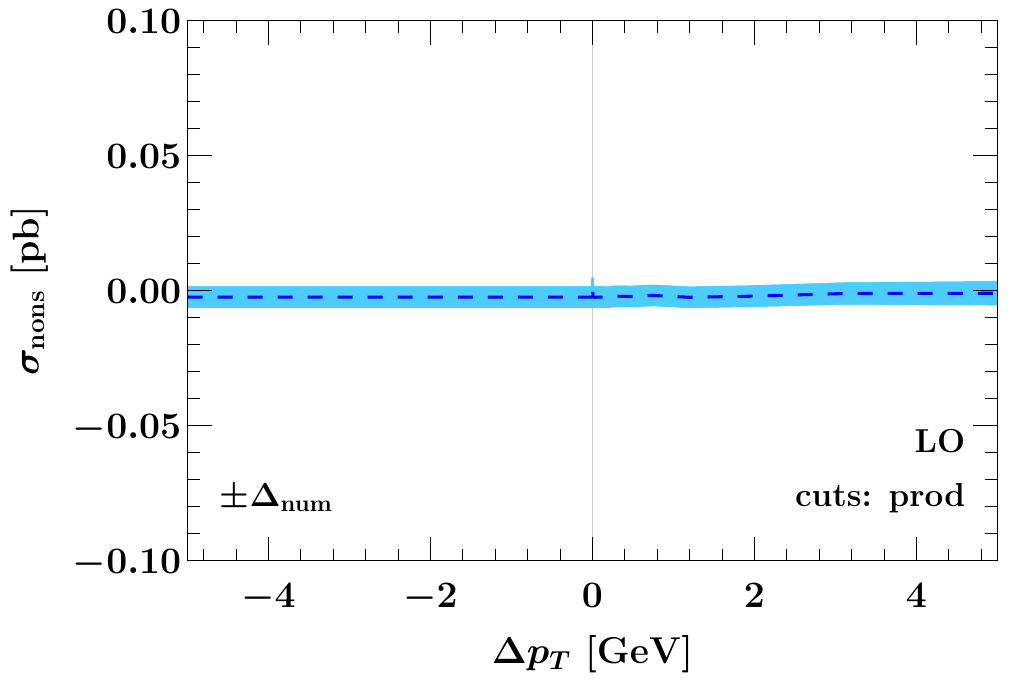}
\hfill
\includegraphics[width=8.15cm]{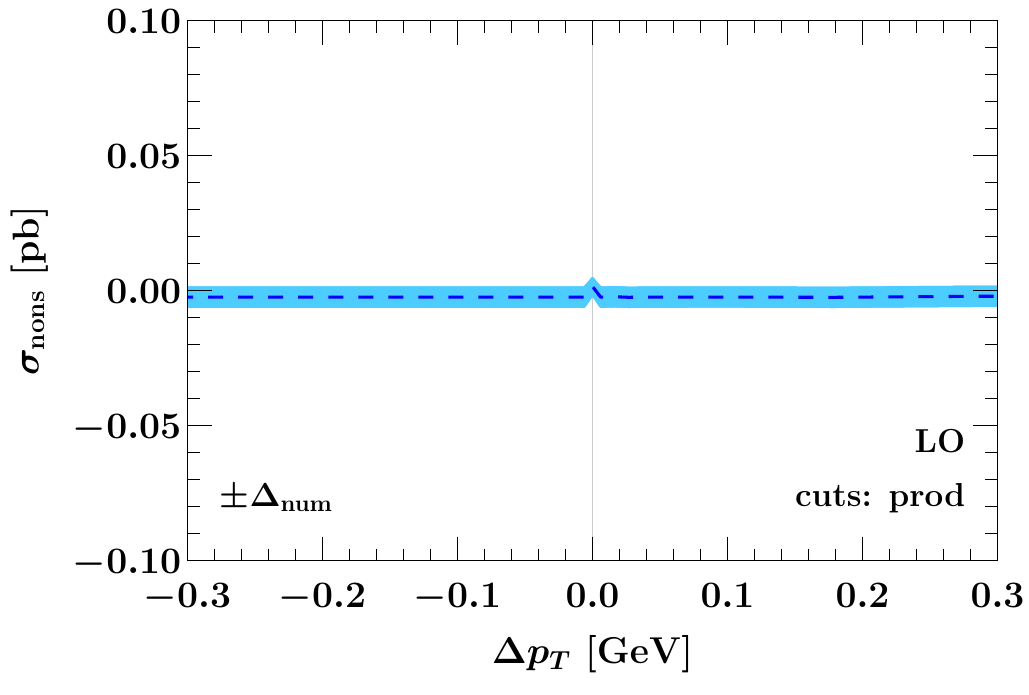}
\\
\includegraphics[width=8.15cm]{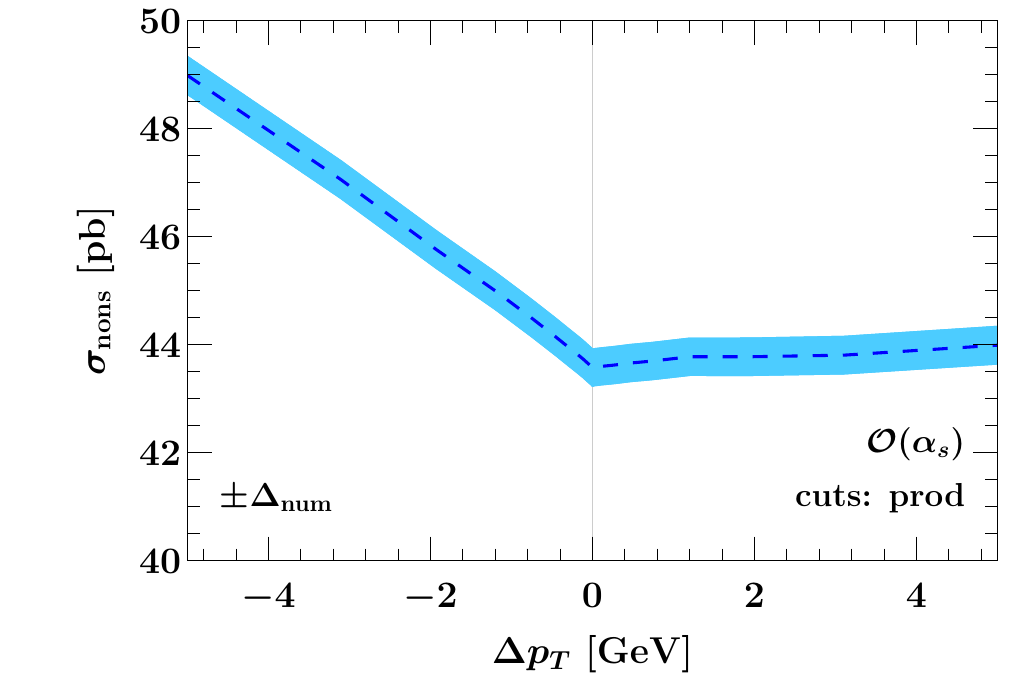}
\hfill
\includegraphics[width=8.15cm]{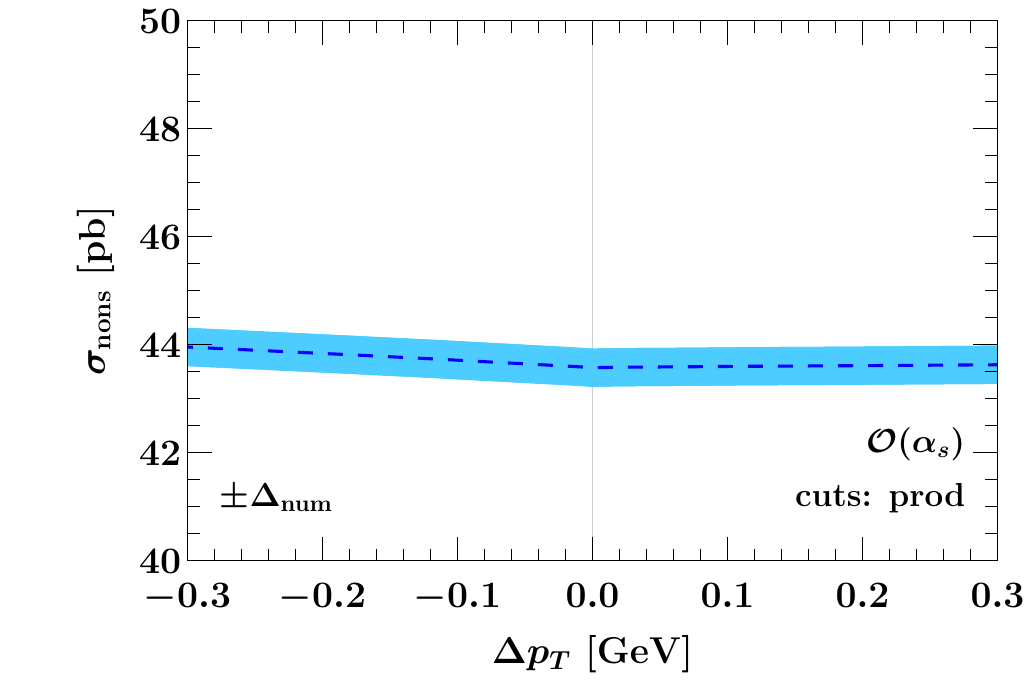}
\\
\includegraphics[width=8.15cm]{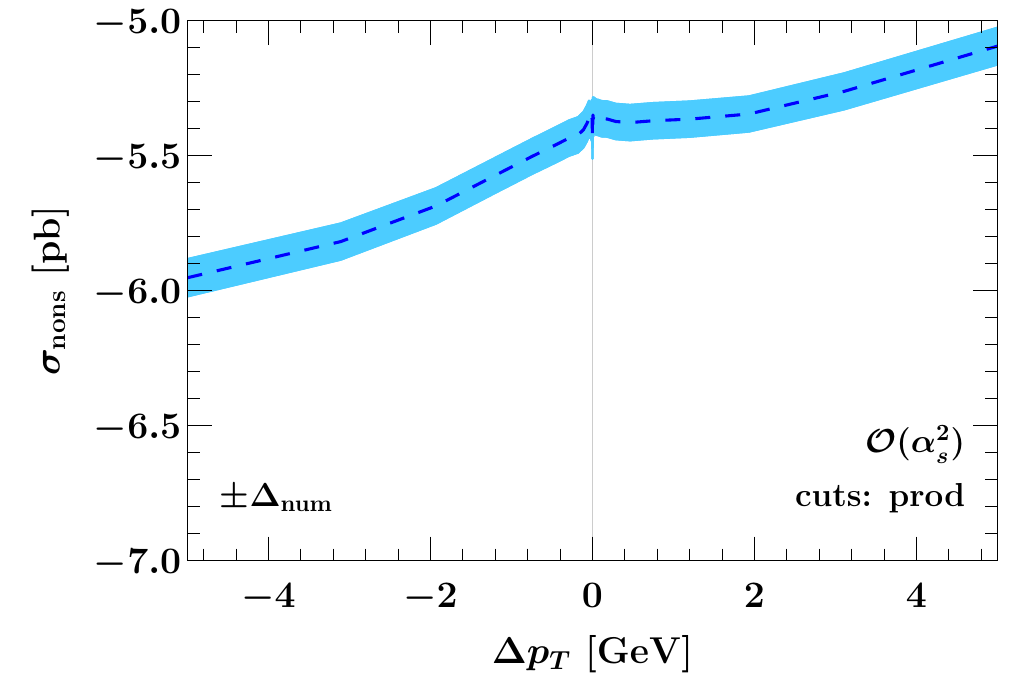}
\hfill
\includegraphics[width=8.15cm]{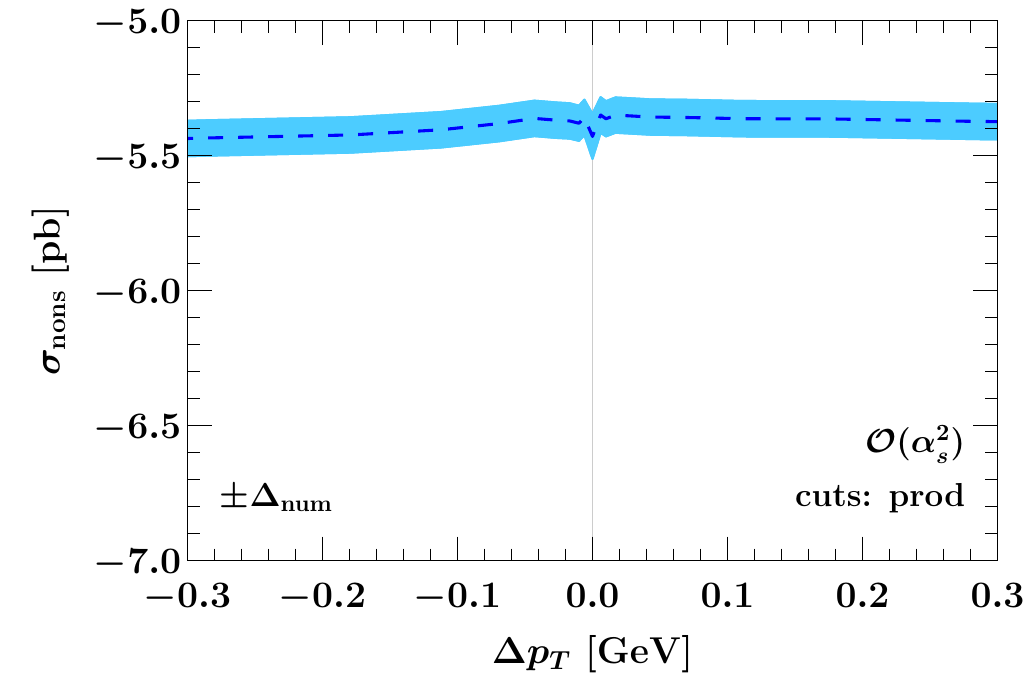}
\caption{\small
\label{fig:nons_prod}
Same as Fig.~\ref{fig:nons_asy} with product cuts according to Eq.~(\ref{eq:cuts_prod}).
}
\end{figure*}

In Fig.~\ref{fig:nons_asy} we show the fixed-order perturbative coefficients of the non-singular cross section including the respective factors of $\alpha_s^n$ in the case of the linear asymmetric cuts in Eq.~(\ref{eq:cuts_asy}).
Within the numerical uncertainty, which is indicated by the blue band
and driven by the \texttt{SCETlib} beam function
interpolation uncertainty, these are all compatible with an asymptotic behavior of two straight lines approaching a constant as $\Delta p_T \to 0$ with different slopes.
The LO non-singular cross section is compatible with zero within a
relative accuracy of $10^{-5}$, validating the \texttt{SCETlib} physics inputs against those used at fixed order (in \texttt{NNLOJET} here).
We stress that the nonsingular cross section in Eq.~(\ref{eq:def_nons})
is independent of the resummation formalism used.
This is because the singular spectrum, whose integral
is subtracted from the total fixed-order cross section in Eq.~(\ref{eq:def_nons}),
is specified by evaluating the underlying hadronic structure functions
in \emph{fixed-order perturbation theory} and expanding them to leading power in $q_T/Q$,
while keeping the exact dependence on $q_T$ in the leptonic phase space integral.
The choice to evaluate the structure functions in the Collins-Soper frame
and the choice to expand in $1/Q$ (instead of the inverse of the transverse mass)
amount to quadratic power corrections.
Due to their lower degree of divergence at the level of the $q_T$ spectrum,
these cannot induce logarithmic dependence on $\Delta p_T$,
and thus these choices (like the choice of $q_T^\mathrm{off}$)
only affect the constant terms and linear slopes in Eq.~(\ref{eq:def_nons}).
For the case of product cuts defined in Eq.~(\ref{eq:cuts_prod}) the non-singular cross sections
at LO, NLO and NNLO are plotted in Fig.~\ref{fig:nons_prod}. 
The only notable feature compared to Fig.~\ref{fig:nons_asy}
is a faster (possibly quadratic) rise of the $\alpha_s^2$ coefficient towards large positive values of $\Delta p_T$,
but this is at the scale of a few tenths of a picobarn.

\subsection{Matched predictions}
\label{sec:matching}

\begin{figure*}
\centering
\includegraphics[width=8.15cm]{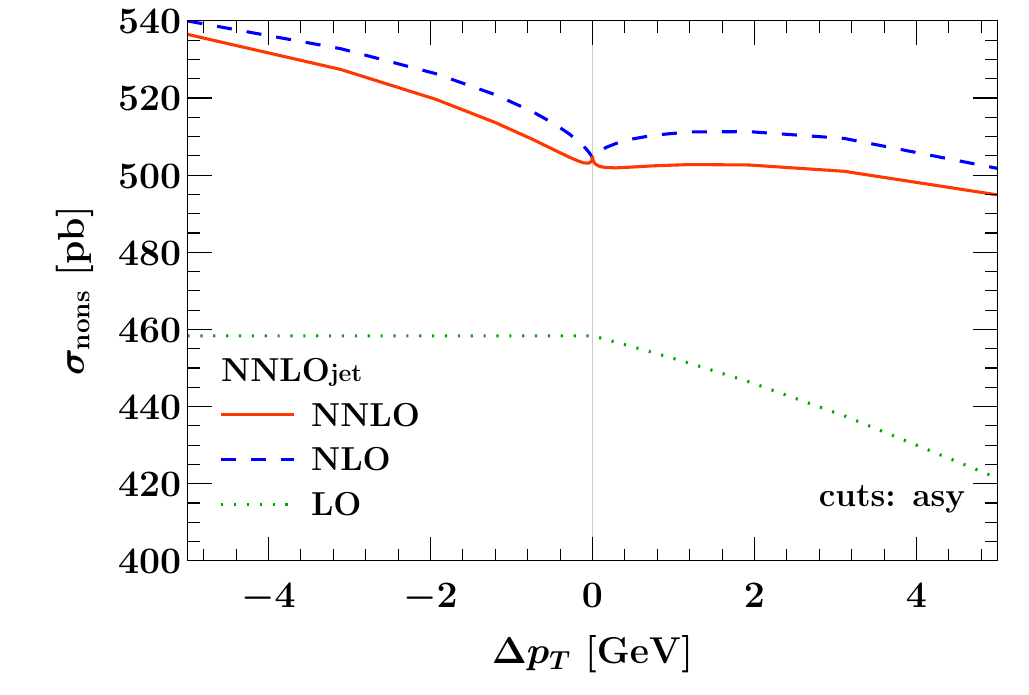}
\\
\includegraphics[width=8.15cm]{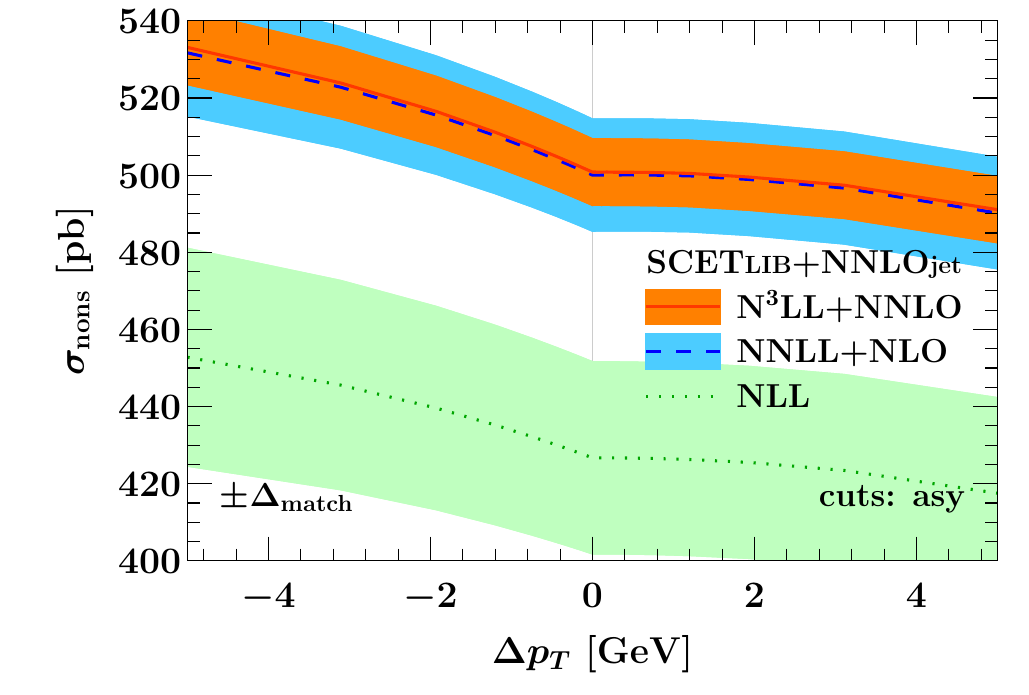}
\hfill
\includegraphics[width=8.15cm]{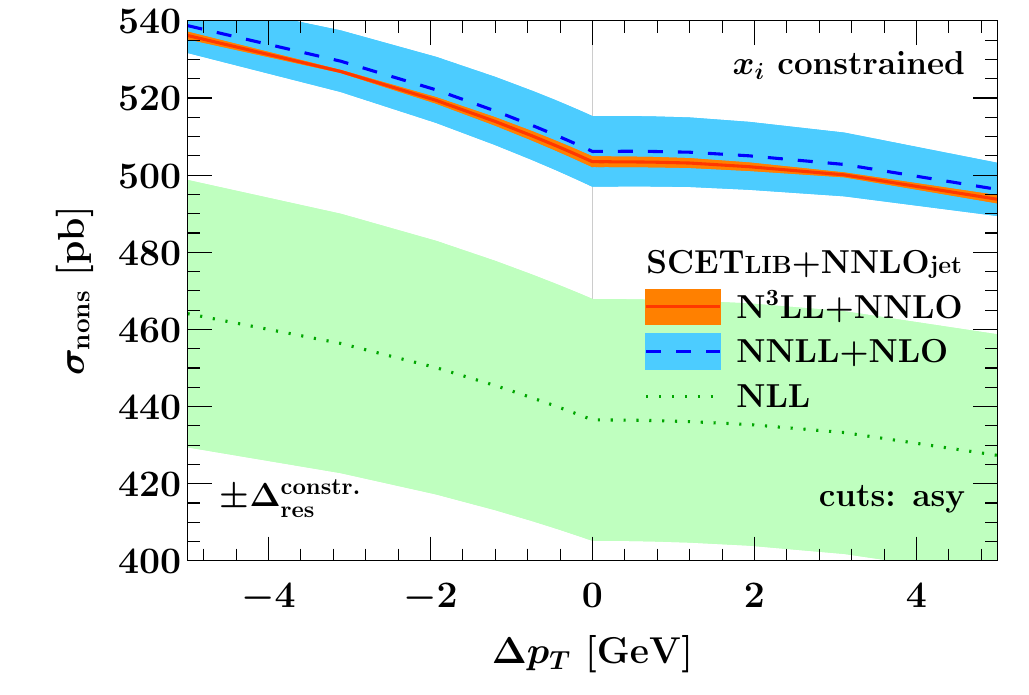}
\\
\includegraphics[width=8.15cm]{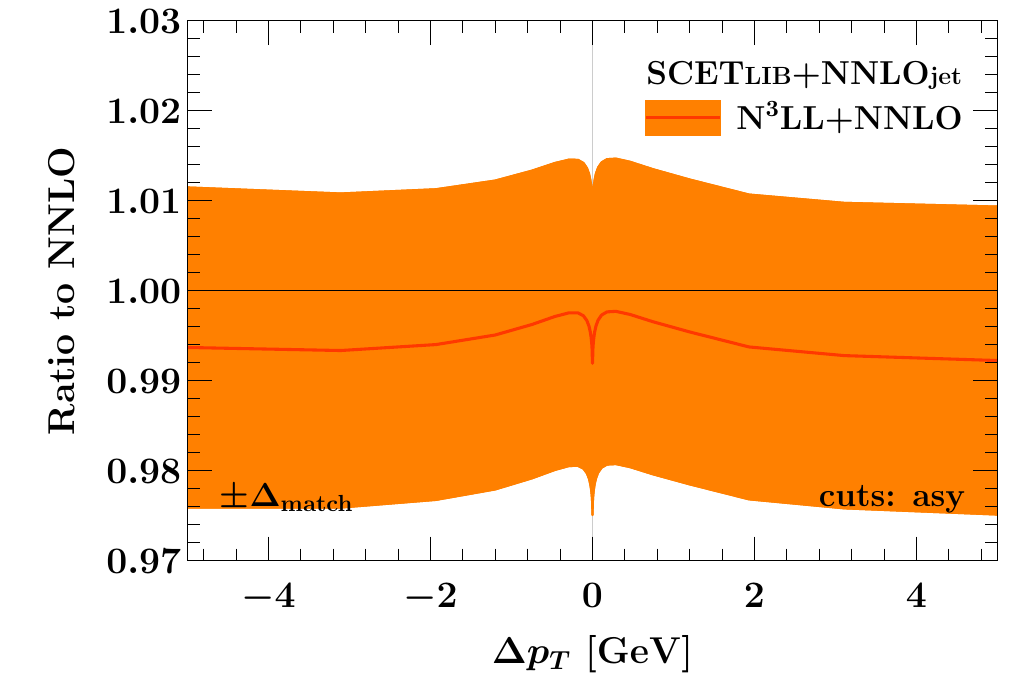}
\hfill
\includegraphics[width=8.15cm]{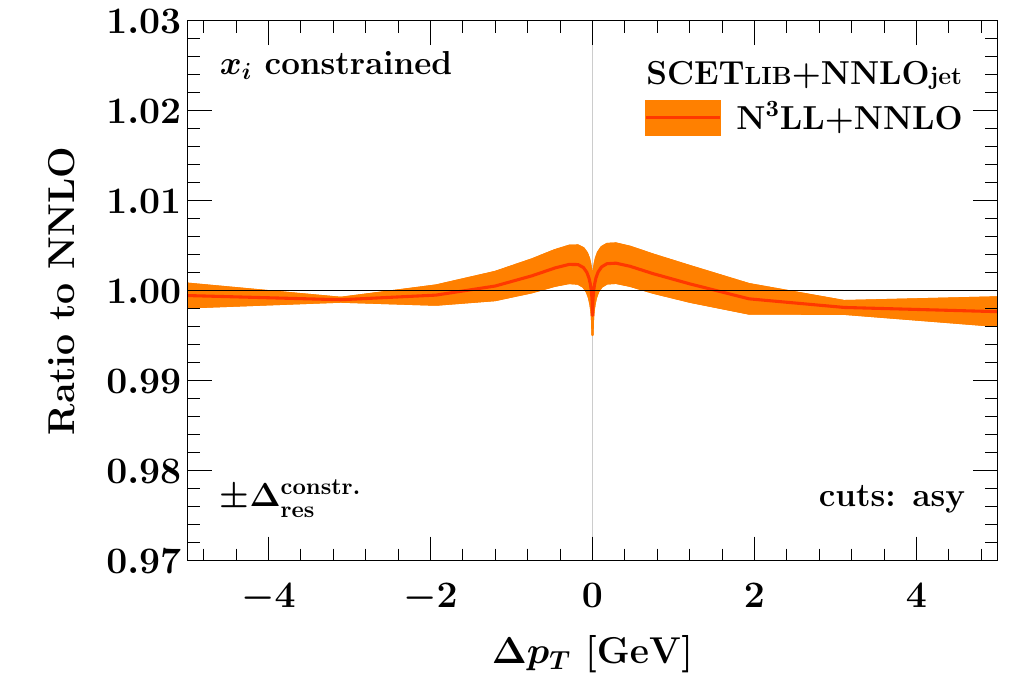}
\caption{
Results for the total fiducial cross section with linear asymmetric cuts according to Eq.~(\ref{eq:cuts_asy}) as a function of $\Delta p_T$ on a wide (left) and zoomed-in scale (right).
The bottom row shows the ratio to the NNLO prediction.
}
\label{fig:convergence_asy}
\end{figure*}
\begin{figure*}
\centering
\includegraphics[width=8.15cm]{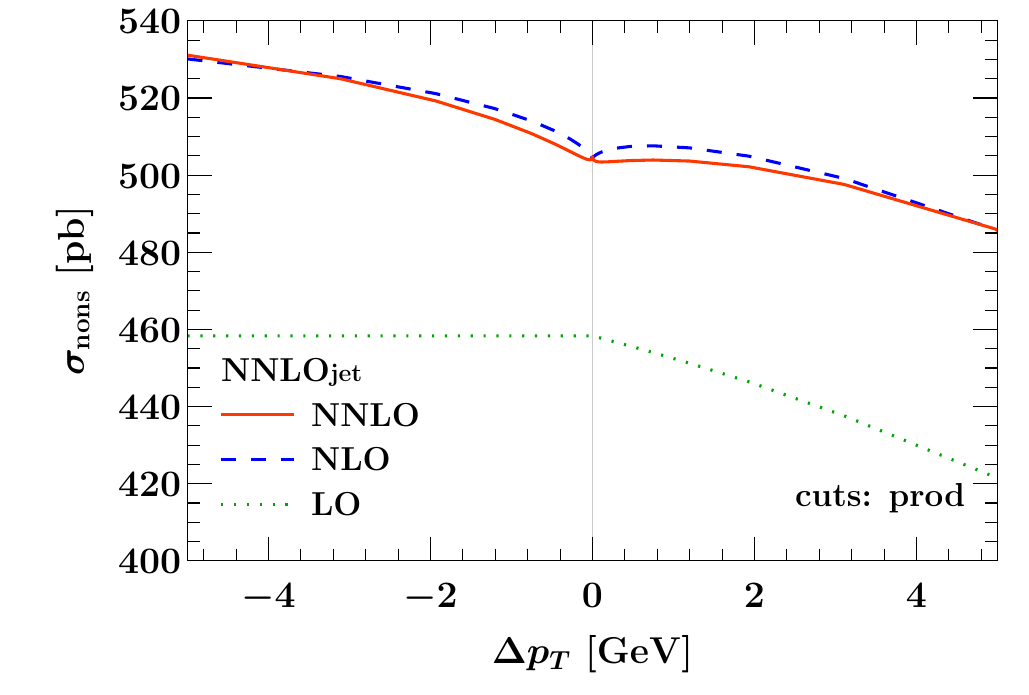}
\\
\includegraphics[width=8.15cm]{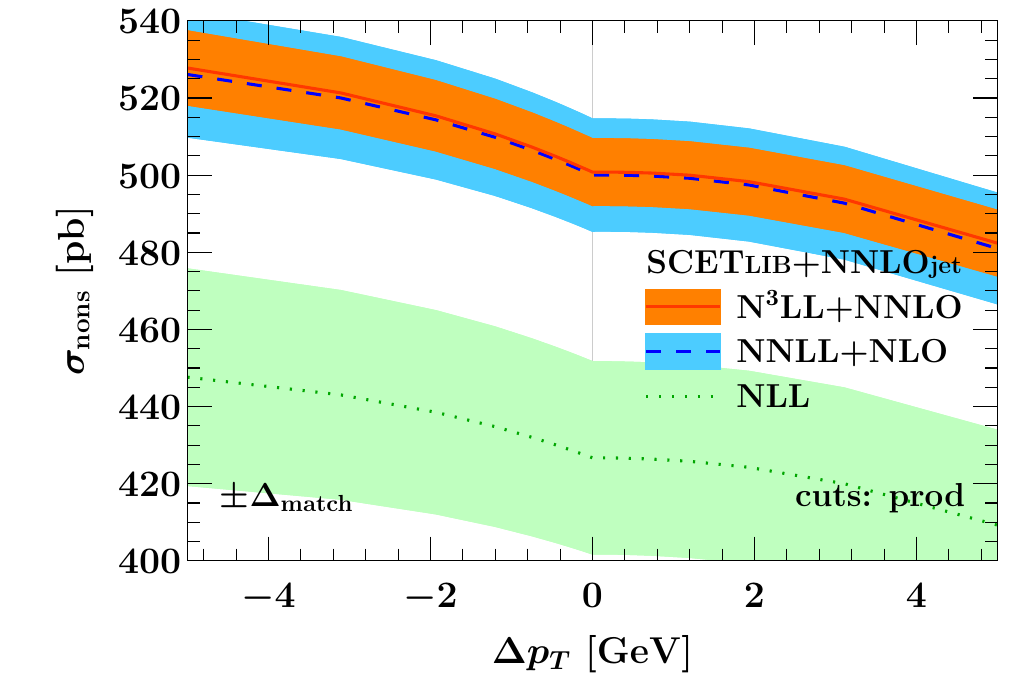}
\hfill
\includegraphics[width=8.15cm]{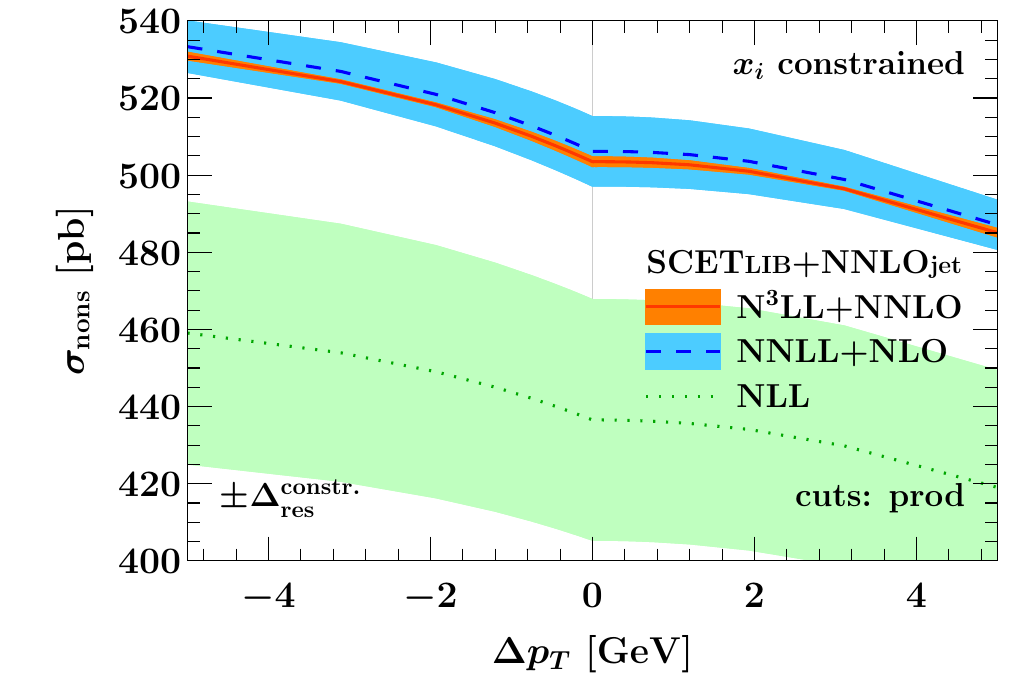}
\\
\includegraphics[width=8.15cm]{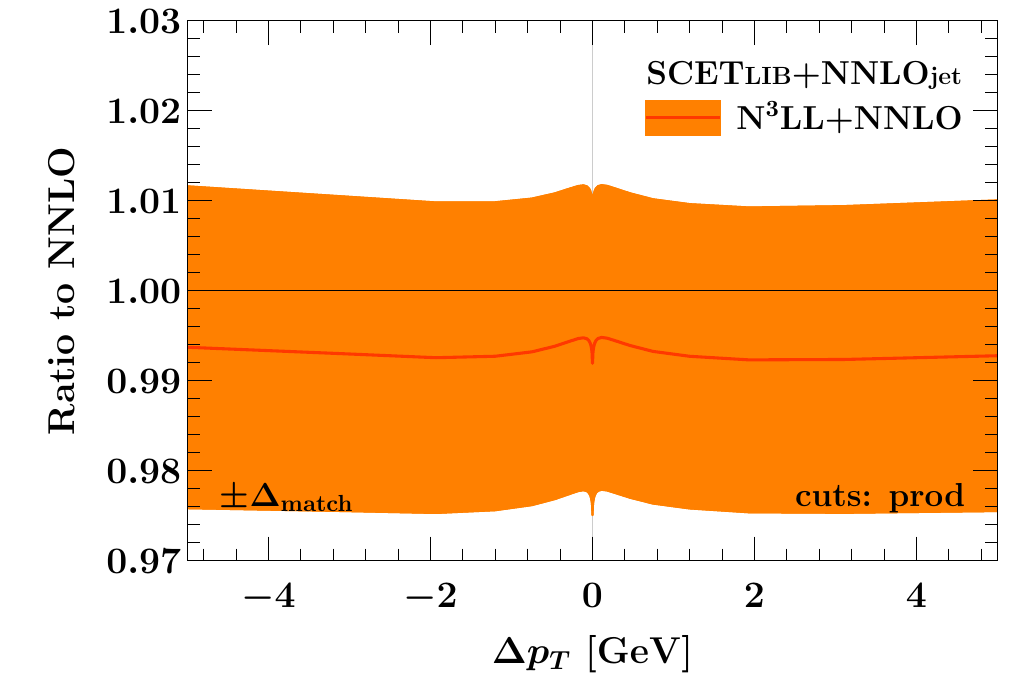}
\hfill
\includegraphics[width=8.15cm]{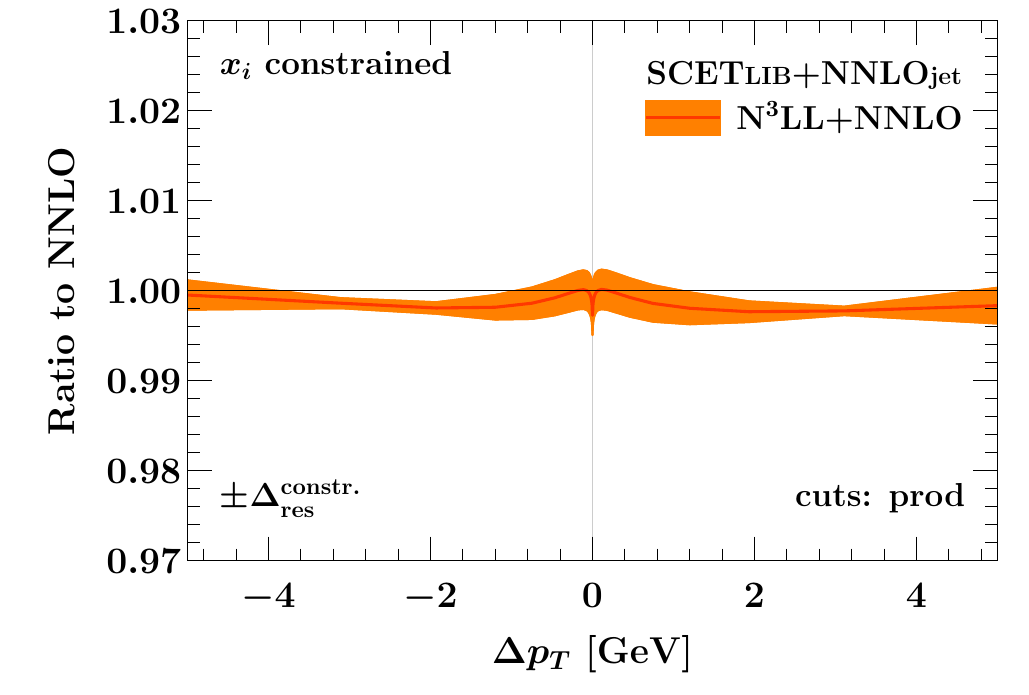}
\caption{
Same as Fig.~\ref{fig:convergence_asy} with product cuts according to Eq.~(\ref{eq:cuts_prod}).
}
\label{fig:convergence_prod}
\end{figure*}

In Fig.~\ref{fig:convergence_asy} we show predictions for the fiducial cross section
at different fixed (top) and resummed and matched orders (bottom).
It is clear that the fixed-order predictions are neither monotonic,
nor do they have a vanishing derivative as $\Delta p_T \to 0^+$. Both of these physical properties are restored by the resummation.
The resummed and matched cross section still feature a discontinuous derivative
at $\Delta p_T = 0$, but this is in fact a physical feature
and a consequence of the definition of $\Delta p_T$ as discussed previously.
The resummed and matched predictions are also monotonically decreasing
and have vanishing derivative as $\Delta p_T \to 0^+$.

The resummation is under good theoretical control,
as evidenced by the matching uncertainty $\Delta_\mathrm{match}$, which we plot
together with the matched predictions in the left column of Fig.~\ref{fig:convergence_asy}.
Here $\Delta_\mathrm{match}$ is estimated by profile scale variations using the formalism and default settings introduced in Ref.~\cite{Ebert:2020dfc},
and features excellent perturbative convergence and coverage.
For later reference, we briefly describe the salient features
of this default matching formalism.
Ref.~\cite{Ebert:2020dfc} employs a hybrid profile scale technique~\cite{Lustermans:2019plv},
where up to a value of $q_T = x_1 Q$
the resummation is fully turned on.
In this first region, the spectrum is obtained from a numerical Fourier transform
of the resummed $b_T$-space cross section. The beam and soft functions are evaluated at the so-called canonical
boundary values for their virtuality and rapidity scales,
\begin{align*} \label{eq:canonical_scales}
\mu_S(b_T) = \mu_B(b_T) = \nu_S(b_T) = \frac{b_0}{b_T}
\,, \qquad
\nu_B = Q
\,,\end{align*}
where $b_0 = 2e^{-\gamma_E}$ and $\gamma_E$ is the Euler-Mascheroni constant.
An identical result for the resummed $b_T$-space cross section is obtained
when choosing canonical $\mu$ and Collins-Soper scales $\zeta$
in TMD factorization.
For $q_T \geq x_1 Q$, the resummation is slowly turned off
by promoting the scales in Eq.~(\ref{eq:canonical_scales})
to be functions of $q_T$
that continuously and monotonically
approach the fixed-order value $\mu_R = \mu_F = Q$
for all values of $b_T$ as $q_T \to Q$.
Specifically, one picks another point $q_T = x_3 Q > x_1 Q$
beyond which all scales are equal to $Q$,
the resummation is fully off,
and the fixed-order result for the spectrum is recovered exactly.
A third intermediate point $q_T = x_2 Q$ with $x_1 < x_2 < x_3$
relates to the precise functional form of the transition
and governs how quickly it departs from the canonical resummation region.
The default settings for these parameters from Ref.~\cite{Ebert:2020dfc},
which were determined quantitatively
by comparing the size of leading-power singular
and the nonsingular terms in the $q_T$ spectrum,
are given by
\begin{align}
(x_1, x_2, x_3) = (0.3, 0.6, 0.9)
\,.\end{align}
The matching uncertainty $\Delta_\mathrm{match}$
is then estimated by varying them as
\begin{align}
(x_1, x_2, x_3) \in V_\mathrm{match} = \{ (0.4, 0.75, 1.1),\, (0.2, 0.45, 0.7),\, (0.4, 0.55, 0.7),\, (0.2, 0.65, 1.1)\}
\,,\end{align}
i.e., one either shifts the transition region up, shifts it down, condenses it, or stretches it out.
Notably, the relative size of singular and nonsingular contributions
from which these default choices were determined
was assessed at the level of the $q_T$ \emph{spectrum} in Ref.~\cite{Ebert:2020dfc}
in order to obtain an optimal prediction for its shape.
By contrast, neither the central value of the $x_i$ nor their variations
were tuned to exactly preserve the fixed-order prediction
for the \emph{integral} of the inclusive $q_T$ spectrum, i.e., the total inclusive cross section.
As an important cross check, we have therefore explicitly verified that the above default resummation and matching setup
nevertheless leads to a net resummation effect compatible with zero when applied to the \emph{inclusive} cross section
within the ATLAS $Q^2$ and $Y$ bins. Explicitly, we find 
\begin{align} \label{eq:resummation_effect_incl_default_matching}
\sigma^\text{incl}_\text{res,NLL}
- \sigma^\text{incl}_\text{sing,LO}
&= (-12.5 \pm_\mathrm{match} 42.7) \, \text{pb}
\,, \nn \\
\sigma^\text{incl}_\text{res,NNLL}
- \sigma^\text{incl}_\text{sing,NLO}
&= (-5.6 \pm_\mathrm{match} 25.5) \, \text{pb}
\,, \nn \\
\sigma^\text{incl}_\text{res,N$^3$LL}
- \sigma^\text{incl}_\text{sing,NNLO}
&= (-5.2 \pm_\mathrm{match} 15.1) \, \text{pb}
\,,
\end{align}
where $\pm_\mathrm{match}$ indicates the matching uncertainty $\Delta_\mathrm{match}$ at the given order.

As an alternative to the default spectrum-level setup from Ref.~\cite{Ebert:2020dfc},
one may instead consider profile scales that are chosen such
that the matching \emph{exactly} preserves the total inclusive cross section
as computed at a given fixed order in perturbation theory.
We stress that because the default matching is implemented through profile scales,
the nonzero resummation effects in Eq.~(\ref{eq:resummation_effect_incl_default_matching})
are by construction also beyond the order at which the fixed-order calculation is truncated.
To do so we maintain the original functional form of the transition as given in Ref.~\cite{Ebert:2020dfc},
but choose the transition points $x_i$ under the constraint
that for the central prediction (which uses Eq.~(\ref{eq:canonical_scales}) in the canonical region $q_T \leq x_1 Q$),
the integral of the matched cross section up to $q_T = x_3 Q$
exactly recovers the fixed-order value.
We further choose to hold $x_1 = 0.3$ fixed for the central prediction
such that the size of the canonical region coincides with the default settings.
By manually scanning the dependence on the other two parameters, we then find that
\begin{align} \label{eq:x_i_constrained}
(x_1, x_2, x_3)_\mathrm{constr.}
= (0.3, 0.9, 1.2)
\end{align}
exactly recovers (for the settings used in this paper, and in particular the PDF set at hand)
the NNLO total inclusive cross section
from the integral of the N$^3$LL$+$NNLO inclusive $q_T$ spectrum.

It is interesting to ask how an uncertainty estimate based on scale variations
can still be obtained for the $q_T$ spectrum in this case,
as well as for derived quantities like fiducial cross sections
that feature sensitivity to resummation effects,
since the matching uncertainty has largely become trivial
by imposing the above constraint.
In particular, these additional variations
should probe possible changes of the shape
of the spectrum in the canonical region itself,
which can also serve as a proxy for the difference
between different resummation formalisms.
To test this we adapt another, separate component
of the total uncertainty estimate for the $q_T$ spectrum from Ref.~\cite{Ebert:2020dfc},
specifically, the so-called resummation uncertainty $\Delta_\mathrm{res}$.
Conventionally, $\Delta_\mathrm{res}$ is estimated
by performing a large set of variations
where the four canonical scales in Eq.~(\ref{eq:canonical_scales})
are varied independently or jointly by factors of $2$
around their central values~\cite{Stewart:2013faa}.
Taking an envelope of these variations
then provides one of the most detailed estimates
of the residual uncertainty on the spectrum in resummed perturbation theory
available within the scale variation paradigm.
While these variations are smoothly turned off as $q_T$ increases
such that the fixed-order prediction in the tail of the spectrum
is unaffected by them,
they again do not necessarily preserve the total integral.
While they probe variations from the canonical Sudakov shape in depth,
they are not necessarily a pure \emph{shape} variation of the spectrum.
To complete our uncertainty estimate in the case where the $x_i$ are constrained,
we  also subject the variations entering $\Delta_\mathrm{res}$
to the integral constraint by adjusting the $x_i$ for each variation.
Restricting to a subset of representative variations
to keep the complexity of the problem manageable, we find
\begin{align}
V_\mathrm{res}^{\rm constr.}
&= \bigl\{
   (\mu_B^\mathrm{down}, \mu_S^\mathrm{down}, 0.55, 1.0, 1.2), \,
   (\mu_B^\mathrm{up},   \mu_S^\mathrm{up},   0.25, 0.65, 1.2), \,
\nn \\
& \qquad
   (\nu_B^\mathrm{down},                      0.35, 0.9, 1.2), \,
   (\nu_B^\mathrm{up},                        0.4, 0.8, 1.2)
\bigr\}
\,,\end{align}
where $\mu_X^{\mathrm{up,down}}$ indicates a soft or beam function
scale variation as described in Ref.~\cite{Ebert:2020dfc},
and the three values quoted are the respective $x_i$.
Note that in this case, we have held $x_3$ fixed at the central value determined earlier
to ensure that all variations in $V_\mathrm{res}^{\rm constr.}$
collapse onto the fixed-order spectrum at the same point,
and instead compensated the changes by mainly varying $x_1$,
i.e., the size of the purely canonical region.

The results of the above exercise are shown in the right column of Fig.~\ref{fig:convergence_asy}.
Compared to the left column, the central values of the matched predictions
are shifted up, largely compensating -- now at the fiducial level --
the small offsets we reported in Eq.~(\ref{eq:resummation_effect_incl_default_matching}).
Note that since we are mainly interested in the behavior at the highest order in perturbation theory,
we have used the same constrained variations from above
at all three orders shown in Fig.~\ref{fig:convergence_asy} for expediency.
This is the reason for the much larger uncertainty bands at lower orders.
Nevertheless, as can be seen from the bottom right panel,
we still find a small net resummation effect
on the central value at N$^3$LL$+$NNLO
that is mainly compatible with zero,
but just outside the refined uncertainty estimate for some values of $\Delta p_T$.
While a complete study of this effect at the next higher order
is beyond the scope of this paper,
we anticipate that the effect will become more significant at higher orders
since the estimate for $\Delta_\mathrm{res}^\mathrm{constr.}$
will be reduced as the residual scale dependence decreases in the resummed cross section,
while the baseline fixed-order prediction picks up two additional logarithms
of $\Delta p_T/Q$ at each order according to Eq.~(\ref{eq:structure_sing_nons_fid_xsecs}).

Another uncertainty intrinsic to the resummed prediction
is given by the impact of non-perturbative transverse momentum dependent (TMD) physics and the Landau pole
in the inverse Fourier transform, which is estimated by variations
of a cutoff parameter $\Lambda_\mathrm{fr}$ defined in Ref.~\cite{Ebert:2020dfc}.
We find that the resulting $\Delta_\mathrm{NP}$ is not even resolved at the current relative numerical uncertainty, which is 
${\cal O}(10^{-4})$.
This is expected because for typical $q_T \sim 25 \,\text{GeV}$ in the baseline symmetric cuts (which dominate the total fiducial cross section here),
the effect of non-perturbative TMD physics is suppressed by $(\Lambda_\mathrm{QCD}/q_T)^2 \sim (500 \,\text{MeV} / 25 \,\text{GeV})^2 \sim 4 \times 10^{-4}$.

Fig.~\ref{fig:convergence_prod} shows our results for the matched cross sections
for the case of the product cuts defined in Eq.~(\ref{eq:cuts_prod}), 
also compared to the \texttt{NNLOJET} fixed-order results.
They are again almost identical to their counterparts for linear asymmetric cuts in Fig.~\ref{fig:convergence_asy}.
A few comments on the product cuts are in order. Ref.~\cite{Salam:2021tbm} focused on constructing $1 \to 2$ fiducial cuts
that would be free of $\mathcal{O}(q_T)$ linear power corrections
at the level of the differential spectrum as $q_T \to 0$, 
which were found to reduce ambiguities in the perturbative series for the case of Higgs production with $\Delta p_T \sim 10$\ GeV. 
Furthermore, the absence of linear power corrections in the acceptance is beneficial for mitigating bias in fixed-order subtraction methods. 
However, the sensitivity of the total fiducial cross section
to resummation effects remains present
through terms of the form $x \ln^n x$ for $x = \Delta p_T/M_Z \ll 1$.
This is evident from our numerical results in Fig.~\ref{fig:prodcuts}, where the unphysical behavior for $\Delta p_T \rightarrow 0$
of the fixed-order cross section persists also for product cuts. We note that this observation does not straightforwardly follow from the definition of  product cuts. Although they trivially coincide with the symmetric cuts for $\Delta p_T \to 0$, they might have displayed a different asymptotic behavior approaching that limit.
In the language of Ref.~\cite{Ebert:2020dfc}, there are two classes of fiducial power corrections, either ``linear'' or ``leptonic'', which do not have to coincide. ``Leptonic'' power corrections in $q_T/\Delta p_T = x q_T/M_Z $ are in general present and lead to sensitivity to low values of $q_T$ at any power in $q_T/\Delta p_T$. Ultimately, the only way to cure the pathological behavior in the $\Delta p_T = 0$ limit is to resort to resummation, due to the presence of $x \ln x$ power corrections for $q_T/M_Z \lesssim x$, as already observed in \cite{Frixione:1997ks}.

\subsection{Rapidity distributions}
\label{sec:rapidity_matched}

\begin{figure*}
\centering
\includegraphics[width=8.15cm]{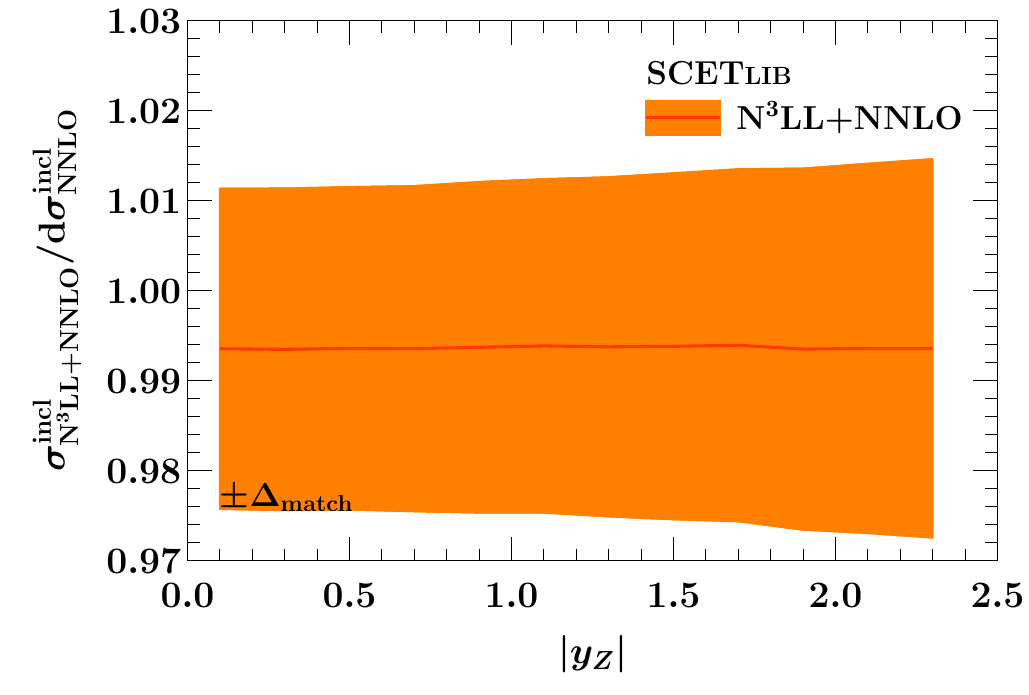}
\hfill
\includegraphics[width=8.15cm]{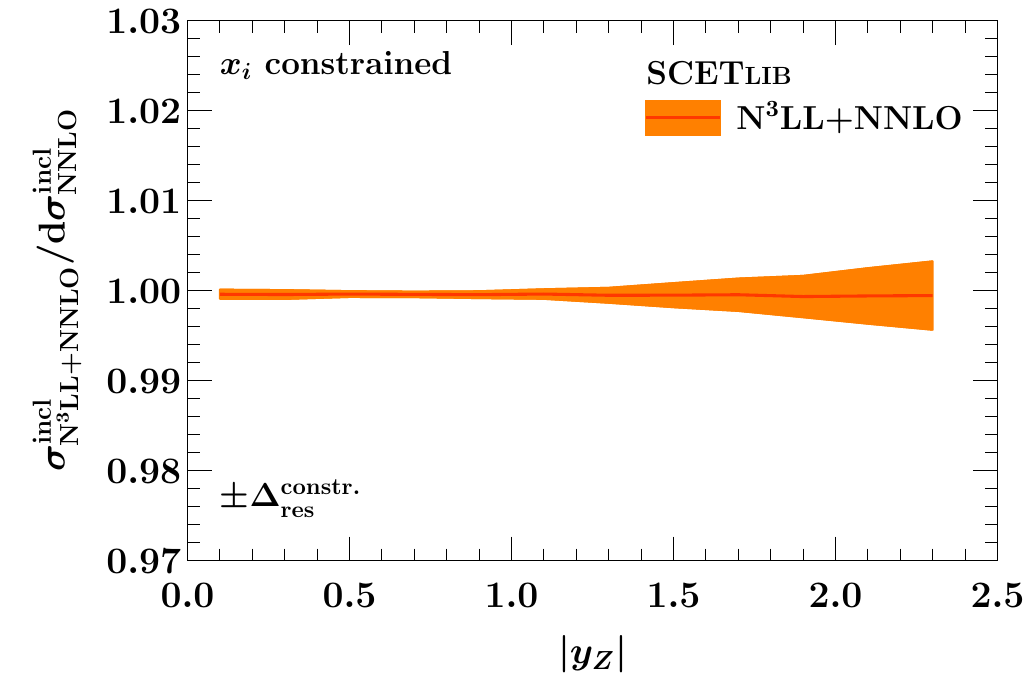}
\caption{
Net resummation effect on the inclusive rapidity spectrum,
using either the default matching setup of Ref.~\cite{Ebert:2020dfc}
with its associated matching uncertainty $\Delta_\mathrm{match}$ (left)
or the integral-preserving matching setup described in Sec.~\ref{sec:matching}
featuring a constrained resummation uncertainty $\Delta_\mathrm{res}^\mathrm{constr.}$ (right).
}
\label{fig:rapidity_spectrum_incl_closure}
\end{figure*}

We are now in a position to assess the size of resummation effects
on the fiducial rapidity distributions studied at fixed order in Sec.~\ref{sec:rapidity}.
As a baseline, we first assess the impact of the resummation
on the inclusive rapidity spectra using the two approaches
described in Sec.~\ref{sec:matching} to perform the matching.
The results at N$^3$LL$+$NNLO are shown
in Fig.~\ref{fig:rapidity_spectrum_incl_closure}.
For the default matching setup of Ref.~\cite{Ebert:2020dfc},
results from which are shown in the left panel,
we find that the integral is compatible with the fixed-order value
within the matching uncertainty for all rapidities,
providing us with a rapidity-differential version
of the check in Eq.~(\ref{eq:resummation_effect_incl_default_matching}).
The results using the constrained profile scales
and the constrained resummation uncertainty introduced in Sec.~\ref{sec:matching}
are shown in the right panel.
Note that the constraint on the transition points $x_i$
was applied at the level of the total inclusive cross section
in the ATLAS $Q$ and central $Y$ bin,
and not point by point in rapidity.
This leads to the finite value
of $\Delta_\mathrm{res}^\mathrm{constr.}/\sigma_\mathrm{LO} \approx 0.4 \%$
at $1.5 \leq |y_Z| < 2.5$, where the effect of the beam function $\mu_B$ variation
changes due to the changing behavior of the underlying PDFs.
While it would in principle be possible to apply the constraint point by point in $y_Z$ (and $Q$),
developing the required automated numerical framework,
e.g.\ along the lines of Ref.~\cite{Bertolini:2017eui},
is beyond the scope of this exercise here.
Here we simply point out that for all $y_Z$,
the integral of the constrained matched prediction
is indeed compatible with the fixed-order result
within the residual $\Delta_\mathrm{res}^\mathrm{constr.}$.

\begin{figure*}
\centering
\includegraphics[width=8.15cm]{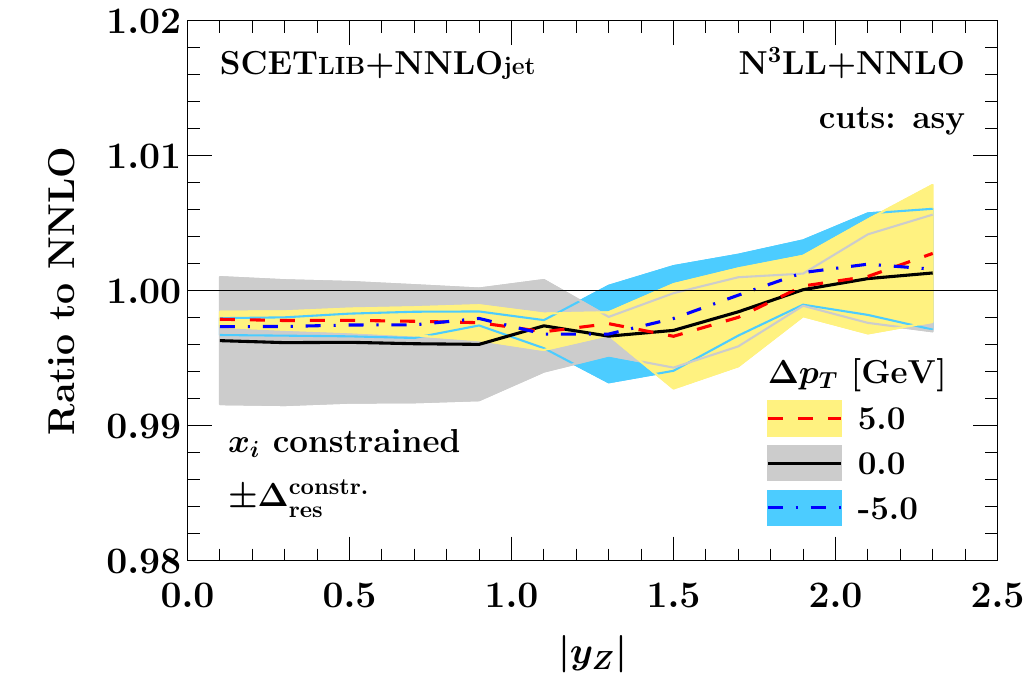}
\hfill
\includegraphics[width=8.15cm]{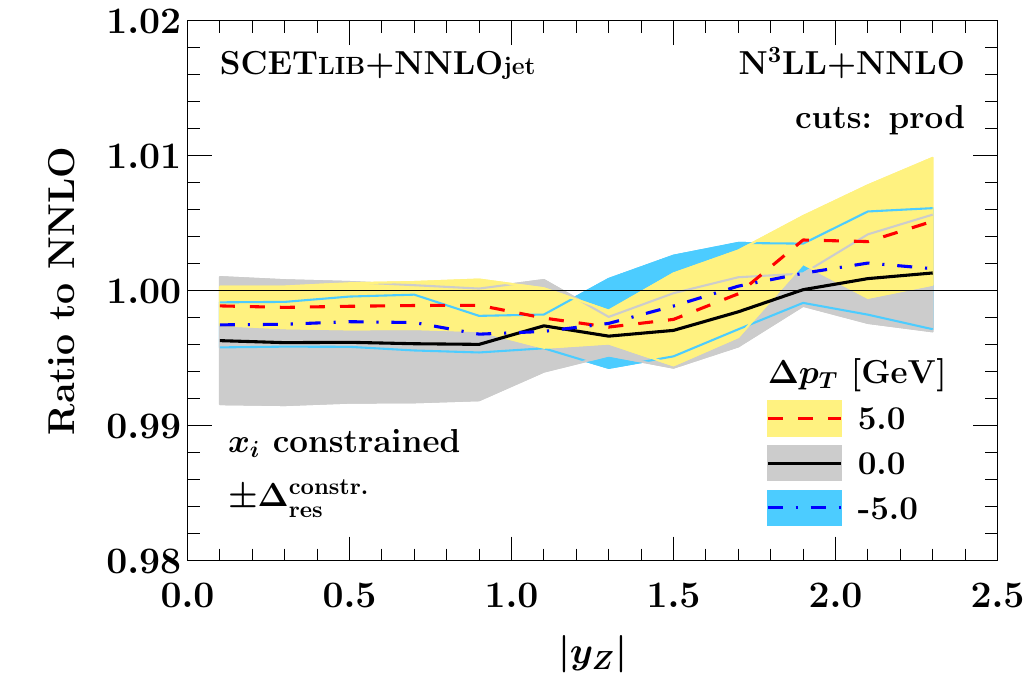}
\caption{
Net resummation effect on the fiducial $Z$ rapidity spectrum
using asymmetric cuts (left) or product cuts (right)
as a function of $|y_Z|$ for representative values of $\Delta p_T$.
For clarity, we only show results for the constrained matching setup.
(The results from the default matching setup are not very instructive at this resolution.)
}
\label{fig:fiducial_y_spectrum_ratio_fo_as_function_of_yZ_matched}
\end{figure*}

Turning to the fiducial $Z$ rapidity spectrum,
we first show the effect of resummation for representative values of $\Delta p_T$
as a function of $y_Z$ in Fig.~\ref{fig:fiducial_y_spectrum_ratio_fo_as_function_of_yZ_matched}.
Our results for $\Delta p_T = 0$ confirm the observations of Ref.~\cite{Amoroso:2022lxw}:
the resummation has a small negative net effect of $-0.4 \%$ at central rapidities,
then starts to rise at the transition point $|y_Z| \sim 1.2$ identified in Sec.~\ref{sec:rapidity},
and eventually contributes a small positive net effect of $+ 0.2\%$ at $|y_Z| = 2.5$.
(Note that in order to read this behavior off from the respective figure in Ref.~\cite{Amoroso:2022lxw},
one must compare the resummation
to the \emph{unbiased} fixed-order result using a recoil prescription.)
The fiducial rapidity spectra for $\Delta p_T = \pm 5\,\mathrm{GeV}$ feature a very similar trend.
We note, however, that the resummation effect at this order
is still compatible with zero for most values of $\Delta p_T$ and $y_Z$ within $\Delta_\mathrm{res}^\mathrm{constr.}$.
To further corroborate these findings, we also compute the net resummation effect
on the fiducial $y_Z$ spectrum for different $\Delta p_T$
using the {\tt RadISH} resummation code~\cite{Re:2021con} and the settings of Ref.~\cite{Chen:2022cgv}. We
find excellent agreement with Fig.~\ref{fig:fiducial_y_spectrum_ratio_fo_as_function_of_yZ_matched}.
This result complements and extends the fixed-order analysis performed in Sect.~\ref{sec:rapidity}, showing that when NNLO is used as the baseline, the resummation of linear power corrections has a small effect.
This suggests that beyond NNLO, the choice of cuts has an effect at the subpercent level, independently of the presence or absence of linear power corrections, and is likely smaller than the N$^3$LO correction, which has been found to be around $-2\%$ for different choices of cuts in~\cite{Chen:2022cgv}.

\begin{figure*}
\centering
\includegraphics[width=8.15cm]{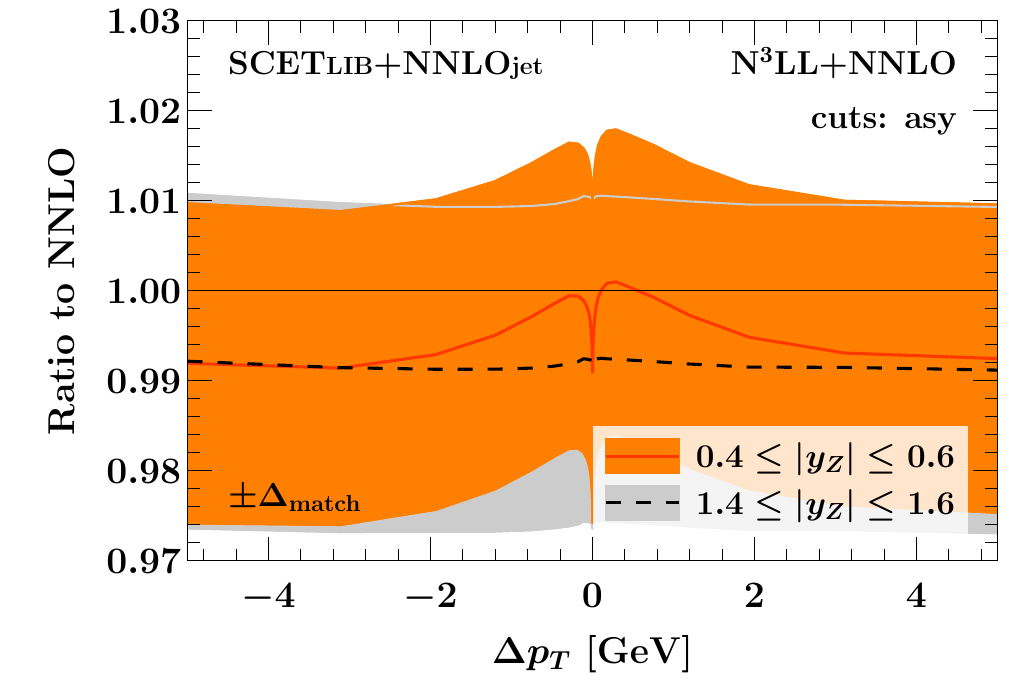}
\hfill
\includegraphics[width=8.15cm]{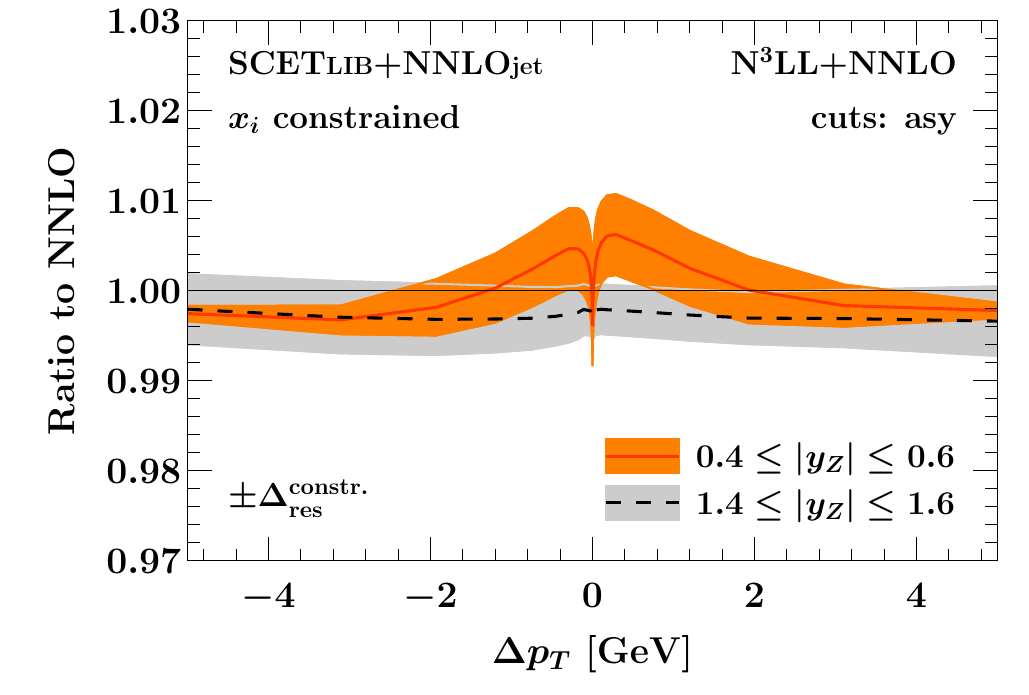}
\\
\includegraphics[width=8.15cm]{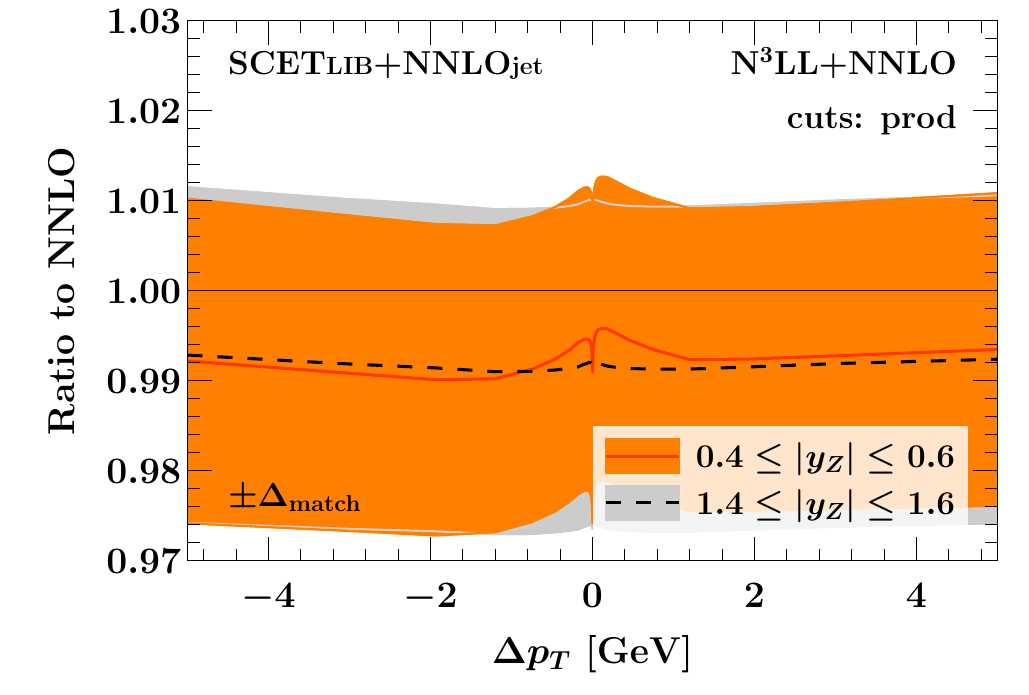}
\hfill
\includegraphics[width=8.15cm]{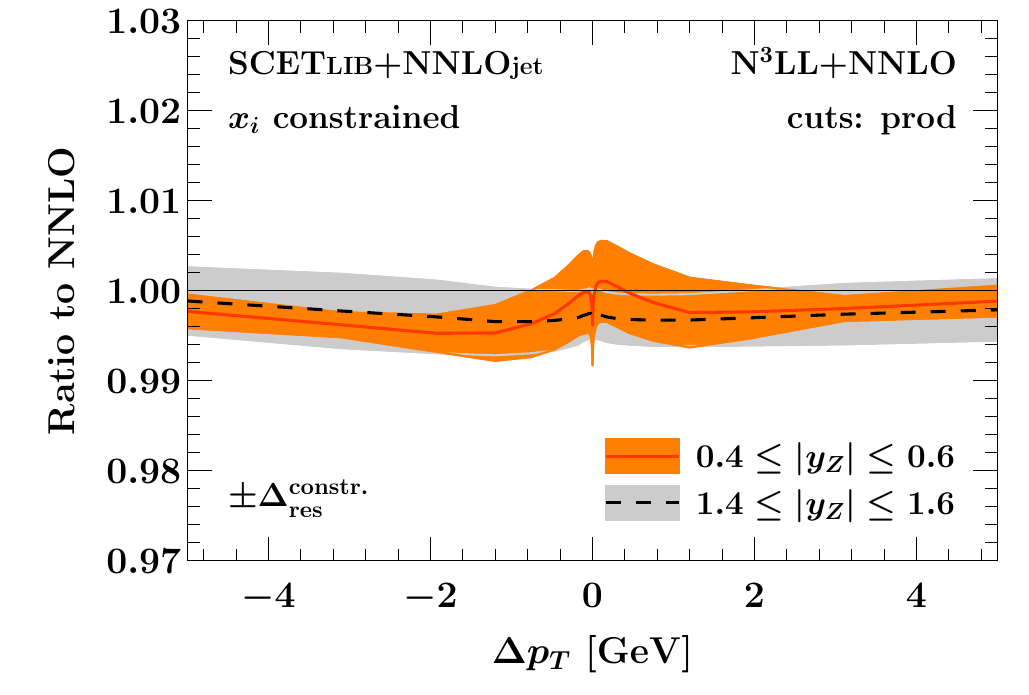}
\caption{
Net resummation effect on the fiducial $Z$ rapidity spectrum
using asymmetric cuts (top) or product cuts (bottom)
as a function of $\Delta p_T$ for two representative bins in $|y_Z|$.
We show results using the default matching setup (left)
and the integral-constrained matching setup (right).
}
\label{fig:fiducial_y_spectrum_ratio_fo_as_function_of_d_pt_matched}
\end{figure*}

In Fig.~\ref{fig:fiducial_y_spectrum_ratio_fo_as_function_of_d_pt_matched}
we show the effect of the resummation on the fiducial rapidity spectrum
in the two representative bins $0.4 \leq |y_Z| \leq 0.6$
and $1.4 \leq |y_Z| \leq 1.6$ considered earlier in Sec.~\ref{sec:rapidity}.
In this case we also show results from the default matching setup for comparison.
We find that in the more forward bin (gray), which lies past the transition point at $|y_Z| \approx 1.2$,
the asymmetric and product cuts both show a very similar behavior as a function of $\Delta p_T$,
with a net resummation effect of $-0.2\%$ (assuming the constrained matching setup)
that is compatible with zero within $\Delta_\mathrm{res}^\mathrm{constr.}$.
This is expected since for these values of $|y_Z|$,
linear power corrections are absent in either case
because the lepton pseudorapidity cuts dominate.
The picture is different for the central rapidity bin (orange),
where the net resummation effect closely resembles
that of the total cross section shown earlier
in Figs.~\ref{fig:convergence_asy} and \ref{fig:convergence_prod}.
In this case, albeit being tiny, the net resummation effect of $\approx -0.3 \%$
in the central $|y_Z|$ bin is not compatible with zero within uncertainties for both sets of cuts
in the region of $\Delta p_T$ between $-5$ and $-2 \, \mathrm{GeV}$.

\section{Experimental resolution}
\label{sec:exp-resolution}

From the previous discussions and numerical results we see that fixed-order codes exhibit an unphysical behavior for symmetric cuts that is remedied by resummation. 
These findings need to be put in perspective of the resolution in the collider experiments, in particular at the LHC.
To quantify this we refer to Fig.~\ref{fig:convergence_asy} (bottom row), where the ratios of the resummed result to the NNLO predictions are shown for the observable under study, 
i.e.\ the cross sections for $pp \to Z/\gamma^*+X \to \ell^+\ell^-  + X$ production at $\sqrt{s}=7$\ TeV integrated over rapidity in the range $y_{\ell\ell} \leq 2.5$, and with fiducial cuts on the decay leptons staggered by $\Delta p_T$.
For large $\Delta p_T$ the difference between the resummed and fixed-order asymptotes to approximately 0.4\%. 
At values for $\Delta p_T$ of a few hundred MeV the difference is 0.2\%. 
Given the good perturbative convergence, we note that this effect is small upon inclusion of the NNLO fixed-order corrections and that the differences between fixed-order and resummed predictions are significantly smaller than the residual theoretical uncertainties from scale variations, cf.\ Fig.~\ref{fig:convergence_asy} (center row).
The observed differences are also smaller than other theoretical uncertainties such as those coming from PDFs, see, e.g.\ Ref.~\cite{Alekhin:2017kpj}.

On the experimental side, the sensitivity to the QCD effects due to symmetric cuts is related to the lepton energy and momentum calibration, and the detector performance.
The ATLAS collaboration reports an experimental resolution on the lepton transverse momenta in the range around 20 to 30\ GeV of a few percent (3 to 4\% for electrons~\cite{ATLAS:2018krz} and 2 to 3\% for muons~\cite{ATLAS:2022jjr}) 
together with a systematic uncertainty in the lepton energy scale calibration in the range between 0.03\% to 0.2\% around the $Z$-boson peak (electrons with transverse momentum close to 45\ GeV)~\cite{ATLAS:2018krz,ATLAS:2022jjr}, 
cf.~\cite{CMS:2020cph} for related studies of the CMS collaboration.

In comparison to the current overall experimental precision reported for the inclusive Drell-Yan cross section, for which systematic uncertainties approaching 2\% are reported~\cite{CMS:2011hqo,ATLAS:2016fij}, the differences from symmetric cuts 
treated at fixed-order in perturbative QCD or resummed to all orders are also small.
For inclusive cross sections the experimental precision is limited by the uncertainties in the luminosity determination, which are typically around 1.5--2.5\%. 
The latest ATLAS luminosity calibration~\cite{ATLAS:2022hro} has been able to achieve 1\% accuracy on the luminosity in Run 3, 
which is expected to be a lower limit on the luminosity uncertainty for Run 3 and the high-luminosity runs.
Our NNLO analysis indicates that the all-order resummation of the logarithms in the fiducial transverse momentum spectrum and the unphysical behavior of fixed-order perturbation theory are unlikely to have an impact on experimental studies of inclusive Drell-Yan cross sections. The findings of this study should eventually be corroborated by a N$^3$LO analysis, although the impact of resummation seems to remain moderate at higher orders (see e.g.\ Ref.~\cite{Chen:2022cgv}).

For differential distributions or normalized cross sections the situation is different, because the effect of the luminosity uncertainty cancels or leads to a coherent shift, and the precision on the shape or ratio of distributions can be much higher. 
In fact, already for the ATLAS measurement~\cite{Aaboud:2016btc} considered in the benchmark study in Sec.~\ref{sec:benchmark} a precision on the $Z$-boson rapidity distribution in the central region at the level of a few per-mille is reported, excluding the luminosity uncertainty. 
Thus, for shapes or ratios of differential distributions $q_T$-resummation can have an impact on the theoretical predictions, cf.~\cite{Amoroso:2022lxw}.
As the dominant uncertainties from the lepton calibrations ($Z$-boson) and from backgrounds and recoil calibration ($W$-boson) are reduced in future measurements of differential distributions for Drell-Yan processes, the differences between fixed-order perturbative QCD and resummed predictions for symmetric fiducial cuts should be revisited.

\section{Conclusions and Discussion}
\label{sec:summary}

Fixed-order QCD perturbation theory for the Drell-Yan process suffers from an instability for symmetric cuts on the transverse momenta of the final-state leptons. This pathology originates from a logarithmically enhanced power correction present for small lepton-pair transverse momentum. It leads to unphysical non-monotonic behavior for the cross section as the difference $\Delta p_T$ between the cuts on the lepton and anti-lepton is varied. Available public codes that compute the fixed-order NNLO QCD corrections to Drell-Yan treat this instability in different ways due to the underlying subtraction scheme in the codes, potentially leading to different theoretical predictions for the cross section and hindering the analysis of experimental data.

In this paper we have performed a detailed analysis of this issue, involving both the careful study of available fixed-order codes, the resummation of QCD perturbation theory to solve this issue, and the consideration of product cuts that have been suggested as a resolution of this issue. We have considered the following codes that compute the NNLO QCD corrections to the Drell-Yan process:
\begin{itemize}
    \item {\tt DYTURBO} (version 1.2), based on the non-local $q_T$ subtraction scheme;
    \item {\tt FEWZ} (version 3.1), based on the local sector decomposition subtraction scheme;
    \item {\tt MATRIX} (version 2.1), based on the non-local $q_T$ subtraction scheme;
    \item  {\tt NNLOJET}, based on the local antenna subtraction scheme. 
\end{itemize}
All codes, whether based on local or non-local subtraction schemes, give consistent results for the case of symmetric cuts. This is a non-trivial result since these techniques treat the $p_T=0$ region very differently. Obtaining the correct fixed-order result with non-local schemes requires a careful treatment of power corrections for small transverse momentum. The agreement between these codes is a testament to the community effort exerted for a proper theoretical treatment of this process, with {\tt DYTURBO}, {\tt FEWZ} and {\tt MATRIX} being readily available to the public.

We also considered product cuts~\cite{Salam:2021tbm}, which replace the separate 
linear asymmetric cuts on the leading and sub-leading lepton transverse momenta with
cuts on their product and the sub-leading lepton instead.
For values of $\Delta p_T \sim 10$\ GeV these cuts mitigate the ambiguities of
the perturbative series present with (a)symmetric cuts, which can be
particularly relevant for processes with Casimir enhancement, notably Higgs
production~\cite{Billis:2021ecs}.  
As expected, we find that their introduction
does not address the pathological behavior for  $\Delta p_T \rightarrow 0$,
which can only be addressed by resummation.  We found that resummation indeed
removes this effect, although a kink in the cross section is still present for
$\Delta p_T=0$, essentially due to the way we defined the observable for this
study. At NNLO the differences induced by resummation are small
for all the values of $\Delta p_T$ considered, ranging from $0$ to a few GeV,
shifting the fixed-order result by sub-percent values that are well below the
current experimental uncertainties. An analogous consideration holds also at the level of the rapidity spectra, in agreement with the findings of Ref.~\cite{Amoroso:2022lxw}.
It remains to be seen whether such
conclusion still holds beyond NNLO, although recent
studies~\cite{Salam:2021tbm,Chen:2022cgv} indicate that the effect of
resummation of linear power corrections remains moderate also at
N$^3$LO. Given the excellent job done by the available fixed-order codes in
reproducing the results of resummation and the substantial effort needed to
implement the relevant modifications when modeling signal and background in
experimental analyses, we foresee only a marginal improvement in adopting a
different set of cuts in the Drell-Yan case for the measurement of fiducial
cross sections. 

\subsection*{Note added:}
The \texttt{NNLOJET} code has been published now in Ref.~\cite{NNLOJET:2025rno}
and is available from \url{https://nnlojet.hepforge.org/}.

\subsection*{Acknowledgments}
This research was supported by the Munich Institute for Astro-, Particle and BioPhysics (MIAPbP) which is funded by the Deutsche Forschungsgemeinschaft (DFG, German Research Foundation) under Germany's Excellence Strategy - EXC-2094 - 390783311.
We also thank the Erwin-Schr\"odinger International Institute for Mathematics and Physics at the University of Vienna for partial support during the Programme {\it Quantum Field Theory at the Frontiers of the Strong Interactions}, July 31 - September 1, 2023. 

Si.Am.~is supported by the Helmholtz Association under the contract W2/W3-123.
L.B.~is funded by the European Union (ERC, grant agreement No. 101044599, JANUS). Views and opinions expressed are however those of the authors only and do not necessarily reflect those of the European Union or the European Research Council Executive Agency. Neither the European Union nor the granting authority can be held responsible for them.
A.K.~is supported by the UNKP-21-Bolyai+ New National Excellence Program of the Ministry for Innovation and Technology from the source of the National Research, Development and Innovation Fund. A.K. kindly acknowledges further financial support from the Bolyai Fellowship programme of the Hungarian Academy of Sciences.
J.M.~is supported by the D-ITP consortium, a program of NWO that is funded by the Dutch Ministry of Education, Culture and Science (OCW).
S.M.~acknowledges support by the DFG grant MO 1801/5-2.
L.R.~has been supported by the SNSF under contract PZ00P2 201878.
F.P.~is supported by the DOE grants DE-FG02-91ER40684 and DE-AC02-06CH11357.
Z.T.~is supported by the Excellence Programme of the Hungarian Ministry of Culture and Innovation under contract TKP2021-NKTA-64

\appendix
\section{Cross section predictions}
\label{sec:appXS}

We present here the benchmark predictions discussed in Sec.~\ref{sec:benchmark} for the 
(pseudo-)rapidity distributions of the decay leptons for  $W^\pm$- and $Z/\gamma^*$-production 
at a center-of-mass energy of $\sqrt{s}=7$\ TeV, corresponding to the fiducial cuts of the ATLAS experiment.
The predictions listed here are all computed with {\tt NNLOJET} and 
results are provided for each perturbative order (LO, NLO, NNLO). 
Note that all predictions use the same NNLO PDF set from ABMP16~\cite{Alekhin:2017kpj}.

We provide predictions for the differential distributions; the cross sections in columns 4, 5 and 6 of Tabs.~\ref{tab:XS-Wp}--\ref{tab:XS-Zfwrd} are already divided by the bin widths.
Note also that we bin in the absolute value of the rapidities. 
This means the normalization is different by a factor of two compared to a calculation that bins the
signed value of rapidity.

\begin{table}[h]
    \centering
    \begin{tabular}{ccc|c|c|c}
    \hline
	min $|y_l|$      & cntr $|y_l|$       &  max $|y_l|$      & $\sigma_{\rm LO}$ [fb]   & $\sigma_{\rm NLO}$ [fb]  & $\sigma_{\rm NNLO}$ [fb] \\
    \hline
        0              & 0.105            & 0.21            & 1144924 (22)          & 1131302 (49)         & 1132372 (218)         \\
        0.21           & 0.315            & 0.42            & 1147528 (23)          & 1133998 (58)         & 1135481 (300)         \\
        0.42           & 0.525            & 0.63            & 1152891 (23)          & 1139771 (59)         & 1141365 (307)         \\
        0.63           & 0.735            & 0.84            & 1160339 (23)          & 1147812 (59)         & 1149361 (327)         \\
        0.84           & 0.945            & 1.05            & 1169448 (23)          & 1157995 (60)         & 1159151 (331)         \\
        1.05           & 1.21             & 1.37            & 1181246 (19)          & 1172429 (44)         & 1173206 (228)         \\
        1.37           & 1.445            & 1.52            & 1189295 (29)          & 1184195 (80)         & 1183306 (473)         \\
        1.52           & 1.63             & 1.74            & 1190632 (24)          & 1190271 (59)         & 1186759 (319)         \\
        1.74           & 1.845            & 1.95            & 1182394 (25)          & 1189693 (61)         & 1183688 (353)         \\
        1.95           & 2.065            & 2.18            & 1155801 (24)          & 1173921 (57)         & 1163509 (326)         \\
        2.18           & 2.34             & 2.5             & 1082093 (22)          & 1117404 (46)         & 1102564 (225)         \\
    \hline
    \end{tabular}
    \caption{\label{tab:XS-Wp}
    Cross sections at LO, NLO and NNLO in QCD in fb for 
    inclusive $pp \to W^+ + X \to l^+ \nu + X$ production at $\sqrt{s}=7$~TeV, 
    subject to the fiducial cuts applied by measured by the ATLAS experiment~\cite{Aaboud:2016btc}. 
    See also Sec.~\ref{sec:benchmark} for the settings. 
    Numbers in round brackets indicate the statistical uncertainty from the Monte Carlo evaluation 
    on the last digits.}
\end{table}
\begin{table}[ht]
    \centering
    \begin{tabular}{ccc|c|c|c}
    \hline
	min $|y_l|$  & cntr $|y_l|$  &  max $|y_l|$ & $\sigma_{\rm LO}$ [fb]  & $\sigma_{\rm NLO}$ [fb]  & $\sigma_{\rm NNLO}$ [fb]  \\
    \hline
        0          & 0.105       & 0.21       & 848830 (14)         & 862143 (28)         & 855642 (122)         \\
        0.21       & 0.315       & 0.42       & 844251 (14)         & 857976 (32)         & 851304 (174)         \\
        0.42       & 0.525       & 0.63       & 835145 (14)         & 849539 (32)         & 843000 (173)         \\
        0.63       & 0.735       & 0.84       & 821844 (14)         & 837206 (32)         & 830928 (165)         \\
        0.84       & 0.945       & 1.05       & 804614 (14)         & 821162 (32)         & 815662 (161)         \\
        1.05       & 1.21        & 1.37       & 777824 (11)         & 795869 (23)         & 790738 (112)         \\
        1.37       & 1.445       & 1.52       & 750484 (16)         & 769521 (43)         & 764996 (221)         \\
        1.52       & 1.63        & 1.74       & 726685 (14)         & 745954 (32)         & 741848 (162)         \\
        1.74       & 1.845       & 1.95       & 697056 (14)         & 716004 (32)         & 712404 (173)         \\
        1.95       & 2.065       & 2.18       & 664666 (14)         & 682198 (32)         & 678928 (156)         \\
        2.18       & 2.34        & 2.5        & 620643 (13)         & 634655 (27)         & 631338 (118)         \\
    \hline
    \end{tabular}
    \caption{\label{tab:XS-Wm}
    Same as Tab.~\ref{tab:XS-Wp} for 
    inclusive $pp \to W^- + X \to l^- \nu + X$ production at $\sqrt{s}=7$~TeV.}
\end{table}
\begin{table}[ht]
    \centering
    \begin{tabular}{ccc|c|c|c}
    \hline
       min $|y_{ll}|$&cntr $|y_{ll}|$&max $|y_{ll}|$ & $\sigma_{\rm LO}$ [fb]     &$\sigma_{\rm NLO}$ [fb]   & $\sigma_{\rm NNLO}$ [fb]   \\
    \hline
        0          & 0.1         & 0.2         & 242501 (4)           & 260269 (6)         & 260299 (11)          \\
        0.2        & 0.3         & 0.4         & 241885 (4)           & 259761 (6)         & 259821 (12)          \\
        0.4        & 0.5         & 0.6         & 240596 (4)           & 258581 (5)         & 258698 (11)          \\
        0.6        & 0.7         & 0.8         & 238771 (4)           & 256990 (6)         & 257210 (11)          \\
        0.8        & 0.9         & 1.0         & 236249 (4)           & 254824 (6)         & 255082 (11)          \\
        1.0        & 1.1         & 1.2         & 227905 (4)           & 251037 (6)         & 249373 (11)          \\
        1.2        & 1.3         & 1.4         & 207158 (4)           & 234939 (5)         & 233888 (12)          \\
        1.4        & 1.5         & 1.6         & 179949 (4)           & 208041 (5)         & 208362 (13)          \\
        1.6        & 1.7         & 1.8         & 147438 (3)           & 173199 (5)         & 174566 (13)          \\
        1.8        & 1.9         & 2.0         & 110916 (3)           & 132203 (5)         & 134005 (14)          \\
        2.0        & 2.1         & 2.2         & 72370 (3)            & 87305 (5)          & 88905 (13)           \\
        2.2        & 2.3         & 2.4         & 34732 (3)            & 42185 (4)          & 43064 (11)           \\
    \hline
    \end{tabular}
    \caption{\label{tab:XS-Z}
    Same as Tab.~\ref{tab:XS-Wp} for 
    central inclusive $pp \to Z + X \to l^+l^- + X$ production at $\sqrt{s}=7$~TeV.}
\end{table}
\begin{table}[ht]
    \centering
    \begin{tabular}{ccc|c|c|c}
    \hline
        min $|y_{ll}|$        &cntr $|y_{ll}|$         &max $|y_{ll}|$         & $\sigma_{\rm LO}$ [fb] &$\sigma_{\rm NLO}$ [fb] & $\sigma_{\rm NNLO}$ [fb] \\
    \hline
        1.2                   & 1.3                    & 1.4                   & 22646 (1)            & 13899 (3)          & 15537 (15)          \\
        1.4                   & 1.5                    & 1.6                   & 45488 (2)            & 36627 (4)          & 36949 (16)          \\
        1.6                   & 1.7                    & 1.8                   & 72941 (2)            & 66461 (4)          & 65760 (15)          \\
        1.8                   & 1.9                    & 2.0                   & 103456 (2)           & 101449 (4)         & 100231 (15)         \\
        2.0                   & 2.1                    & 2.2                   & 134086 (2)           & 13815 (4)          & 13692 (15)          \\
        2.2                   & 2.3                    & 2.4                   & 161961 (2)           & 17286 (4)          & 17205 (13)          \\
        2.4                   & 2.6                    & 2.8                   & 157264 (2)           & 17023 (3)          & 16929 (9)           \\
        2.8                   & 3.0                    & 3.2                   & 77063 (1)            & 80140 (2)          & 78526 (8)           \\
        3.2                   & 3.4                    & 3.6                   & 20432 (1)            & 20227 (1)          & 19374 (6)           \\
    \hline
    \end{tabular}
    \caption{\label{tab:XS-Zfwrd}
    Same as Tab.~\ref{tab:XS-Wp} for 
    forward inclusive $pp \to Z + X \to l^+l^- + X$ production at $\sqrt{s}=7$~TeV.}
\end{table}

\newpage
\clearpage

The predictions for the distributions in the electron pseudo-rapidity for the electron charge asymmetry 
measured by the D{\O} experiment in $W^\pm$-boson production at $\sqrt{s}=1.96$\ TeV at the Tevatron~\cite{D0:2014kma}. 
The fiducial cuts and the settings are listed in Sec.~\ref{sec:benchmark}.
The predictions are computed with {\tt NNLOJET}, use the same NNLO PDF set from ABMP16~\cite{Alekhin:2017kpj} for each perturbative order (LO, NLO, NNLO) and are again provided  
for the differential distributions, i.e., the cross sections 
(columns 4, 5 and 6 in Tabs.~\ref{tab:XS-Wp-asy} and \ref{tab:XS-Wm-asy}) are already divided by the bin widths.
\begin{table}[h]
\centering
\begin{tabular}{ccc|c|c|c}
\hline
min $y_l$      & cntr $y_l$       &  max $y_l$      & $\sigma_{\rm LO}$ [fb]     & $\sigma_{\rm NLO}$ [fb]   & $\sigma_{\rm NNLO}$ [fb]  \\
\hline
-3.2  & -2.95  & -2.7    & 20546 (1)      & 20997  (3)      & 20789 (7)     \\ 
-2.7  & -2.55  & -2.4    & 47756 (1)      & 50253  (5)      & 50219 (17)    \\ 
-2.4  & -2.3   & -2.2    & 68277 (3)      & 72825  (8)      & 73124 (29)    \\ 
-2.2  & -2.1   & -2.0    & 84664 (3)      & 91041  (9)      & 91660 (30)    \\ 
-2.0  & -1.9   & -1.8    & 100108 (3)     & 108305 (9)      & 109294 (33)   \\ 
-1.8  & -1.7   & -1.6    & 114218 (3)     & 124151 (10)     & 125558 (33)   \\ 
-1.6  & -1.4   & -1.2    & 132796 (2)     & 144935 (6)      & 146797 (18)   \\ 
-1.2  & -1.1   & -1.0    & 149451 (3)     & 163439 (11)     & 165639 (35)   \\ 
-1.0  & -0.9   & -0.8    & 159754 (4)     & 174717 (11)     & 177081 (36)   \\ 
-0.8  & -0.7   & -0.6    & 169663 (4)     & 185463 (12)     & 187936 (38)   \\ 
-0.6  & -0.5   & -0.4    & 179264 (4)     & 195794 (12)     & 198292 (38)   \\ 
-0.4  & -0.3   & -0.2    & 188493 (4)     & 205663 (12)     & 208161 (40)   \\ 
-0.2  & -0.1   & 0.0     & 197178 (4)     & 215042 (13)     & 217563 (42)   \\ 
0.0   & 0.1    & 0.2     & 204955 (5)     & 223597 (13)     & 225922 (44)   \\ 
0.2   & 0.3    & 0.4     & 211345 (5)     & 230892 (13)     & 233051 (44)   \\ 
0.4   & 0.5    & 0.6     & 215610 (5)     & 236284 (14)     & 238207 (47)   \\ 
0.6   & 0.7    & 0.8     & 216713 (5)     & 238739 (14)     & 240426 (48)   \\ 
0.8   & 0.9    & 1.0     & 213411 (5)     & 236982 (14)     & 238314 (48)   \\ 
1.0   & 1.1    & 1.2     & 204339 (5)     & 229334 (14)     & 230561 (46)   \\ 
1.2   & 1.4    & 1.6     & 176867 (3)     & 202614 (7)      & 203859 (22)   \\ 
1.6   & 1.7    & 1.8     & 136734 (4)     & 160146 (11)     & 161689 (37)   \\ 
1.8   & 1.9    & 2.0     & 105594 (3)     & 125021 (9)      & 126689 (30)   \\ 
2.0   & 2.1    & 2.2     & 75804 (3)      & 90298  (8)      & 91681 (23)    \\ 
2.2   & 2.3    & 2.4     & 50319 (2)      & 60105  (5)      & 61155 (17)    \\ 
2.4   & 2.55   & 2.7     & 26925 (1)      & 32183  (3)      & 32744 (7)     \\ 
2.7   & 2.95   & 3.2     & 7572  (0)      & 9011   (1)      & 9178  (2)     \\
\hline
\end{tabular}
    \caption{\label{tab:XS-Wp-asy}
    Cross sections at LO, NLO and NNLO in QCD in fb for 
    inclusive $p{\bar p} \to W^+ + X \to l^+ \nu + X$ production at $\sqrt{s}=1.96$~TeV,
    subject to the fiducial cuts applied by the D{\O} experiment~\cite{D0:2014kma}.
    See also Sec.~\ref{sec:benchmark} for the settings. 
    Numbers in round brackets indicate the statistical uncertainty from the Monte Carlo evaluation 
    on the last digits.}
\end{table}
\begin{table}[ht]
\centering
\begin{tabular}{ccc|c|c|c}
\hline
min $y_l$      & cntr $y_l$       &  max $y_l$      & $\sigma_{\rm LO}$ [fb]     & $\sigma_{\rm NLO}$ [fb]   & $\sigma_{\rm NNLO}$ [fb]  \\
\hline
-3.2  & -2.95  & -2.7    & 7572  (0)      & 9013   (1)      & 9177  (2)      \\ 
-2.7  & -2.55  & -2.4    & 26923 (1)      & 32179  (3)      & 32767 (6)      \\ 
-2.4  & -2.3   & -2.2    & 50319 (2)      & 60117  (5)      & 61104 (16)     \\ 
-2.2  & -2.1   & -2.0    & 75806 (3)      & 90296  (6)      & 91699 (22)     \\ 
-2.0  & -1.9   & -1.8    & 105591 (3)      & 125011 (7)      & 126677 (29)    \\ 
-1.8  & -1.7   & -1.6    & 136730 (3)      & 160142 (9)      & 161692 (35)    \\ 
-1.6  & -1.4   & -1.2    & 176866 (3)      & 202613 (6)      & 203843 (22)    \\ 
-1.2  & -1.1   & -1.0    & 204318 (4)      & 229345 (11)     & 230505 (47)    \\ 
-1.0  & -0.9   & -0.8    & 213404 (4)      & 236976 (11)     & 238445 (47)    \\ 
-0.8  & -0.7   & -0.6    & 216709 (4)      & 238721 (11)     & 240343 (45)    \\ 
-0.6  & -0.5   & -0.4    & 215604 (4)      & 236278 (11)     & 238212 (44)    \\ 
-0.4  & -0.3   & -0.2    & 211355 (4)      & 230888 (10)     & 233115 (45)    \\ 
-0.2  & -0.1   & 0.0     & 204953 (4)      & 223595 (10)     & 225952 (44)    \\ 
0.0   & 0.1    & 0.2     & 197181 (4)      & 215043 (10)     & 217516 (42)    \\ 
0.2   & 0.3    & 0.4     & 188503 (4)      & 205688 (9)      & 208133 (39)    \\ 
0.4   & 0.5    & 0.6     & 179268 (4)      & 195782 (9)      & 198264 (37)    \\ 
0.6   & 0.7    & 0.8     & 169669 (4)      & 185482 (9)      & 187997 (37)    \\ 
0.8   & 0.9    & 1.0     & 159752 (3)      & 174711 (8)      & 177061 (36)    \\ 
1.0   & 1.1    & 1.2     & 149455 (3)      & 163430 (8)      & 165686 (35)    \\ 
1.2   & 1.4    & 1.6     & 132819 (2)      & 144938 (5)      & 146791 (17)    \\ 
1.6   & 1.7    & 1.8     & 114222 (3)      & 124137 (7)      & 125506 (34)    \\ 
1.8   & 1.9    & 2.0     & 100112 (3)      & 125021 (7)      & 109334 (30)    \\ 
2.0   & 2.1    & 2.2     & 84665  (3)      &  90292 (7)      & 91686  (29)    \\ 
2.2   & 2.3    & 2.4     & 68282  (2)      & 72826  (7)      & 73109  (28)    \\ 
2.4   & 2.55   & 2.7     & 47758  (1)      & 50315  (4)      & 50373  (17)    \\ 
2.7   & 2.95   & 3.2     & 20544  (1)      & 20947  (2)      & 20778  (7)     \\ 
\end{tabular}
    \caption{\label{tab:XS-Wm-asy}
    Same as Tab.~\ref{tab:XS-Wp-asy} for 
    inclusive $p{\bar p} \to W^- + X \to l^- \nu + X$ production at $\sqrt{s}=1.96$~TeV.}
\end{table}

\newpage
\clearpage

\section{DYTURBO inputs}
\label{sec:appB}

The input values used for the computations with {\tt DYTURBO} for $Z$-boson production with central leptons 
at $\sqrt{s}=7$\ TeV in Sec.~\ref{sec:benchmark}.

{\scriptsize
\begin{verbatim}
# Process settings
sroot        = 7e3  # Center-of-mass energy
ih1          = 1    # Hadron 1: 1 for proton, -1 for antiproton
ih2          = 1    # Hadron 2: 1 for proton, -1 for antiproton
nproc        = 3    # Process: 1) W+; 2) W-; 3) Z/gamma*

# Perturbative order
fixedorder_only = true
order           = 2  # QCD order: 0) LO(+LL), 1) NLO(+NLL), 2) NNLO(+NNLL), 3) N3LO(+N3LL)

#PDF settings
LHAPDFset    = ABMP16_5_nnlo # PDF set from LHAPDF
LHAPDFmember = 0                    # PDF member

# Functional form of QCD scales
fmuren = 0     # Functional form of the renormalisation scale
fmufac = 0     # Functional form of the factorisation scale
fmures = 0     # Functional form of the resummation scale 

# QCD scale settings
kmuren = 1         # Scale factor for the renormalisation scale
kmufac = 1         # Scale factor for the factorisation scale
kmures = 1         # Scale factor for the resummation scale

# EW scheme
ewscheme = 1 #1: Input: Gf, wmass, zmass;
conv2fixw = true # Convert Z and W masses and widths from running to fixed width 

# CKM matrix
Vud = 0.97427
Vus = 0.2253
Vub = 0.00351
Vcd = 0.2252
Vcs = 0.97344
Vcb = 0.0412

# qt-subtraction cut-off. 
xqtcut = 0.008  # cutoff on qt/m
qtcut = 0.      # cutoff on qt

qtfpc = 1e-4 # FPC cutoff on qt/m

#cut off on invariant mass between emitted and radiator in V+jet
mcutoff = 1e-3

# Lepton cuts
makecuts = true

# charged leptons cuts
lptcut = 20
lycut = 2.5 # absolute rapidity cut

# integration types and settings for costh phi_lep phase space
cubaint   = false   # integration with Cuba Suave
trapezint = false  # trapezoidal rule for the phi_lep integration and semi-analytical for costh
quadint   = true  # quadrature rule for the phi_lep integration and semi-analytical for costh

suavepoints  = 1000000 # number of points for suave integration, newpoints is set to suavepoints/10;
nphitrape    = 1000    # number of steps for trapezoidal rule of phi_lep integration
phirule      = 4       # quadrature rule of phi_lep integration
phiintervals = 40      # number of segments for quadrature rule of phi_lep integration
ncstart      = 200     # starting sampling for the costh semi-analytical integration 
                       # (common settings for the trapezoidal and quadrature rules)

# Output settings
output_filename = results_zypeakcc_FEWZ  # output filename
texttable   = true         # dump result table to text file (including pdf variations)
redirect    = false        # redirect stdout and stderr to log file (except for gridverbose output)
unicode     = true         # use unicode characters for the table formatting
silent      = false        # no output on screen (except for gridverbose output)
makehistos  = true         # fill histograms
gridverbose = false        # printout number of events to keep job alive when running on grid

# normalise cross sections by bin width
ptbinwidth = false
ybinwidth  = false
mbinwidth  = false

# qt, y, m bins
qt_bins = [ 0 7000 ]
y_bins  = [  0.0 0.2 0.4 0.6 0.8 1.0 1.2 1.4 1.6 1.8 2.0 2.2  2.4 ]
m_bins  = [ 66 116 ]


\end{verbatim}
}

\section{FEWZ runcard}
\label{sec:appC}

The runcard used for the computations with {\tt FEWZ} (version 3.1) for $Z$-boson production with central leptons 
at $\sqrt{s}=7$\ TeV in Sec.~\ref{sec:benchmark}.

{\scriptsize
\begin{verbatim}

=============================================
'CMS collision energy (GeV)    = ' 7000d0
=============================================
'Factorization scale  (GeV)    = ' 91.1876d0
'Renormalization scale  (GeV)  = ' 91.1876d0
=============================================
'Z production (pp=1,ppbar=2)   = ' 1
=============================================
Alpha QED (0) is for photon-induced channels (which use photon PDFs); set to zero to turn off these channels
'Alpha QED (0)                 = ' 0.007297352568d0
'Alpha QED (Mz)                = ' 0.007756146746d0
'Fermi constant (1/Gev^2)       = ' 1.16637d-5
=============================================
'Lepton mass (GeV)             = ' 1.05d-1
'W mass (GeV)                  = ' 80.403d0
'W width (GeV)                 = ' 2.141d0
'Z mass (GeV)                  = ' 91.1876d0
'Z width (GeV)                 = ' 2.4952d0
'Top mass (GeV)                = ' 170.4d0
'Higgs mass (GeV)              = ' 125d0
=============================================
Only QED corrections is on if the input scheme is manual
Input scheme: 0. Manual input; 1. Gmu scheme; 2. AlphaMz scheme
'Which input scheme:           = ' 1
'sin^2(theta_w)                = ' 0.22255d0
'up quark charge               = ' 0.6666667d0
'down quark charge             = ' -0.3333333d0
'lepton charge                 = ' -1d0
'up quark vector coupling      = ' 0.4091d0
'down quark vector coupling    = ' -0.7045d0
'lepton vector coupling        = ' -0.11360d0
'up quark axial coupling       = ' -1d0
'down quark axial coupling     = ' 1d0
'lepton axial coupling         = ' 1d0
=============================================
Vegas Parameters
'Relative accuracy (in %)           = ' 0d0
'Absolute accuracy                  = ' 0d0
'Number of calls per iteration      = ' 1000000
'Number of increase calls per iter. = ' 500000
'Maximum number of evaluations      = ' 1000000000 
'Random number seed for Vegas       = ' 211
=============================================
'QCD Perturb. Order (0=LO, 1=NLO, 2=NNLO) = ' 2
'EW Perturb. Order (0=LO, 1=NLO)    = ' 0
'Z pole focus (1=Yes, 0=No)	= ' 1
'EW control (leave 0 to keep all on) = ' 0 
'Turn off photon (1=Yes, 0=No, disabled if weak corr. is on) = ' 0
=============================================
'Lepton-pair invariant mass minimum = ' 66d0
'Lepton-pair invariant mass maximum = ' 116d0
'Transverse mass minimum           = ' 0d0
'Transverse mass maximum           = ' 7000d0
'Z pT minimum                       = ' 0d0
'Z pT maximum                       = ' 7000d0
'Z rapidity minimum                 = ' -2.4d0
'Z rapidity maximum                 = ' 2.4d0
'Lepton pT minimum                  = ' 20d0
'Lepton pT maximum                  = ' 7000d0
'Anti-lepton pT minimum             = ' 20d0
'Anti-lepton pT maximum             = ' 7000d0
'pT min for softer lepton           = ' 0d0
'pT max for softer lepton           = ' 7000d0
'pT min for harder lepton           = ' 0d0
'pT max for harder lepton           = ' 7000d0
Taking absolute value of lepton pseudorapidity?
'(yes = 1, no = 0) 		    = ' 1
'Lepton pseudorapidity minimum      = ' 0d0
'Lepton pseudorapidity maximum      = ' 2.5d0
Taking absolute value of anti-lepton pseudorapidity?
'(yes = 1, no = 0) 		    = ' 1
'Anti-lepton pseudorapidity minimum = ' 0d0
'Anti-lepton pseudorapidity maximum = ' 2.5d0
Taking absolute value of soft lepton pseudorapidity?
'(yes = 1, no = 0)                  = ' 1
'Softer lepton pseudorapidity min   = ' 0d0 
'Softer Lepton pseudorapidity max   = ' 10000d0
Taking absolute value of hard lepton pseudorapidity?
'(yes = 1, no = 0)                  = ' 1
'Harder lepton pseudorapidity min   = ' 0d0
'Harder Lepton pseudorapidity max   = ' 10000d0
PHOTON RECOMBINATION-----------------------------
'DeltaR sep. for photon recomb.     = ' 0d0
'Minimum pT for observable photon   = ' 10d0
'Maximum eta for observable photon  = ' 2.5d0
PHOTON CUTS--------------------------------------
'Minimum Number of Photon           = ' 0
'Maximum Number of Photon           = ' 1
JET DEFINITION-------------------------------
Jet Algorithm & Cone Size ('ktal'=kT algorithm, 'aktal'=anti-kT algorithm, 'cone'=cone)
'ktal, aktal or cone		    = ' ktal
'Jet algorithm cone size (deltaR)   = ' 0.4d0
'DeltaR separation for cone algo    = ' 1.3
'Minimum pT for observable jets     = ' 20d0
'Maximum eta for observable jets    = ' 4.5d0
JET CUTS--------------------------------------
'Minimum Number of Jets		    = ' 0
'Maximum Number of Jets		    = ' 2
'Min. leading jet pT                = ' 0d0
ISOLATION CUTS-------------------------------
'Lep-Anti-lep deltaR minimum        = ' 0.0d0
'Lep-Anti-lep deltaPhi min	    = ' 0.0d0
'Lep-Anti-lep deltaPhi max	    = ' 4.0d0
'Lep-Jet deltaR minimum             = ' 0.0d0
'Lep-Photon deltaR minimum          = ' 0.0d0
=============================================
Cut on Z rapidity for well-defined Collins-Soper Angles at pp Collider
'Z rapidity cutoff for CS frame     = ' 0.0d0
=============================================
(See manual for complete listing)
'PDF set =                        ' 'ABMP16_5_nnlo'
'Turn off PDF error (1=Yes, 0=No)    = ' 0
(Active for MSTW2008 only, if PDF error is on:)
(Compute PDF+as errors: 1; just PDF errors: 0)
'Which alphaS                       = ' 0
(Active for MSTW2008 only; 0: 90 CL for PDFs+alphas, 1: 68 CL)
'PDF+alphas confidence level        = ' 1
=============================================

\end{verbatim}
}

\section{MATRIX runcard}
\label{sec:appD}

The runcard used for the computations with {\tt MATRIX} (version 2.1) for $Z$-boson production with central leptons 
at $\sqrt{s}=7$\ TeV in Sec.~\ref{sec:benchmark}.

{\scriptsize
\begin{verbatim}
##########################
# MATRIX input parameter #
##########################

#----------------------\
# General run settings |
#----------------------/
process_class   =  pp-emep+X   #  process id
E               =  3500.       #  energy per beam
coll_choice     =  1           #  (1) PP collider; (2) PPbar collider
photon_induced  =  0           #  switch to turn on (1) and off (0) photon-induced contributions
enhance_tails   =  1           #  switch to improve statistics in tail of distributions (a factor of two slower)


#----------------\
# Scale settings |
#----------------/
scale_ren         =  91.1876   #  renormalization (muR) scale
scale_fact        =  91.1876   #  factorization (muF) scale
dynamic_scale     =  0         #  dynamic ren./fac. scale
                               #  0: fixed scale above
                               #  1: invariant mass (Q) of system (of the colourless final states)
                               #  2: transverse mass (mT^2=Q^2+pT^2) of system (of the colourless final states)
factor_central_scale = 1       #  relative factor for central scale (important for dynamic scales)
scale_variation   =  1         #  switch for muR/muF uncertainties (0) off; (1) 7-point (default); (2) 9-point variation
variation_factor  =  2         #  symmetric variation factor; usually a factor of 2 up and down (default)


#------------------------------\
# Order-dependent run settings |
#------------------------------/
# LO-run
run_LO             =  1        #  switch for LO cross section (1) on; (0) off
LHAPDF_LO          =  ABMP16_5_nnlo  #  LO LHAPDF set
PDFsubset_LO       =  0        #  member of LO PDF set
precision_LO       =  1.e-4    #  precision of LO cross section

# NLO-run
run_NLO_QCD        =  1        #  switch for NLO QCD cross section (1) on; (0) off
run_NLO_EW         =  0        #  switch for NLO EW cross section (1) on; (0) off 
LHAPDF_NLO         =  ABMP16_5_nnlo  #  NLO LHAPDF set
PDFsubset_NLO      =  0        #  member of NLO PDF set
precision_NLO_QCD  =  3.e-4    #  precision of NLO QCD cross section
precision_NLO_EW   =  3.e-4    #  precision of NLO EW correction
NLO_subtraction_method = 1     #  switch to use (2) qT subtraction (1) Catani-Seymour at NLO

# NNLO-run
run_NNLO_QCD       =  1        #  switch for NNLO QCD cross section (1) on; (0) off 
add_NLO_EW         =  0        #  switch to add NLO EW cross section to NNLO run (1) on; (0) off
                               #  note: can be added only if also running NNLO
LHAPDF_NNLO        =  ABMP16_5_nnlo  #  NNLO LHAPDF set
PDFsubset_NNLO     =  0        #  member of NNLO PDF set
precision_NNLO_QCD =  1.5e-4    #  precision of NNLO QCD cross section
precision_added_EW =  3.e-4    #  precision of NLO EW correction in NNLO run
switch_qT_accuracy =  1        #  switch to improve qT-subtraction accuracy (slower numerical convergence)
                               #  0: lowest value of r_cut = 0.0015 varied up to 0.01 (default)
                               #  1: lowest value of r_cut = 0.0005 varied up to 0.01
                               #  2: lowest value of r_cut = 0.0001 varied up to 0.01 (only if extrapolate_binwise=1;
                               #     output of fixed-r_cut result remains 0.0005, while 0.0001 used for extrapolation)
                               #     for Drell-Yann it is recommended to turn on power_corrections 
                               #     rather than use switch_qT_accuracy
power_corrections  =  1        #  switch to include leading power corrections in qT-subtraction through recoil
                               #  (not recommended for processes involving photons and heavy quarks)
power_corrections_pT0 = 20.    #  characteristic transverse momentum pT0 used to optimise the generation
                               #  of the phase space for the integration of the power corrections. It should be set
                               #  to the minimum requirement on the transverse momentum of the 2-body final state
                               #  (for Drell-Yan for instance this should be the minimum transverse momentum 
                               #  of the leptons)
extrapolate_binwise = 1        #  switch for bin-wise r_cut extrapolation of distributions
                               #  (note: increases written output for distributions by factor of 8)


#----------------------------\
# Settings for fiducial cuts |
#----------------------------/
# Jet algorithm
jet_algorithm = 3              #  (1) Cambridge-Aachen (2) kT (3) anti-kT
jet_R_definition = 0           #  (0) pseudo-rapidity (1) rapidity
jet_R = 0.4                    #  DeltaR

# Photon recombination (lepton dressing)
photon_recombination = 1       #  switch for photon recombination (1) on; (0) off; must be on for EW runs
photon_R_definition = 1        #  (0) pseudorap; (1) rapidity
photon_R = 0.1                 #  DeltaR: photon combined with charged particle when inside this radius

# Jet cuts
define_pT jet = 25.            #  requirement on jet transverse momentum (lower cut)
define_eta jet = 4.5           #  requirement on jet pseudo-rapidity (upper cut)
define_y jet = 1.e99           #  requirement on jet rapidity (upper cut)
n_observed_min jet = 0         #  minimal number of observed jets (with cuts above)
n_observed_max jet = 99        #  maximal number of observed jets (with cuts above)

# Lepton cuts
define_pT lep = 20.            #  requirement on lepton transverse momentum (lower cut)
define_eta lep = 2.5           #  requirement on lepton pseudo-rapidity (upper cut)
define_y lep = 1.e99           #  requirement on lepton rapidity (upper cut)
n_observed_min lep = 2         #  minimal number of observed leptons (with cuts above)
n_observed_max lep = 99        #  maximal number of observed leptons (with cuts above)

# Negatively-charged lepton cuts
define_pT lm = 0.              #  requirement on negatively-charged lepton transverse momentum (lower cut)
define_eta lm = 1.e99          #  requirement on negatively-charged lepton pseudo-rapidity (upper cut)
define_y lm = 1.e99            #  requirement on negatively-charged lepton rapidity (upper cut)
n_observed_min lm = 0          #  minimal number of observed negatively-charged leptons (with cuts above)
n_observed_max lm = 99         #  maximal number of observed negatively-charged leptons (with cuts above)

# Positively-charged lepton cuts
define_pT lp = 0.              #  requirement on positively-charged lepton transverse momentum (lower cut)
define_eta lp = 1.e99          #  requirement on positively-charged lepton pseudo-rapidity (upper cut)
define_y lp = 1.e99            #  requirement on positively-charged lepton rapidity (upper cut)
n_observed_min lp = 0          #  minimal number of observed positively-charged leptons (with cuts above)
n_observed_max lp = 99         #  maximal number of observed positively-charged leptons (with cuts above)

####################
# User-defined cuts
# (only used if defined in 'MATRIX/prc/$process/user/specify.cuts.cxx')
#
user_switch M_leplep = 1            #  switch to turn on (1) and off (0) cuts on lepton-lepton invariant mass
user_cut min_M_leplep = 66.         #  requirement on lepton-lepton invariant mass (lower cut)
user_cut max_M_leplep = 116.        #  requirement on lepton-lepton invariant mass (upper cut)

user_switch R_leplep = 0            #  switch to turn on (1) and off (0) cuts on lepton-lepton separation
user_cut min_R_leplep = 0.          #  requirement on lepton-lepton separation in y-phi-plane (lower cut)

user_switch lepton_cuts = 0         #  switch to turn on (1) and off (0) cuts on leptons
user_cut min_pT_lep_1st = 25.       #  requirement on hardest lepton transverse momentum (lower cut)
user_cut min_eta_lep_1st = 0.       #  requirement on hardest lepton pseudo-rapidity (lower cut)
user_cut max_eta_lep_1st = 1.e99    #  requirement on hardest lepton pseudo-rapidity (upper cut)
user_cut min_pT_lep_2nd = 15.       #  requirement on second-hardest lepton transverse momentum (lower cut)
user_cut min_eta_lep_2nd = 0.       #  requirement on second-hardest lepton pseudo-rapidity (lower cut)
user_cut max_eta_lep_2nd = 1.e99    #  requirement on second-hardest lepton pseudo-rapidity (upper cut)
user_cut min_eta_one_lep = 0.       #  requirement on one of the two leptons (lower cut)
user_cut max_eta_one_lep = 1.e99    #  requirement on one of the two leptons (upper cut)
user_cut min_eta_other_lep = 0.     #  requirement on the other lepton (lower cut)
user_cut max_eta_other_lep = 1.e99  #  requirement on the other lepton (upper cut)

####################
# Fiducial cuts
# (defined via general interface)
#

#-----------------\
# MATRIX behavior |
#-----------------/
max_time_per_job = 24          #  very rough time(hours) one main run job shall take (default: 24h)
                               #  unreliable when < 1h, use as tuning parameter for degree of parallelization
                               #  note: becomes ineffective when job number > max_nr_parallel_jobs
                               #        which is set in MATRIX_configuration file
switch_distribution = 1        #  switch to turn on (1) and off (0) distributions
save_previous_result = 1       #  switch to save previous result of this run (in result/"run"/saved_result_$i)
save_previous_log = 0          #  switch to save previous log of this run (in log/"run"/saved_result_$i)
#include_pre_in_results = 0    #  switch to (0) only include main run in results; (1) also all extrapolation (pre) runs;
                               #  crucial to set to 0 if re-running main with different inputs (apart from precision)
                               #  note: if missing (default) pre runs used if important for precision
                               #  (separately for each contribution)
reduce_workload = 0            #  switch to keep full job output (0), reduce (1) or minimize (2) workload 
                               #  on slow clusters
random_seed = 0                #  specify integer value (grid-/pre-run reproducible)

\end{verbatim}
}

\section{Set-up for \texttt{SCETlib}}
\label{sec:appE}

The relevant physical settings and cuts on $Q \equiv m_{ll}$ and $Y \equiv Y_{l l}$
used for the matched predictions in Sec.~\ref{sec:resummation}
are given by the \texttt{SCETlib} input file below.
The cuts on the lepton transverse momenta are described in the main text.
Note that we always cut on the lepton pseudorapidities
$\eta_{l_1,l_2} \leq 2.5$ in addition.
{\scriptsize
\begin{verbatim}
[Calculation_settings]
recoil_scheme = collins_soper
profile_functional_form = slope
muf_max = 7000.
lambda = 1.
transition_points: [0.3, 0.6, 0.9]
muFO_fixed = 91.1876

[QCD]
nf = 5
alphas_mu0 = 0.1147
mu0 = 91.1876
alphas_order = n3ll
pdf_set = ABMP16_5_nnlo
pdf_member = 0
Ecm = 7000.

[Electroweak]
alphaem = 0.007565201876112782
# = 1./0.13218417913704487887E+03
sin2_thw = 0.22301322532678335975
mZ = 91.1876
GammaZ = 2.4952

[Process]
boson = Z

[Grid_Q]
min = 46
max = 150
steps = 1
bins = yes

[Grid_Y]
min = -2.5
max = 2.5
steps = 1
bins = yes

[Grid_qT]
min = 0.
max = 150.
steps = 1
bins = yes
\end{verbatim}
}


\end{document}